\documentclass[twocolumn]{aastex701}

\usepackage{enumitem}
\usepackage{amsmath}
\usepackage{graphicx}
\usepackage{subcaption}
\usepackage{alphabeta}
\usepackage{sidecap}
\usepackage{upgreek}

\newcommand{\f}{\frac}
\newcommand{\p}{\partial}

\newcommand{\nuin}{\nu_{\rm in}}
\newcommand{\nuinmax}{|\nuin|_{\rm max}}
\newcommand{\ainit}{a_{\rm init}}
\newcommand{\risco}{r_{\rm ISCO}}
\newcommand{\rchar}{r_{\rm char}}
\newcommand{\rchnine}{r_{\rm char,9}}
\newcommand{\alphach}{\upalpha_{\rm char}}
\newcommand{\rhard}{r_{\rm hard}}
\newcommand{\thard}{t_{\rm hard}}
\newcommand{\rg}{{\rm R_g}}
\newcommand{\agw}{a_{\rm GW}}
\newcommand{\agwrg}{a_{\rm GW}}
\newcommand{\agwcm}{a_{\rm GW[cm]}}
\newcommand{\anine}{a_{\rm GW,9}}
\newcommand{\aninerg}{a_{\rm GW,9}}
\newcommand{\aninecm}{a_{\rm GW,9[cm]}}
\newcommand{\alphagw}{{\upalpha}_{\rm GW}}
\newcommand{\betagw}{\beta_{\rm GW}}
\newcommand{\msun}{{\rm M}_{\odot}}
\newcommand{\inv}{^{-1}}

\newcommand{\dadt}{\dot a}
\newcommand{\dadtrc}{\dadt(\rchar)}
\newcommand{\ctin}{\texttt{A0}}
\newcommand{\cagw}{\texttt{B0}}
\newcommand{\ctinstar}{\texttt{Astar}}
\newcommand{\cagwstar}{\texttt{Bstar}}
\newcommand{\ctingas}{\texttt{Agas}}
\newcommand{\cagwgas}{\texttt{Bgas}}
\newcommand{\aninevar}{\texttt{-agw9var}}
\newcommand{\alphvar}{\texttt{-{\alpha}var}}
\newcommand{\betavar}{\texttt{-{\beta}var}}
\newcommand{\nuvar}{\texttt{-nuvar}}
\newcommand{\toutvar}{\texttt{-toutvar}}
\newcommand{\rchninevar}{\texttt{-rch9var}}

\newcommand{\tauingw}{\tau_{\rm in+gw}}
\newcommand{\dadtmax}{\left|\dadt\right|_{\rm max}}
\newcommand{\thub}{t_{\rm Hub}}

\newcommand{\rev}{}

\shorttitle{An ``inside-out" SMBH binary inspiral model}
\shortauthors{L. Blecha}

\begin{document}

\title{An ``inside-out" approach to modeling supermassive black hole binary inspiral}

\author[orcid=0000-0002-2183-1087]{Laura Blecha}
\affiliation{University of Florida, Department of Physics}
\email[show]{lblecha@ufl.edu}

\begin{abstract}
The inspiral and merger of two supermassive black holes (SMBHs) releases immense energy in low-frequency gravitational waves (GWs). Recent pulsar timing array (PTA) observations of the stochastic nHz GW background (GWB) are consistent with a SMBH binary origin. GW data from PTAs and from the upcoming Laser Interferometer Space Antenna (LISA) can probe late-stage binary evolution where electromagnetic constraints are scarce. However, the complexity of the relevant astrophysics necessitates a well optimized approach. I argue that a key physical quantity to constrain with PTAs is the orbital semi-major axis at which binary inspiral transitions from the astrophysical to the GW-dominated regime ($\agw$). This quantity should be treated as a free parameter in analysis of the GWB. Using this premise, I present a simple analytic framework for modeling SMBH binary inspiral in an “inside-out” fashion. At orbital separations slightly larger than $\agw$, a power-law scaling for the astrophysical inspiral timescale is assumed, while the outermost phase (prior to the PTA regime) is simply modeled via a delay time. I show that the GWB spectral shape and amplitude are most sensitive to $\agw$ (normalized to $10^9\,\msun$, equal-mass binaries and expressed in gravitational units), with weaker dependence on the inner power-law index and the outer delay time. The GWB is largely insensitive to the mass and mass-ratio scaling of $\agw$ and to the boundary between the inner and outer astrophysical inspiral regimes. I compare with models for gas- and stellar-driven binary inspiral and discuss implications for LISA and for GW source parameter inference.
\end{abstract}


\section{Introduction}

Supermassive black hole binaries (SMBHBs) form after the merger of two galaxies and should therefore be common throughout the Universe. However, electromagnetic (EM) evidence for SMBHBs remains very scarce \citep[e.g.,][]{burke11, eracleous12, charisi16, sesana18, bogdanovic22}. Pulsar timing arrays (PTAs) recently \rev{found evidence} for a stochastic, nanoHertz gravitational-wave  background \citep[GWB;][]{agazie23a,epta23, reardon23, xu23} that is consistent with a cosmic population of SMBHBs \citep{agazie23e, epta24astro}. 

Using the 15-yr data set of the North American Nanohertz Observatory for Gravitational Waves \citep[NANOGrav][]{agazie23a}, \citet[][hereafter NG15Astro]{agazie23e} presented constraints on SMBHBs. Although this marks a major advance in understanding SMBH evolution, significant challenges remain. The NANOGrav 15-yr data appear to favor more massive and/or numerous SMBHBs than inferred from EM observations of the galaxy stellar mass function (GSMF), the SMBH-galaxy scaling relations, and the galaxy merger rate. Numerous explanations have been proposed, including revisions to the high-mass tail of the GSMF \citep{sato23, liepold24}, redshift evolution of the SMBH-galaxy relations \citep{matt26a, matt26b}, and GWs arising from physics beyond the Standard Model \citep[][]{afzal23}. In addition, the NG15Astro parameter posteriors are broad and in some cases prior-dominated. \citet{laal25} recently developed a normalizing flow emulator that outperforms previous methods for connecting the GWB to underlying physical models. Future PTA data sets that can better characterize the nHz GWB, in conjunction with improvements to SMBHB population models, will enhance the physical information extracted from PTA observations. 

Low-frequency GW observations are uniquely well suited to probe late-stage SMBHB inspiral and are thus complementary to EM constraints. \rev{In particular, if SMBHBs are driven through the low-frequency PTA regime by astrophysical processes (e.g., interactions with their stellar or gaseous environment), the GWB amplitude at those frequencies will be attenuated relative to the case of purely GW-driven inspiral. The steepness of the GW-driven inspiral rate ($\propto a^4$) indicates that at some point prior to merger, SMBHBs decouple from their astrophysical environment and transition to purely GW-driven inspiral. Thus, the shape of the low-frequency GWB encodes the transition from the astrophysical to the GW regime \citep[e.g.,][]{kocsis11,sampson15,chen2017,kelley17a,arzoumanian2018}. This low-frequency attenuation or turnover is sometimes characterized via a broken power-law or other analytic form for the GWB spectral shape \citep{sampson15,chen2017,arzoumanian2018}.} The focus of this work is designing a practical approach to modeling SMBHB inspiral that \rev{highlights key physical parameters controlling both the shape and amplitude of the GWB, and that} is optimized for parameter inference with PTAs.

A key issue is the highly uncertain timescales for SMBHBs to form, inspiral through the PTA frequency band (where they are powerful GW emitters), and merge. Central to this is the problem of dynamic range---the roughly ten orders of magnitude in separation that SMBHs must traverse from galaxy infall to SMBH merger. SMBHs follow their host galaxies as a galaxy merger begins on $\sim 100$s of kpc scales. After the galaxy nuclei coalesce, SMBHs inspiral further via dynamical friction until they form a gravitationally-bound system at separations $\lesssim 1$ pc (for low-mass SMBHBs) to $\gtrsim10$-100 pc (for massive PTA binaries). At this stage, the time for the SMBHB to merge via GW emission alone is much greater than \rev{the age of the Universe} ($\thub$), even for massive binaries. Further interactions with the astrophysical environment are needed to reach the GW regime on $\sim$mpc scales.

\citet{begelman80} showed that SMBHB merger timescales depend strongly on the stellar and gaseous environment of their host galaxies, and that some SMBHBs might ``stall" at $\sim$ parsec (pc) scales for more than $\thub$---the so-called ``final-parsec problem." While many viable solutions to the final-parsec problem have been presented \cite[e.g.,][]{escala05, holley06, berczik06, mayer07, khan13, holley15}, SMBHB inspiral rates remain highly uncertain and could range from $\lesssim10^6$ yr to $\gg\thub$ \cite[e.g.,][]{volonteri03, sesana09, barausse12, colpi14, kelley17a, ricarte18, katz20}. This limits our ability to constrain SMBHB populations using PTAs and to make predictions for the Laser Interferometer Space Antenna \citep[LISA][]{amaro23}. Planned for launch in the next decade, LISA will be sensitive to mHz GWs from mergers of lower-mass SMBHBs out to $z\sim20$ \citep{amaro17}. 

SMBHB inspiral is often called SMBHB hardening, and a ``hard" binary is one for which $v_{\rm orb}$ exceeds the stellar velocity dispersion $\sigma_*$, such that individual stellar scattering events will shrink the binary orbit. Following convention, I use ``inspiral" and ``hardening" interchangeably to refer to the shrinking of binary orbits, regardless of whether stars or other processes drive the evolution. The simplest SMBHB evolution models assume that all SMBHBs efficiently reach the GW regime after galaxy mergers, through unspecified means, and that only GW emission drives SMBHB inspiral through the relevant GW frequency bands \citep[e.g.,][]{phinney01, wyithe03a, sesana08, roebber16}. The galaxy merger rate 
can be taken from a semi-analytic model or derived from empirical constraints.

Studies of astrophysically driven SMBHB inspiral often focus on either gaseous or stellar processes. Collisionless $N$-body simulations model SMBHB evolution via scattering events with stars in the binary's ``loss cone." Stalling occurs if stars are ejected from the loss cone faster than they are replenished, in the absence of other hardening mechanisms \citep[e.g.,][]{quinlan96, yu02, merritt05, sesana06}. 
Triaxiality, asymmetry, and rotation of stellar potentials can refill the loss cone and drive SMBHBs to the GW regime \cite[e.g.,][]{berczik06, holley06, khan11, khan13, holley15, gualandris16}. 

A similarly rich body of literature explores SMBHB inspiral driven  by circumbinary gas disks, often using (magneto)hydrodynamics simulations \citep[e.g.,][]{armnat02, escala05, mayer07, macfadyen08}.
The efficiency of circumbinary disk hardening is sensitive to the detailed structure and properties of the accretion disk, as well as binary mass ratio and eccentricity \citep[e.g.,][]{shi12, farris14, lai23}. 
Binaries may even undergo ``softening" (orbital expansion) rather than hardening in some cases \citep[e.g.,][]{munoz19, moody19, munoz20, tiede20}. Simulations also show the existence of an equilibrium orbital eccentricity that depends on mass ratio 
\citep{roedig11, dorazio21, zrake21, siwek23b}.

In addition to their dynamical effects, circumbinary gas disks can fuel gas accretion onto SMBHs, producing electromagnetic (EM) counterparts to GW sources 
\citep[e.g.,][]{milphi05, noble12, tang18, bogdanovic22}. 
Detection of a multi-messenger SMBHB system would have far-reaching implications for astrophysics and cosmology \cite[e.g.,][]{amaro23}. In unequal-mass systems, the secondary (less massive) SMBH accretes preferentially more, driving the binary mass ratio toward unity and increasing the GW signal \citep[e.g.,][]{farris14, duffell20, siwek23a, comerford25}. This can also enhance EM variability signatures associated with binary orbital motion \cite[e.g.,][]{dorazio15, duffell20}.

Triple SMBH interactions provide another channel for SMBHB hardening, as chaotic or resonant three-body interactions can greatly reduce the merger time \citep[][]{hofloe07, bonetti16, bonetti18a, bonetti18b, bonetti19}. Even if all binaries stall, \citet{bonetti18b} find that SMBH inspiral mediated by triple interactions should produce an observable GWB. 
Triple interactions and GW recoil can also eject SMBHs from galactic nuclei, further complicating the association between galaxy mergers and SMBH mergers \citep[e.g.,][]{volonteri07, hofloe07, holley08, blecha11, satheesh25}.

Ideally, cosmological SMBHB population models will account for multiple hardening processes, including dynamical friction, stellar scattering, gas drag, and GW emission. Such frameworks have been constructed using SMBHB host properties from either simulations \rev{\citep[e.g.,][]{kelley17a, kelley17b, katz20, li23, siwek24, sah24, harris26, francois26}} or semi-analytic models \citep[e.g.,][these also include SMBH triples]{izquierdo22, bonoli25}. Hydrodynamics simulations provide detailed information about galactic environment and internal structure, but they are computationally expensive, subject to resolution limits, and rely on sub-grid or post-processing SMBH prescriptions. Semi-analytic models are much more computationally efficient but are even more reliant on semi-analytic prescriptions for the processes governing galaxy and SMBH evolution. 

\rev{Recent work has developed cosmological simulation techniques that include on-the-fly sub-grid evolution of SMBHB hardening and GW emission. KETJU directly integrates the dynamics of SMBHs and stellar particles, including post-Newtonian corrections, in the GADGET-4 code \citep{rantala17, mannerkoski23}. RAMCOAL is a subgrid model implemented in the RAMSES hydrodynamical simulation code that evolves SMBHB inspiral from tens of parsec scales down to coalescence; it also includes a prescription for triple SMBH evolution \citep{li2025_ramcoal, li2026_ramcoal}. } 

\rev{E}ven simple SMBHB models have many free parameters. Those controlling large-scale evolution can be constrained with EM observations (e.g., the GSMF, galaxy merger rates, \& BH-galaxy relations), but observations of bound SMBHBs are very scarce \citep[][]{rodriguez06, burke11}. Thus, SMBHB evolution through the PTA frequency band is the regime that is least accessible to EM observations. 

The PTA evidence for a stochastic, nHz GW background provides an unprecedented opportunity to characterize the SMBHB population. In NG15Astro, the SMBHB population synthesis code \texttt{holodeck} was used to analyze the NANOGrav 15-yr data set. Six parameters were varied: two controlling the GSMF, two controlling the $M_{\rm BH}$-$M_{\rm bulge}$ relation, and two controlling SMBHB inspiral. The need for many realizations of SMBHB populations in order to build reliable parameter posteriors, and the need to avoid overfitting the data, places stringent constraints on the complexity of viable SMBHB inspiral models \citep{laal2026}.

NG15Astro introduced an analytic fixed-timescale, double-power-law  model for SMBHB hardening (hereafter the ``2PL" hardening model), motivated by the simulation-based models of \citet{kelley17a, kelley17b, kelley18}. 
The disadvantage of this approach is that the fixed total timescale inextricably ties small-scale inspiral to evolution on much larger scales, well outside the PTA regime. 

More broadly, careful selection of parameters to which PTAs are most sensitive will be crucial to maximize SMBHB constraints from the GWB. I argue for an approach that focuses on SMBHB hardening at small scales, corresponding to observable GW frequencies. I present an analytic model for SMBHB hardening that achieves this goal while maintaining similar form and simplicity as the 2PL model of NG15Astro. \rev{As in NG15Astro, circular orbits are assumed throughout this work; extensibility to eccentric orbits is discussed in Section~\ref{ssec:ecc} and will be explored further in future work.} 

I highlight a quantity to which the GWB amplitude and spectral shape are particularly sensitive: $\agw$, the orbital semi-major axis at which binary inspiral transitions from the astrophysical to the GW-dominated regime. I argue that $\agw$ should  be treated as a free parameter to be constrained with PTAs. 
This model centers on the fact that the innermost SMBHB inspiral regime determines how much GW energy is emitted at each PTA frequency, while the evolution at larger separations determines which binaries reach the PTA regime, and when. \rev{This approach is distinct from spectral models in which the break frequency is treated as a key parameter \citep[][]{sampson15,chen2017,arzoumanian2018}. $\agw$ is a physical quantity associated with each SMBHB inspiral, and I will show that this parameter strongly impacts both the shape and amplitude of the GWB. In addition, this approach is agnostic to the environmental processes driving inspiral prior to the transition to the GW regime. The model requires only the orbital separation of the transition, the environmental hardening rate at larger separations within the PTA band, and the delay time between galaxy merger and reaching the PTA regime.}
Because I treat this ``outer" inspiral regime  simply as a means of connecting the ``inner" regime to a cosmological SMBHB population at large scales, I refer to this as an ``inside-out" model for SMBHB inspiral. 

This paper is organized as follows. In Sections~\ref{ssec:gwonly} \&~\ref{ssec:2PL}, respectively, I review relevant aspects of analytic prescriptions for SMBHB inspiral via GW emission only and using the 2PL model. Section~\ref{ssec:newhard} presents the inside-out SMBHB inspiral model and describes the model variants used in this work. Section~\ref{ssec:paramspace} outlines criteria that limit the physical regions of parameter space. Sections~\ref{ssec:fiducial} \&~\ref{ssec:cagw} present the results for two fiducial models, and Sections~\ref{ssec:a9} \&~\ref{ssec:alphabetavar} describe model dependence on $\agw$. Section~\ref{ssec:nuinvar} describes dependence on the power-law scaling of the inner astrophysical hardening rate, $\nuin$, while Section~\ref{ssec:outer} discusses the parameters that connect the inner and outer astrophysical regimes. Section~\ref{ssec:litcomparison} discusses connections to models for gas- and stellar-driven hardening. Caveats and applications beyond PTA science are discussed in Section~\ref{sec:discuss}, and I summarize and conclude in Section~\ref{sec:conclude}. Throughout this paper a WMAP9 cosmology is assumed with $\Omega_m=0.288$, $\Omega_b=0.0472$, and \rev{$H_0=69.33$} km s$^{-1}$ Mpc$^{-1}$.

\section{Analytic prescriptions for SMBH binary hardening}
\label{sec:hardmodels}
I summarize relevant aspects of two analytic prescriptions for SMBHB evolution and then present the inside-out hardening model that is central to this work.

\subsection{GW-only binary hardening}
\label{ssec:gwonly}
Although it is well-known that SMBHBs must interact with their astrophysical environment to reach the GW regime, the simplicity of modeling SMBHB inspiral via GW emission alone makes this a widely used method in the literature. Such calculations follow the logic of \citet{phinney01} starting from a number density of merger remnants, using the fact that the GW energy emitted from all mergers must sum to the cosmic GW energy density (per log frequency interval). 
This is the most essentially ``inside-out" approach to modeling SMBHB hardening, and it is the origin of the $h_c(f) \propto f^{-2/3}$ power-law scaling of the characteristic GW strain used in idealized models of the GWB. (The corresponding pulsar-timing-residual power spectral density scales as $f^{-13/3}$.) 

This approach does not require knowledge of how SMBHBs evolve to the GW regime, except that they must reach it much faster than via GWs alone. 
Because it works backwards from a merger event rate, however, this approach breaks down if the GWB includes emission from SMBHBs that will not merge by $z=0$. In other words, binaries cannot contribute to the merger number density if they enter the low-frequency edge of the observed GWB with a time to merger of $\tau_{\rm GW} > \thub$ (or more precisely, if they have $\tau_{\rm GW} > t_{\rm look}(z)$, the lookback time for a binary forming at redshift $z$). A GWB calculation that includes GW emission from such binaries is thus inherently inconsistent.

\begin{figure}
    \includegraphics[width=\columnwidth]{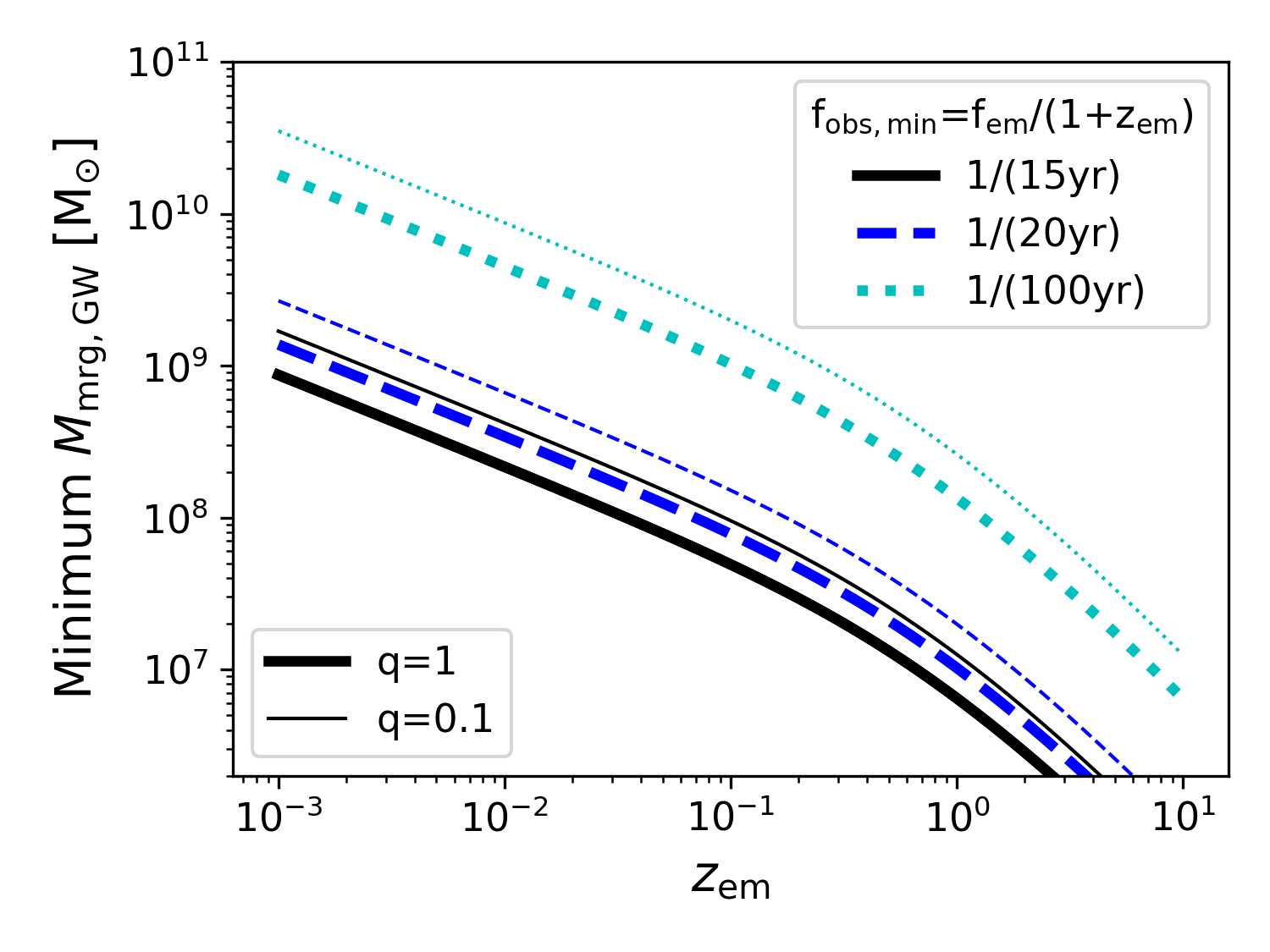}
\caption{Minimum required mass for circular SMBHBs to merge by $z=0$, starting from $z_{\rm em}(f_{\rm obs,min})$, via GW emission alone. Curves are shown for $f_{\rm obs,min}= (15\,{\rm yr})\inv$, $(20\,{\rm yr})\inv$, \& $(100\,{\rm yr})\inv$. They indicate the SMBHB masses that would be included in a self-consistent GWB calculation based on a SMBHB merger rate as in \citet{phinney01}, if all binaries are evolved from the low-frequency edge of the PTA band via GW emission alone. Lower-mass binaries cannot contribute to a self-consistent GWB calculation {\em any} frequency $>f_{\rm obs,min}$, unless astrophysical hardening is included along with GW-driven hardening.}
\label{fig:gwonly}
\end{figure}
It is useful to quantify the limits in which this GW-only calculation for SMBHB inspiral remains self-consistent at least within the range of observable PTA frequencies---arguably the minimal requirement for a valid ``inside-out" model of SMBHB hardening for PTA data. Consider a circular SMBHB with total mass $M=m_1+m_2$, mass ratio $q\equiv m_2/m_1<1$, and symmetric mass ratio $\eta\equiv q/(1+q)^2$ 
that has an initial separation $a$ and is emitting GWs at a frequency $f_{\rm em}=2f_{\rm orb}$. Driven by GW emission only, this binary will inspiral to merger within a time \citep{peters64}
\begin{align}
\tau_{\rm GW} &= \f{5 a^4 c^5}{\rev{256\,} G^3 M^3 \eta} \nonumber \\
&= \rev{8.16\times10^3} \: {\rm yr} \: M_9^{-5/3} \left(4\eta\right)^{-1} \left( \f{f_{\rm em}}{{\rm yr}^{-1}} \right)^{-8/3},
\label{eqn:tmrg}
\end{align}
where $M_9 \equiv M/(10^9\msun)$. Thus, circular SMBHBs at a redshift $z_{\rm em}$ will merge by $z=0$ only if
\begin{equation}
    M \!>\! \f{\rev{2.2}\times10^7\msun}{(4\eta)^{3/5}}\!\left(\f{\thub}{t_{\rm look}(z_{\rm em})}\right)^{\!\!3/5}\!\!\!\rev{\left( \f{f_{\rm obs}(1+z_{\rm em})}{(20\,{\rm yr})^{-1}} \right)^{\!\!-8/5}}\!,
\label{eqn:gwonly_mass_limit}
\end{equation}
where $t_{\rm look}(z_{\rm em})$ is the lookback time for a source at redshift $z_{\rm em}$, and \rev{$f_{\rm obs}\equiv f_{\rm em}/(1+z)$}. Setting $f_{\rm obs}=f_{\rm obs,min}=1/T_{\rm obs}$ in Equation~\ref{eqn:gwonly_mass_limit}
provides an estimate of the binary masses that could be self-consistently included in a (circular) GW-only inspiral model for a GWB observed over a time $T_{\rm obs}$. Figure~\ref{fig:gwonly} shows the minimum mass versus redshift \rev{for $T_{\rm obs}=$ 15, 20, \& 100 yr. 
Below this mass,} circular binaries driven by GW emission alone, from an orbital period of $\tau_{\rm orb} = 2/f_{\rm obs}$, cannot contribute to the GW-only GWB calculation at any frequency, or to SMBHB merger rates. Put another way, if circular binaries below this mass contribute nonnegligibly to the GWB, even just at the highest frequencies, they must pass through the observable lower-frequency bands faster than if they were driven by GW emission alone.

In practice, GW-only calculations of the GWB often use the number density of galaxy mergers, rather than the SMBH mergers themselves. If no additional delay is imposed, this means that binaries are brought to the GW regime on artificially short timescales. Self-consistency {\em within} the PTA regime in this case requires only that SMBHBs reach the high-frequency edge of the PTA band by $z=0$, rather than reaching merger by $z=0$. However, the difference in the constraints imposed by Equation~\ref{eqn:gwonly_mass_limit} is negligible, because $\tau_{\rm GW}$ will always be much smaller at the high-frequency edge of the PTA band than at the low-frequency edge.

In NG15Astro, \rev{binaries at $z_{\rm em}\sim 0.4$-$0.5$ have the highest fractional contribution to the GWB, weighted by their strain squared. At these redshifts, $M\sim 2$-$3\times10^7\, \msun$ (with $q=1$) is the minimum SMBHB mass that can be self-consistently included in a GW-only calculation of the GWB. (The minimum is $M\sim 4$-$5\times10^7\, \msun$ for $q=0.1$ binaries).
The NG15Astro analysis also finds that binaries $\lesssim$ a few $\times 10^7\, \msun$ contribute $\gtrsim0.1\%$} to the high-frequency GWB (weighted by $h_c^2$). 
This indicates that a GW-emission-only model for SMBHB hardening can be considered a valid model for current GWB data, insofar as precision at the \rev{$\sim 0.1\%$} level is not required. However, such a model cannot be used for lower-mass SMBH evolution (i.e., LISA predictions using PTA data).

\begin{SCfigure*}
\includegraphics[width=0.68\textwidth]{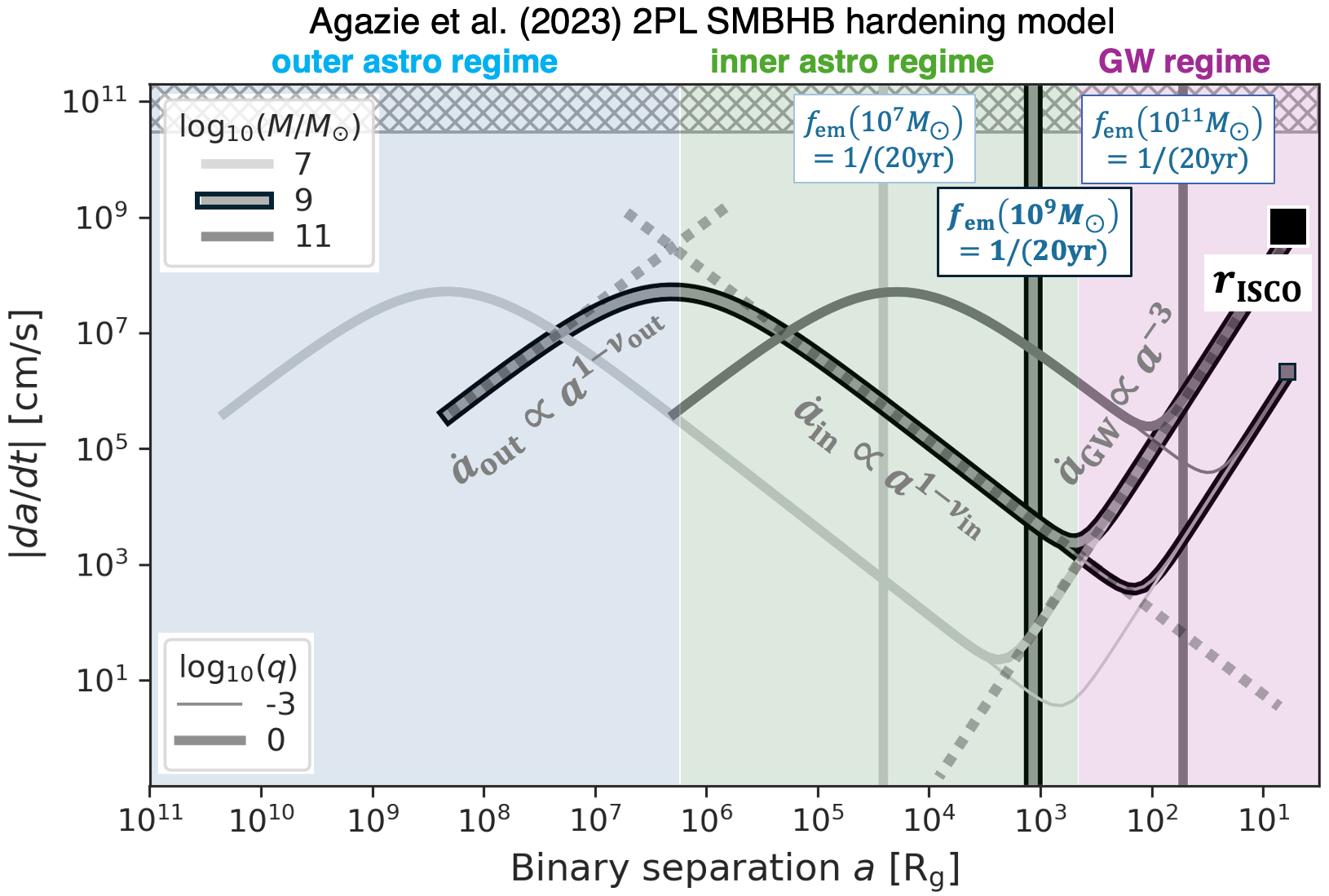}
\caption{\rev{Schematic illustrating key features of the 2PL SMBHB inspiral model used in NG15Astro. The hardening rate vs. separation is shown for select masses and mass ratios, for parameter values consistent with NG15Astro ($a_{\rm init}=10 {\rm kpc}$, $\rchar=100 {\rm pc}$, $\nu_{\rm out}=+2.5$, $\nuin=-0.5$, $\tau_{\rm tot}=1.0 {\rm Gyr}$). Binary separation decreases from left to right (the direction of forward time). The overall normalization of the hardening rate is set by $\tau_{\rm tot}$, which is often dominated by the outer astrophysical inspiral regime, many orders of magnitude outside the PTA regime.\\ } \\ \label{fig:twoPLschematic}}
\end{SCfigure*}

As the observation time of PTAs increases, the mass threshold for valid GW-only hardening models will increase $\propto T_{\rm obs}^{8/5}$. A GWB calculation using GW-only inspiral for a hypothetical  100-yr PTA data set would be self-consistent only for \rev{$M>3$-$4\times 10^8\, \msun$, if the GWB is dominated by $q=1$ binaries at $z_{\rm em}\sim0.4$-$0.5$ (with a higher mass threshold for $q<1$ or $z_{\rm em}<0.4$).}
In other words, a 100-yr PTA data set would require a SMBHB model that either a) includes astrophysically driven hardening mechanisms, or b) has zero GW contribution {\em at any frequency} from $\lesssim$ \rev{$10^9\, \msun$} SMBHs, if only GW emission is assumed to drive SMBHB hardening within the PTA band. Eccentric binaries merge faster and emit GWs at higher-frequency harmonics than assumed in Equation~\ref{eqn:tmrg}, but because GW emission efficiently circularizes binary orbits, invoking nonnegligible eccentricity again requires astrophysical hardening processes.

The above discussion highlights that current PTA observations are to a small extent already pushing beyond the regime in which the \citet{phinney01} GWB calculation using only GW-driven SMBHB inspiral is self-consistent, and the magnitude of this issue will only grow with time. This is separate from the limitations of a pure power-law GWB model arising from the discrete nature of sources and from the intrinsic non-Gaussianity of the GWB signal \citep[][]{sesana08, sesana09, ravi12}. It is also separate from, but related to, the important question of how astrophysical hardening modifies the shape and amplitude of the GWB spectrum, and it further motivates the need for a SMBHB hardening model that accounts for astrophysics without introducing an intractable number of free parameters.

\subsection{The 2PL hardening model} 
\label{ssec:2PL}
Tracing SMBHB evolution from galaxy merger to SMBHB merger vastly increases model complexity. The SMBHB hardening prescription used in NG15Astro is a simple parameterization of this kind of ``outside-in" model \citep[based on][]{kelley17a, kelley17b}, and I briefly review it here. The GW-only hardening rate is \citep{peters64}:
\begin{align}
\dadt_{\rm GW} &\equiv \f{da}{dt}_{\rm GW} = - \f{64 G^3}{5 c^5} \f{\eta M^3}{a^3} g(e), \label{eqn:adotgw} \\
g(e) &= (1-e^2)^{-7/2} \left ( 1 + \f{73}{24} e^2 + \f{37}{96} e^4 \right).
\end{align}
The astrophysically driven hardening is prescribed by the phenomenological 2PL model (Equation \ref{eqn:adotphenom2pl}), with power-law indices $\nu_{\rm out}$ and $\nu_{\rm in}$, respectively, describing the dominant behavior at separations above and below a characteristic value $\rchar$: 
\begin{align}
\dadt_{\rm 2PL} &= H \left ( \f{a}{\rchar} \right)^{1-\nuin} \left ( 1 + \f{a}{\rchar} \right )^{\nuin - \nu_{\rm out}}. \label{eqn:adotphenom2pl}
\end{align}
The total hardening rate is the sum of the GW and the phenomenological hardening rates (Eqn. \ref{eqn:adottot2pl}): 
\begin{align}
\dadt &=  \dadt_{\rm 2PL} + \dadt_{\rm GW}. \label{eqn:adottot2pl}
\end{align}
\rev{Figure~\ref{fig:twoPLschematic} shows a schematic of the hardening rate vesus separation for the 2PL model, using parameter values consistent with NG15Astro.}

One can also define $\thard\equiv a/(-\dadt)$ as the instantaneous hardening timescale. Throughout this work, I use $\thard$ to express instantaneous hardening timescales and $\tau$ to express integrated inspiral times. I also assume that $\dadt$ is always negative---i.e., that binary orbits always shrink (harden) rather than expand (soften). Thus, larger hardening rates (faster inspiral) correspond to more negative values of $\dadt$. Because I extensively discuss the maximum physical hardening rates, I use the notation $\left|\dadt\right|$ as needed to avoid confusion.
At $a\ll\rchar$, the hardening timescale $t_{\rm hard,2PL} \propto a^{\nuin}$, and at $a\gg\rchar$, $t_{\rm hard,2PL} \propto a^{\nu_{\rm out}}$. In the GW phase, $t_{\rm hard,GW}\propto a^4$.

The normalization $H$ in Eqn. \ref{eqn:adotphenom2pl} is defined such that 
\begin{eqnarray}
\tau_{\rm tot} \equiv \int_{\ainit}^{\risco} \f{da}{\dadt}    
\end{eqnarray}
is constant for all binaries, where $\risco$ is the innermost stable circular orbit. The initial separation at which binaries are assumed to form, $\ainit$, also has a single value for all binaries. The 2PL model therefore has five free parameters: $\nu_{\rm out}$, $\nuin$, $\rchar$, $\ainit$, \& $\tau_{\rm tot}$. 

A drawback to this approach is that inspiral rate through the inner astrophysical and GW regimes is strongly constrained by binary evolution at large separations, far outside the PTA regime. This is because the overall normalization of the hardening rate, $H$, is set by parameters that predominantly affect early binary inspiral. The total time $\tau_{\rm tot}$ is dominated by the early stages of inspiral for the model parameter values favored in NG15Astro, and $\ainit$ and $\nu_{\rm out}$ by definition concern only this early phase. By using such a model, in which the outer and inner evolution is intertwined via parameters such as $\tau_{\rm tot}$, the information provided by PTAs about the inner phase is diluted by the highly uncertain outer phase of SMBHB evolution.

For the default value of $\nu_{\rm out}=+2.5$ used in NG15Astro, the slowest SMBHB hardening rate often occurs at the largest separation. This exacerbates the problem, because when $t_{\rm hard,2PL}(\ainit) \sim \tau_{\rm tot}$, there is often essentially one valid solution for $H$ over all SMBHB masses and mass ratios. This suppresses variation in hardening rates through the PTA regime, which is not ideal for performing parameter inference. This issue is even more important given that the PTA regime is precisely the realm least accessible to EM observations. 
To maximize the information gleaned from PTAs, we should parameterize SMBHB inspiral in a manner that isolates and optimizes sensitivity to the PTA frequency regime. We should also account for the fact that the hardening rate outside this frequency regime cannot be constrained directly with PTAs. In the next section, I present a model that satisfies these requirements.

\begin{SCfigure*}
\includegraphics[width=0.7\textwidth]{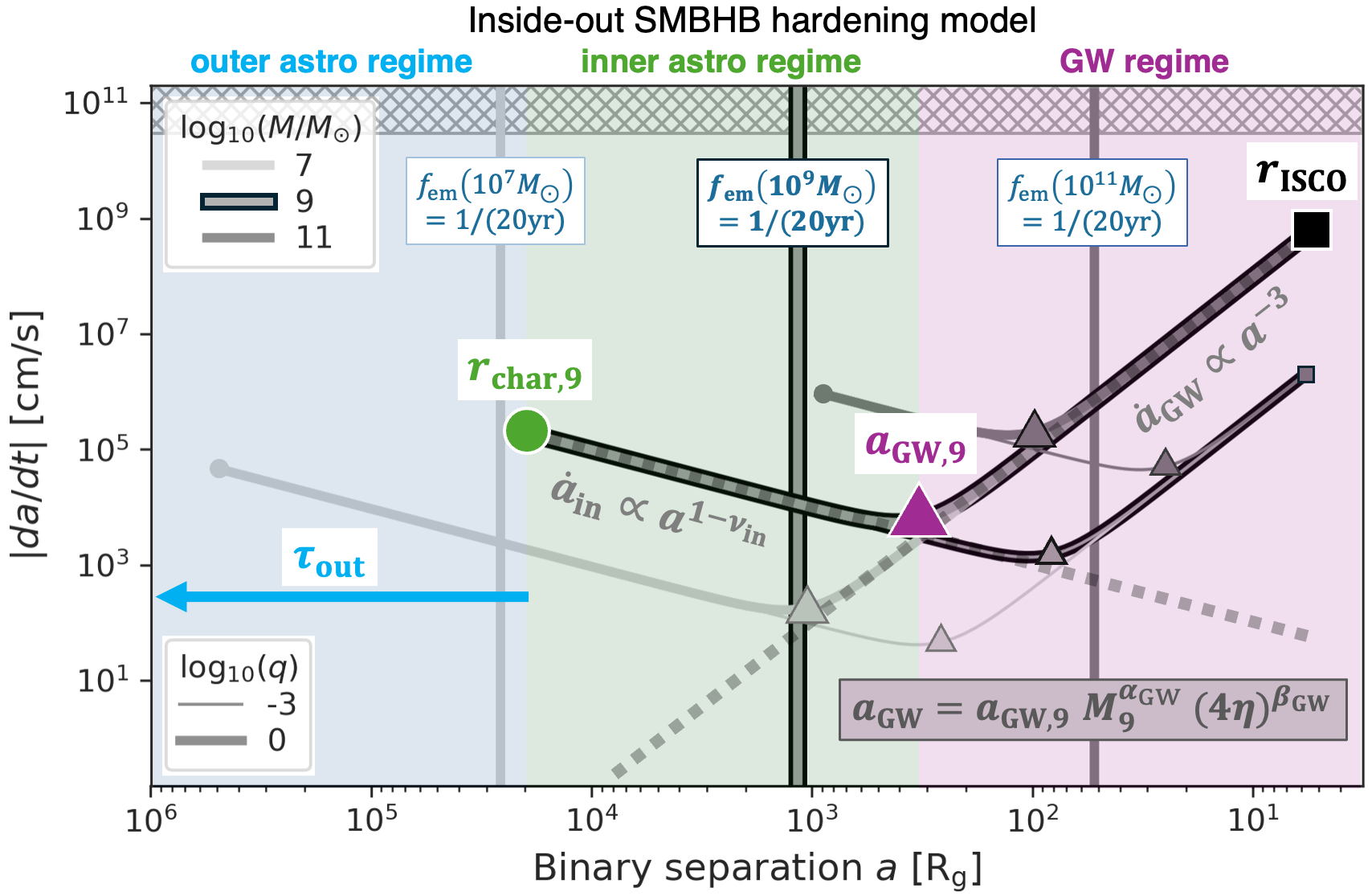}
\caption{Schematic illustrating key features of the inside-out SMBHB inspiral model. The hardening rate vs. separation is shown for the \ctin\ model (defined below), for select masses and mass ratios as in Figure~\ref{fig:twoPLschematic}. $\agwrg\!\equiv\aninerg\, M_9^{\alphagw} (4\eta)^{\betagw}$  is the transition from the astrophysical to the GW regime (triangles). $\rchar$ ($\propto M_9^{\alphach}$ when in $\rg$ units) is the boundary between the inner and outer astrophysical regimes (circles); it is chosen to lie outside the PTA band. \rev{At all $a<\rchar$, $\dadt$ is the sum of the inner$+$GW hardening rates.} The outer regime is modeled via a timescale $\tau_{\rm out}$ rather than an explicit form for $\dadt$. \\ \\ \label{fig:IOschematic}}
\end{SCfigure*}

\subsection{``Inside-out" SMBHB hardening }
\label{ssec:newhard}
The orbital semi-major axis $\agw$ at which a SMBHB transitions from astrophysically to GW-driven hardening can be defined as that at which the astrophysical hardening timescale equals the GW timescale $t_{\rm hard,GW}(\agw)$.
Because of the steep scaling of $t_{\rm hard,GW} \propto a^4$, the total merger time and the GW energy emitted at each frequency are very sensitive to this parameter. Binaries with very large $\agw$ will stall, while those with very small $\agw$ will contribute minimally to the GWB.
As in NG15Astro, I focus only on circular binaries (such that $g(e)=1$ in Equation~\ref{eqn:adotgw}). The extensibility of this model to include eccentric orbits is discussed in Section~\ref{ssec:ecc}.
For reasonable parameter choices, $\agw$ for $M\sim10^9\msun$ BHs tends to lie near or within the the PTA band (see Figures~\ref{fig:IOschematic}-\ref{fig:fid_dadt}).

I propose that $\agw$ should be considered a key free parameter to be constrained by PTA observations of the GWB. Beginning with this premise, I construct an ``inside out" model for SMBHB hardening. Figure~\ref{fig:IOschematic} illustrates the hardening model components, which are described in detail below. The model retains the ansatz of the 2PL model that the hardening rate follows a single power law in the ``inner" astrophysically driven regime, with a power-law index $\nuin$.

The inner hardening regime must be connected to the cosmological SMBHB formation rate at larger separations. 
For the purposes of constraining SMBHB populations with PTAs, the exact hardening rate needs to be modeled only out to the largest binary separations accessible to PTAs. Unlike the 2PL model, I do not explicitly model the `outer' hardening rate but rather define $\tau_{\rm out}$ as the time elapsed from binary formation ($a=\ainit$) until $a=\rchar$. This formalism does not require $\ainit$ to be specified explicitly. A drawback is that $\tau_{\rm out}$ cannot be directly constrained with EM observations of dual active galactic nuclei (AGN), but such constraints would at any rate rely on the uncertain active fraction of SMBH pairs. My model could be readily extended to include explicit binary evolution in the outer phase, but at the expense of adding more parameters.

I assume that $\tau_{\rm out}$ has a constant value for all binaries, though one could add dependence on mass, mass ratio, redshift, or environmental parameters. 
$\rchar$ is constrained to be larger than the orbital separation of binaries corresponding to the minimum GW frequency probed by PTA observations, such that $\dadt$ is explicitly modeled within the PTA band.

This inside-out hardening model is specified by:
\begin{eqnarray}
\dadt_{\rm in} = \dadt_{\rm GW}\left(\agw\right)\: \left ( \f{a}{\agw} \right)^{1-\nuin}, \\
\dadt\left(a < \rchar\right) = \dadt_{\rm in} + \dadt_{\rm GW}
\end{eqnarray}    
where $\dadt_{\rm GW}$ is given by Equation~\ref{eqn:adotgw}, and $\dadt(a>\rchar)$ is not modeled explicitly. The total inspiral time from binary formation to merger is
\begin{eqnarray}
\tau_{\rm tot} \equiv \tau_{\rm out}\: + \:\tauingw, \\ 
\tauingw \equiv \int_{\rchar}^{\risco} \f{da}{\dadt(a<\rchar)},   
\end{eqnarray}
such that a binary forming at a lookback time of $t_{\rm look}$ will reach $\rchar$ at the time $t_{\rm look} - \tau_{\rm out}$. 

The simplest choice for the parameter $\agw$ would be to assume a constant value for all binary masses and mass ratios. However, because gravitational length units scale linearly with mass ($\rg \equiv G M / c^2$), one obtains very different results if $\agw$ is set to a constant in physical length units versus a constant in gravitational units. 
If $a_{\rm GW}$ has constant value in physical units of pc for all binaries, then $\dadt(\agw) \propto M^3$ and $t_{\rm hard,in}(\agw) \propto M^{-3}$, such that both quantities vary by 24 orders of magnitude across the SMBH mass range of $10^4$-$10^{12}\msun$ considered in this work. In other words, if this model represented reality, only the most massive SMBHBs could merge within $\thub$, because $\tauingw<\thub$ for low-mass BHs would imply unphysical (or even superluminal) hardening rates for massive SMBHs. 

Alternatively, if $\agw$ has a constant value in gravitational units, then $\dadt(\agw) \propto M^0$  and $t_{\rm hard,in}\propto M$. This eliminates the problem of steep mass scaling, though it imposes the opposite (linear) scaling for the inspiral timescales (i.e., more massive binaries have longer inspiral timescales from a fixed $\rchar$). 

In light of these considerations, and in the absence of empirical constraints, I abandon the notion of a constant value of $\agw$ for all SMBHBs (in either physical or gravitational units) and instead introduce new parameters for the scaling of $\agw$ with $M$ and $\eta$.  
I define $\anine$ as a free parameter representing the binary separation at which hardening transitions to GW-dominated for an equal-mass, circular binary with $M=10^9\msun$ and then define $\agw$ according to
\begin{align}
    \agwrg 
    = \aninerg  \: M_9^{\alphagw} \: (4\eta)^{\betagw} 
\end{align}
Here $\agw$ and $\anine$ are defined in gravitational units. In physical length units of cm this yields 
\begin{align}
    \agwcm = \aninerg \left (\f{G \,10^9\msun}{c^2}\right)  M_9^{\alphagw+1} \:(4\eta)^{\betagw}.
\end{align}
Throughout this paper, $\agw$ and $\aninerg$ refer to quantities in gravitational units unless the explicit notation $\agwcm$ or $\aninecm$ is used.  Plugging this into Equation~\ref{eqn:adotgw} yields
\begin{align}
     \dadt_{\rm in}(\rev{\agwcm}) = -\f{16\,(G \,10^9\msun)^3}{5\,c^5 \rev{\aninecm}^{3}} 
       M_9^{-3\alphagw} \: (4\eta)^{1-3\betagw}.    
\end{align}
The hardening timescale at $\agw$ is
\begin{align}
t_{\rm hard,in}(\rev{\agwcm}) = \f{5\,c^5 \rev{\aninecm}^4}{16 (G\,10^9\msun)^3}  M_9^{4\alphagw+1} (4\eta)^{4\betagw-1}.
\end{align}

Thus, $\alphagw=-1$ yields no mass scaling of $\agwcm$ but has cubic and inverse cubic mass scaling of the hardening rate and timescale, respectively, at $\agw$. $\alphagw=0$ yields a constant inner hardening rate for all masses, a constant $\agwrg$ (in units of $\rg$), and linear mass scaling of the hardening timescale.
While in principle $\alphagw$ could take any value, steep mass scalings will strongly limit the range of SMBHB masses that can merge within $t_{\rm H}$, as well as the range of SMBHB masses for which $|\dadt|<\dadtmax$ (defined in Section~\ref{ssec:paramspace}). The mass scalings of $\agw$, $\dadt(\agw)$, and $t_{\rm hard,in}(\agw)$ are all less than linear for $-1/3 < \alphagw < 0$. Similarly, $0 < \betagw < 1/2$ yields less-than-linear scaling of these quantities with $\eta$. 

I focus on two different cases here:
\begin{enumerate}
    \item $\alphagw=-1/4$, $\betagw=+1/4$: this corresponds to an inner hardening timescale with no mass or mass ratio dependence, such that in $\rg$ units, $\agwrg \propto M^{-1/4}\eta^{1/4}$. Both $\agwcm$ and $\dadt(\agw)$ then scale as $\propto M^{3/4} \eta^{1/4}$. With the choice $\nuin=0$, the inner hardening timescale is also constant for all $a$. I refer to this as the ``\ctin" model and treat it as the primary fiducial model in the subsequent analysis.
    \item $\alphagw=0$, $\betagw=0$: this yields a constant $\agwrg$ for all masses and mass ratios, a constant  $\dadt(\agw)$, and an inner hardening timescale that scales linearly with $M$ (and is independent of $\eta$). I refer to this (with $\nuin=0$) as the ``\cagw" model.
\end{enumerate}

Note that the 2PL hardening model \rev{with $\nuin=0$} has a similar scaling of $\agw$ with $M$ and $\eta$ as the \ctin\ model, because of its constant overall inspiral timescale. The scaling in that case does not exactly follow $\alphagw=-1/4$, $\betagw=+1/4$, because $\agw$ has some dependence on the outer hardening phase, but it is close\rev{. When $\nuin=0$, the mass scaling deviates from $\alphagw=-1/4$ by factors of $< 10^{-3}$, and the mass ratio scaling deviates from $\betagw=+1/4$ by $< 1\%$ for $q\lesssim0.03$, increasing up to $\sim 30\%$ deviation at $q\sim1$. For nonzero values of $\nuin$, the mass scaling in the 2PL model varies much more, ranging from a power-law index of $\gtrsim -1/2$ (for $\nuin=-1$) to $\sim +1/2$ (for $\nuin=+2$), with the strongest variation at the highest masses.} 

I make one final addition to the inside-out hardening model: the separation $\rchar$ at which the inner hardening regime begins is also allowed to scale with $M$. This choice is partly pragmatic, because it increases the parameter space of hardening models that are valid across a wide range of binary masses (see Section~\ref{ssec:paramspace}). I define
\begin{eqnarray}
    \rchar = \rchnine \: M_9^{\alphach+1},
\end{eqnarray}
where $\rchar$ and $\rchnine$ are both defined here in {\em physical} rather than gravitational units, unlike $\agw$. $\rchnine$ is the start of the inner hardening regime for an equal-mass, $10^9\msun$ binary, and $\alphach$ is defined such that $\alphach=-1$ yields a constant $\rchar$ (in physical units) for all binary masses. If $\rchar$ is assumed to scale with the transition to the ``hard" binary regime ($\rchar \propto \rhard$), then $\rchar \propto G M / \sigma_*^2$ for stellar scattering models \citep{quinlan96}. If a scaling $M\propto \sigma_*^4$ is also assumed, then $\rchar \propto M^{+1/2}$, corresponding to $\alphach=-1/2$.

One can also consider the separation at which (circular) binaries reach a given GW frequency: $a \propto M^{+1/3}$ ($\alphach=-2/3$). Choosing $\alphach=-2/3$ has the advantage that, as long as $\rchnine$ is larger than the separation  at which $10^9 \msun$ binaries enter the PTA regime, binaries of {\em all} masses are guaranteed to start their evolution outside the PTA regime. In practice, I find that $\alphach-1/2$ and $\alphach=-2/3$ yield very similar results, so I adopt $\alphach=-2/3$ as the fiducial value for all models in this work. The fiducial value of $\rchnine$ is chosen to be 1 pc, which is comfortably outside the low-frequency edge of the PTA band even for a hypothetical 100-yr data set.

To summarize, the inputs for this SMBHB hardening model are $\agw$,  $\nuin$, $\rchar$, and $\tau_{\rm out}$. Three parameters specify $\agw$ to control its scaling with mass and mass ratio ($\anine$, $\alphagw$, \& $\betagw$), while  two parameters specify the mass scaling of $\rchar$ ($\rchnine$ \& $\alphach$). Table~\ref{tab:models} gives parameter values for the fiducial models \ctin\ and \cagw, along with the fiducial parameters for ``stellar-like" and ``gas-like" models discussed in Section~\ref{ssec:litcomparison}, and variants of these fiducial models featuring systematic parameter sweeps (see Sections~\ref{ssec:fiducial}-\ref{ssec:outer}.
In Appendix~\ref{sec:model1}, I outline an alternate version of this model in which $\dadtrc$ (for equal-mass, circular binaries) is assumed to be a constant input parameter, such that $\nuin$ is instead derived from $\dadtrc$.

I implement this inside-out hardening model in the SMBHB population synthesis code \texttt{holodeck}\footnote{\href{https://github.com/nanograv/holodeck}{https://github.com/nanograv/holodeck}}, which is used for analysis of cosmological SMBHB populations and GWB spectra throughout this work.

\begin{deluxetable*}{lccccccl}
    \tablewidth{0pt} 
    \tablecaption{The table gives the model names and fiducial parameter values for the hardening models discussed in this work. All models shown assume $\alphach=-2/3$. The ``Fiducial Models" indicate the primary 6 sets of model parameters used in this work. The ``Model Variants" are used to systematically vary a single parameter across the range of values shown while keeping all other parameters fixed at the fiducial values for a given fiducial model. For example, the \ctin\aninevar\ parameter sweep varies $\aninerg$ from $10^{1.5}$-$10^{3.75} \rg$ while keeping all other parameters fixed at the fiducial \ctin\ values. The last column notes the section(s) of the paper where each model or model variant is primarily discussed.
    \label{tab:models} }
    \tablehead{
    \colhead{Name} & \colhead{$\log_{\rm 10}\left(\f{\aninerg}{\rg}\right)$} & \colhead{$\alphagw$} & \colhead{$\betagw$} & \colhead{$\nuin$} & \colhead{$\log_{\rm 10}\left(\f{\tau_{\rm out}}{\rm Gyr}\right)$} & \colhead{$\log_{\rm 10}\left(\f{\rchnine}{\rm pc}\right)$} & \colhead{Section(s)} }   
    \startdata
    \multicolumn{7}{c}{Fiducial Models}\\ \hline
    \ctin\ & $2.5$ & $-\f{1}{4}$ & $+\f{1}{4}$ & 0 & 0.0 & 0.0 & \S~\ref{ssec:fiducial} \\
    \cagw\ & $2.5$ & 0 & 0 & 0 & 0.0 & 0.0 & \S~\ref{ssec:cagw} \\
    \ctinstar\ & $3.5$ &$-\f{1}{4}$ & $+\f{1}{4}$& $-1$  &0.0 & 0.0  & \S~\ref{sssec:stellar},~\ref{ssec:nuinvar} \\
    \cagwstar\ & $3.5$ & 0 & 0 & $-1$  &0.0 &0.0 & \S~\ref{sssec:stellar},~\ref{ssec:nuinvar} \\
    \ctingas\ & $2.0$ & $-\f{1}{4}$ & $+\f{1}{4}$ & $+2$ & 0.0&0.0 & \S~\ref{sssec:gas},~\ref{ssec:nuinvar}\\
    \cagwgas\ & $2.0$ & 0 & 0 & $+2$ & 0.0&0.0  & \S~\ref{sssec:gas},~\ref{ssec:nuinvar}\\ \hline
    \multicolumn{7}{c}{Model Variants}\\ \hline
    \texttt{*-agw9var}  & $[1.5, 3.75]$ & & & & & & \S~\ref{ssec:a9}\\
    \texttt{*-{\alpha}var}  & & $[-\f{1}{2}, 0]$& & & & & \S~\ref{ssec:alphabetavar}\\
    \texttt{*-{\beta}var}  & & & $[-\f{1}{2}, +\f{1}{2}]$ & & & & \S~\ref{ssec:alphabetavar} \\
    \texttt{*-nuvar}  & & & &$[-1,+2]$ & & & \S~\ref{ssec:nuinvar}\\
    \texttt{*-toutvar}  & & & & & $[-1.0, 1.0]$ & & \S~\ref{ssec:outer}\\
    \texttt{*-rch9var}  & & & & & & $[-1.0, 2.0]$ & \S~\ref{ssec:outer}
    \enddata
\end{deluxetable*}

\subsection{Constraints on physical regions of parameter space}
\label{ssec:paramspace}

\begin{figure*}
    \centering
    \includegraphics[width=0.59\linewidth]{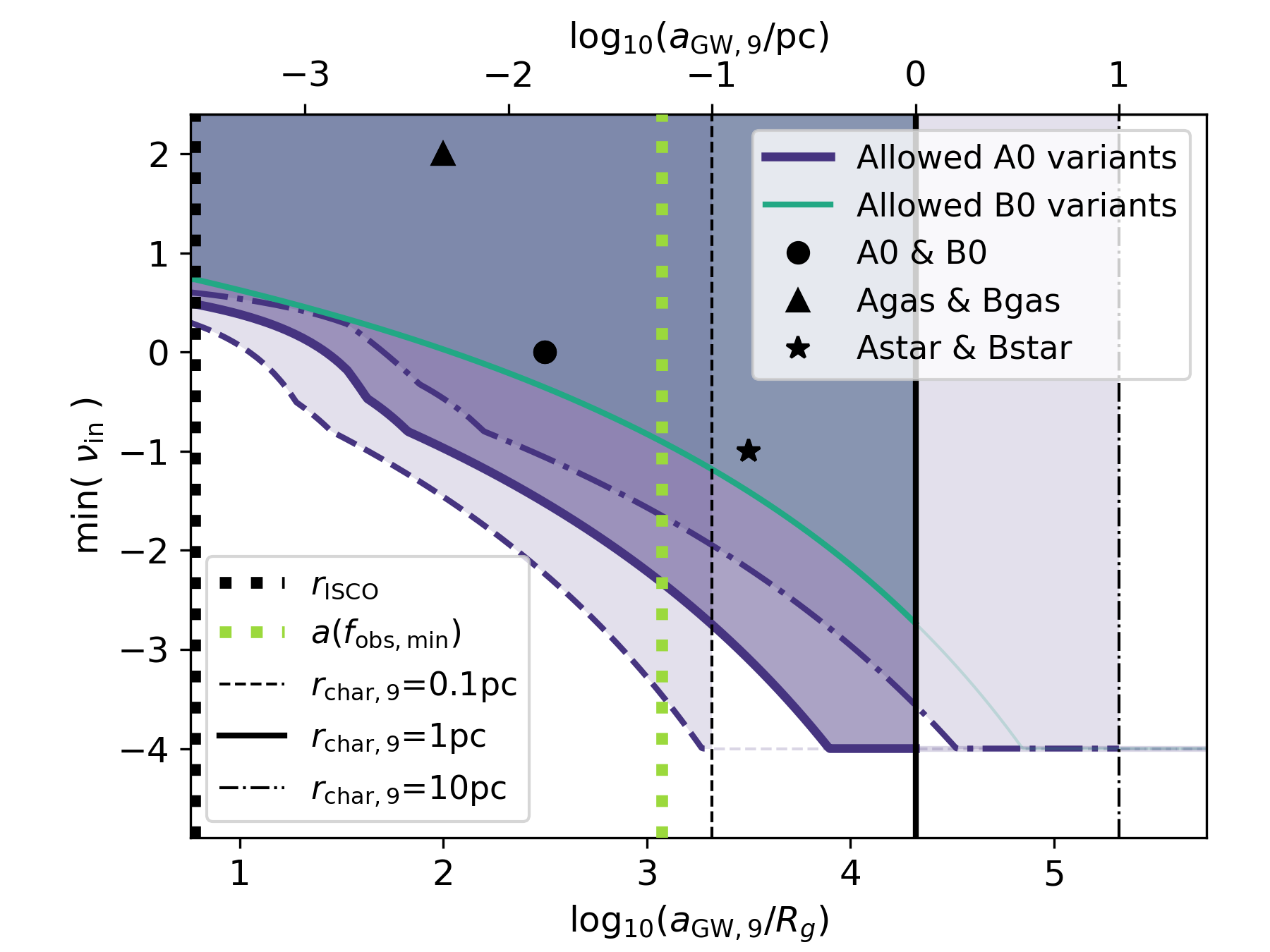}
    \includegraphics[width=0.39\linewidth]{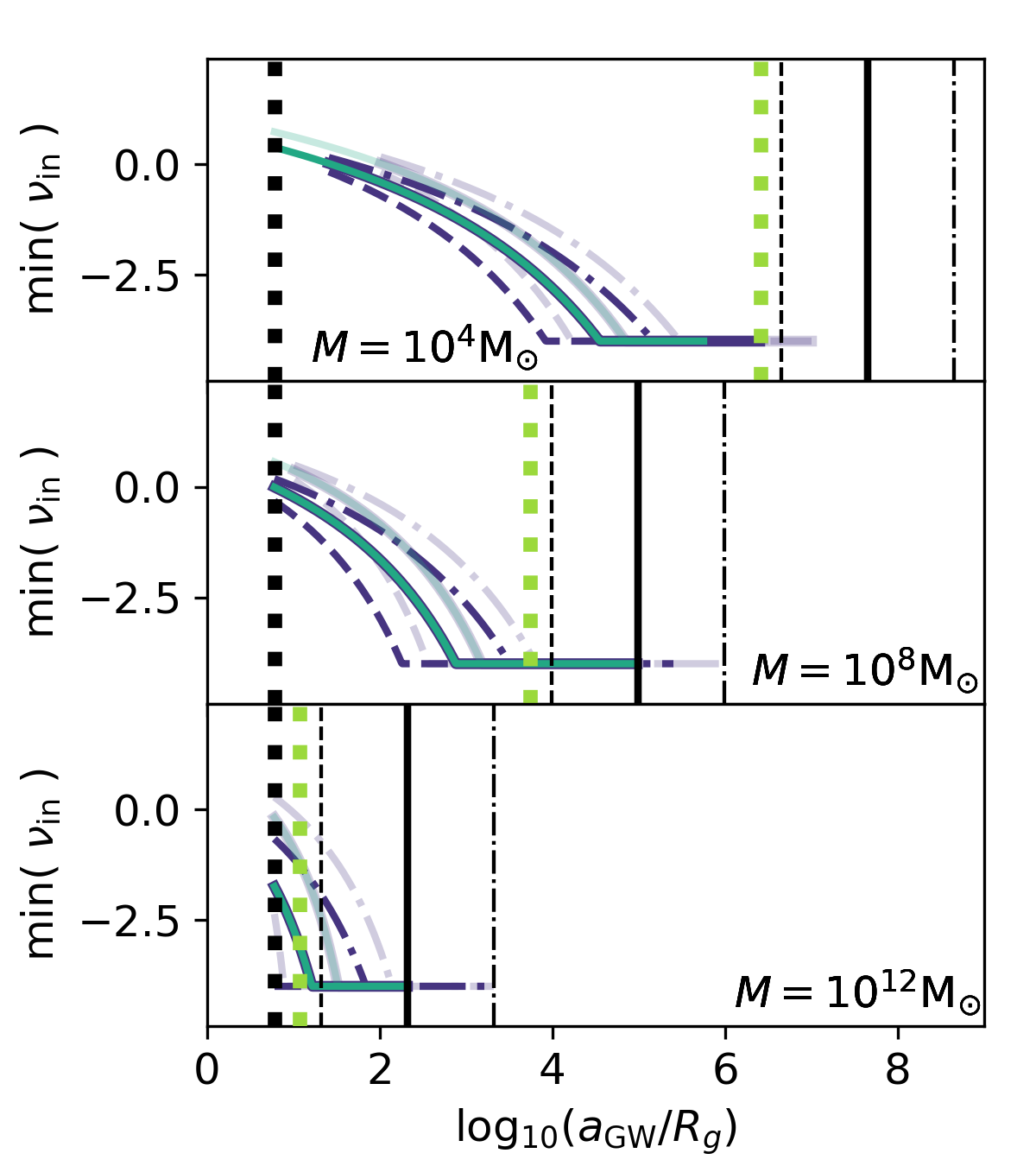}
    \caption{{\em Left panel:} Shaded regions denote the allowed parameter space  ($\nuin, \aninerg$) for select SMBHB inspiral models. Purple curves denote the minimum $\nuin$ for a given $\aninerg$ for the \ctin\ model with $\rchnine=$ 0.1pc (dashed), 1 pc (solid; fiducial value), \& 10 pc (dot-dashed). These $\rchnine$ values are also marked by vertical black lines. The green solid curve shows the minimum $\nuin$ for the \cagw\ model with $\rchnine=1.0$pc. Lower $\nuin$ would produce $|\dadt|>|\dadt_{\rm max}|$ for at least some binary masses in the range $M=[10^4,10^{12}]\msun$. The absolute minimum is $-\nuinmax=-4$.  Larger positive $\nuin$ values are allowed (up to $\nuinmax=4$) but will yield increasingly long inspiral timescales. Binaries with $\anine>\rchar$ do not have an ``inner" hardening phase prior to entering the GW regime. The black dotted line denotes the lower bound of $\anine=\risco$. The yellow-green dotted line denotes the separation at which $M=10^9\msun$ circular binaries emit GWs at $f_{\rm obs,min}=1/(20{\rm yr})=1.6\times10^{-9}$ Hz. A consistent GWB calculation requires $\rchar>a(f_{\rm obs,min})$. The black circle denotes the fiducial models ($\nuin=0,\: \aninerg=10^{2.5}$), the black triangle denotes the ``gas-like" models ($\nuin=2,\: \aninerg=10^{2}$), and the black star denotes the ``star-like" models ($\nuin=-1,\: \aninerg=10^{3.5}$). {\em Right panel:} The minimum $\nuin$  versus $\agw$ is shown for three different binary masses, for the same models as in the left panel. $q=1$ binaries are shown in bold curves, and $q=0.001$ binaries are shown in light-shaded curves. $\risco$, $\rchar$, \& $a_{\rm obs}(f_{\rm obs,min})$ are also shown for each mass.  Note that the $x$-axis in the right panels is $\agwrg$, not $\aninerg$\rev{. The ``bumps" in the \ctin\ curves in the left panel owe to the $M$ \& $\eta$ scaling of $\dadtrc$, such that the minimum $\nuin$ for a given $\aninerg$ is set by low-$M$, low-$\eta$ binaries at large $\aninerg$ and by high-$M$, high-$\eta$ binaries at small $\aninerg$.} Overall, a relatively narrow range of the model parameter space produces binary populations that do not exceed $\dadtmax$ while still merging most systems by $z=0$.}
    \label{fig:allowedparams}
\end{figure*}

A disadvantage of analytic power-law models for SMBHB hardening is that no limits are naturally imposed on the maximum rate of hardening, possibly leading to unphysical results. (Notably, the same shortcoming affects models that ignore astrophysical SMBHB hardening entirely and assume that all binaries efficiently reach the GW regime.) Indeed, the hardening rate can in some portions of parameter space exceed the speed of light.

To avoid unphysical hardening rates, I modify the above models by imposing a ``speed limit" on the hardening rate, $\dadtmax$. Here I simply take $\dadtmax = c$. While hardening rates near this limit are unrealistic at separations much larger than $\risco$, choosing this value provides a firm upper limit for exploring the allowed parameter space. Within the observable PTA band, a \rev{less extreme} limit for $|\dadt|$ would be the binary orbital speed, $\dadtmax=v_{\rm orb}$, calculated at each binary separation or at $\risco$.  \rev{$\dadtmax=v_{\rm orb}$ is still a very high inspiral rate compared to that expected from models of gas- or stellar-driven hardening; such rapid changes in orbital separation would require strong dynamical perturbations from, e.g., chaotic three-body SMBH interactions.} I verify that the models presented here have $|\dadt|\,\rev{\ll}\, v_{\rm orb}$ for all binaries contributing non-negligibly to the GWB. \rev{(The only cases in which $|\dadt|$ approaches $v_{\rm orb}$ are for $\sim 10^4\msun$ SMBHBs in the \cagw\ and \cagwstar\ models, and then only for $a$ near $\rchar$.) } In practice, SMBHBs that spend much time at \rev{very high} hardening rates will not contribute \rev{much} to an observable GWB anyway, because they will have very little time to emit GWs before merging. \rev{Care should be used when interpreting the results of any GWB analysis that points to hardening rates near $\dadtmax$ for the binaries dominating the background.} 

Setting $(a_{\rm init}-\rchar)/\tau_{\rm out} < c$ for the outer astrophysical inspiral phase again provides a firm but unrealistic upper limit. For the initial separation of $a_{\rm init}=10$ kpc used in NG15Astro, this yields $\tau_{\rm out} \gtrsim 30$ kyr. A more realistic estimate of the minimum $\tau_{\rm out}$ would be the dynamical time of the host galaxy at $a_{\rm init}$, or one that yields $\left<\left|\dadt_{\rm out}\right|\right> \sim \sigma_*$. I show in Section~\ref{ssec:outer} that the shape and amplitude of the GWB is not very sensitive to the choice of $\tau_{\rm out}$ when it is less than a Gyr.  

Although no maximum $|\dadt|$ was imposed on the 2PL hardening model used in NG15Astro, the hardening rates are well below $c$ for the favored model parameters. For their fiducial choice of $\nu_{\rm out}=+2.5$, $\rchar=100$ pc, and $\ainit=10$ kpc, only models with $\nuin\gtrsim+1.25$, $\nuin\lesssim-1$, and/or $\tau_{\rm tot} < 0.01$ Gyr have peak hardening rates $|\dadt|>c$. Only values of $-1.5<\nuin<0.5$ and $\tau_{\rm tot}>0.1$ Gyr were considered in NG15Astro, and models with $\nuin<-1$ lie in the tail of the posterior distribution. Moreover, in all NG15Astro models with $\nuin<1$, the peak hardening rates occur at $\rchar=100$ pc, which is well outside the PTA regime. Unrealistic hardening rates at $\rchar$ are equivalent to underestimating the ``outer" inspiral time ($\tau_{\rm out}$). Superluminal hardening rates are achieved only for $M\lesssim10^4\msun$ (for $\nuin\lesssim-1$) and  $M\lesssim10^6\msun$ (for $\nuin\lesssim-1.5$)---a mass regime that contributes negligibly to the PTA signal. Thus, the NG15Astro results were unaffected by the lack of a maximum allowed $|\dadt|$ in the 2PL hardening model.

I do not require that $\risco < \agw < \rchar$ for all binaries. Binaries with $\agw \gtrsim \rchar$ will have no inner hardening regime (i.e., GW-only inspiral for $a<\rchar$). Binaries with $\agw \lesssim \risco$ would have essentially no GW regime---in other words, astrophysics would dominate all the way to merger in these cases. Such binaries will produce very little GW emission during inspiral. 

These edge cases motivate an additional constraint on the hardening model parameter space: $|\nuin|$ should be capped at a maximum value $|\nuin|_{\rm max}$ to prevent unrealistic ``jumps" in the hardening rate. This is most important for the alternate parameterization described in Appendix~\ref{sec:model1}, where $\nuin$ is derived from $\dadtrc$ and could cause numerical instabilities if it approaches infinity. However, 
$\nuin>+4$ is problematic for a different reason: because $t_{\rm hard,GW}\propto a^4$, $\nuin>+4$ corresponds to an inconsistent scenario in which GWs dominate at $a>\agw$ and the inner power-law dominates at $a<\agw$---the opposite of how these phases are defined. In practice, most binaries will have $\tauingw>\thub$ for $\nuin>+4$, and some will have $|\dadt|>c$ for $\nuin<-4$. I therefore choose $|\nu_{\rm in}|_{\rm max} = 4$.

When this hardening model is used for GW calculations, $\rchar$ should be larger than the binary separations corresponding to the lowest relevant GW frequencies. In other words, because $\dadt$ is not explicitly modeled in the outer hardening phase, the inner  phase must begin before binaries enter the PTA band, or the resulting GWB will be missing GW strain from the early stages of inspiral.  A circular $10^9\msun$ binary at $z=0$ reaches $f_{\rm obs}=(100\,{\rm yr})\inv$ at $a=0.2$ pc and $f_{\rm obs}=(20\,{\rm yr})\inv$ at $a=0.06$ pc. If $\alphach=-2/3$, choosing $\rchnine>a(f_{\rm em}=f_{\rm obs,min})$ ensures that $\rchar>a(f_{\rm obs,min})$ for all binary masses and redshifts.

Thus, the criteria that define physically reasonable portions of the hardening model parameter space are:
\begin{enumerate}[noitemsep]
    \item $|\dadt| < \dadtmax$
    \item $|\nuin| < |\nu_{\rm in}|_{\rm max}$ 
    \item $\rchar > a(f_{\rm obs,min})$ 
\end{enumerate}
where the last criterion strictly applies only when used for GW calculations.

Criterion (1) can be written in terms of input parameters as follows. First, note that this criterion is satisfied at all binary separations if it is satisfied at $\rchar$. The hardening rate at $\rchar$ is
\begin{align}
    \dadt\left(\rchar\right) &= \dadt_{\rm GW}\left(\rchar\right) + \dadt_{\rm in}\left(\rchar\right)  \nonumber \\
    &= - \f{64 G^3}{5 c^5} \f{\eta M^3}{\rchar^3}  \left [ 1 + \left( \f{\rchar}{\rev{\agwcm}} \right)^{4-\nuin} \right ]. 
\end{align}
For set values of $\agw$ and $\rchar$, the criterion $|\dadtrc|<\dadtmax$, combined with criterion (2), sets a minimum allowed value of $\nuin$: 
\begin{align}
    \nuin &> {\rm max}\left( \vphantom{\f{x^x}{x_x}} -|\nu_{\rm in}|_{\rm max} \right., \nonumber \\ 
    &\left . \rev{1 - \f{\log_{\rm 10}\left[\dadt_{\rm max}/\dadt_{\rm GW}(\agwcm)-1\right]}{\log_{\rm 10}\left[\rchar/\agwcm\right]}} \right). \label{eqn:nuin_min}
\end{align}
Note that $\nuin$ has no maximum value except that imposed by criterion (2). If we want to instead fix $\nuin$ (and $\agw$) and find the permitted range of $\rchar$, criterion (1) does not yield a general analytic solution for the maximum $\rchar$ in terms of $\nuin$. However, a conservative upper limit on $\rchar$ can be set by noting that $|\dadt_{\rm GW}\left(\rchar\right)| < \rev{|\dadt_{\rm GW}\left(\agw\right)|}$, often by many orders of magnitude, such that
\begin{align}
    \f{64 G^3}{5 c^5} &\f{\eta M^3}{\rev{\agwcm}^3}  \left [ 1 \rev{+} \left( \f{\rchar}{\rev{\agwcm}} \right)^{1-\nuin} \right ]  < \dadtmax\\
    \f{\rchar}{\rev{\agwcm}} &< \left( \f{5c^5 \dadtmax \agw^3}{64 G^3 \eta M^3}\rev{-}1\right)^{1/(1-\nuin)},\:\: (\nuin<1)\nonumber\\
\end{align}
is a useful limit on $\rchar$ when an analytic expression is desired.  
(The $\dadtmax$ criterion does not impose any limits on $\rchar$ if $\nuin \geq 1$.)

The second criterion  $|\nuin| < |\nuin|_{\rm max}$ is already expressed in terms of input parameters for this model, and it is almost always a less stringent constraint on $\nuin$ than Equation~\ref{eqn:nuin_min} (see Figure~\ref{fig:allowedparams}). Appendix~\ref{sec:model1} outlines  these criteria for the alternate model parameterization using $\dadtrc$ as an input parameter instead of $\nuin$. 

Because $\agw$ and $\rchar$ depend on mass, the valid values of $\rchnine$ and $\nuin$ also vary with mass. $\rchnine$ and $\nuin$ must be valid across the entire range of masses and mass ratios considered in order to use these models for SMBHB parameter inference.

Figure \ref{fig:allowedparams} visualizes the model parameter space allowed by these physical constraints. 
Because $|\dadt|$ increases sharply as the GW-dominated inspiral proceeds, smaller values of $\aninerg$ lead to stricter limits on $\nuin$. Larger values of $\rchar$ also restrict $\nuin$, because $\left|\dadt_{\rm in}\right|$ is maximum at $\rchar$ for all $\nuin<1$. $\nuin$ can be arbitrarily large (up to $\nuinmax=4$), though this will yield very long inspiral timescales.

The fiducial \ctin\ model (purple solid line) requires $\nuin\gtrsim0.5$ when $\anine=\risco$, a limit that expands to $\nuin\gtrsim-5.5$ at $\rchnine=1$pc. In the \cagw\ model (green solid line), 
$\nuin$ is more constrained than in the \ctin\ model. 
Variations in $\rchnine$ do not alter the minimum $\nuin$ very much when $\aninerg$ is small, but these curves diverge as $\aninerg$ increases. For $\aninecm\geq\rchnine$, there is no inner hardening phase---the binary transitions directly from the outer phase to the GW phase. However, since $\aninecm$ and $\rchnine$ have different mass scaling unless $\alphagw=\alphach$, some binary masses may be in this GW-only regime even if $\aninecm<\rchnine$. For the fiducial \ctin\ model, all binaries masses and mass ratios considered ($M=10^4-10^{12}\msun$ and $q=10^{-3}-1$) have $\agwcm<\rchar$ and thus have an inner hardening phase. In contrast, the most massive SMBHBs in the fiducial \cagw\ model are already in the GW regime at $\rchar$.

 \begin{figure*}
    \centering
    \includegraphics[width=0.48\textwidth]{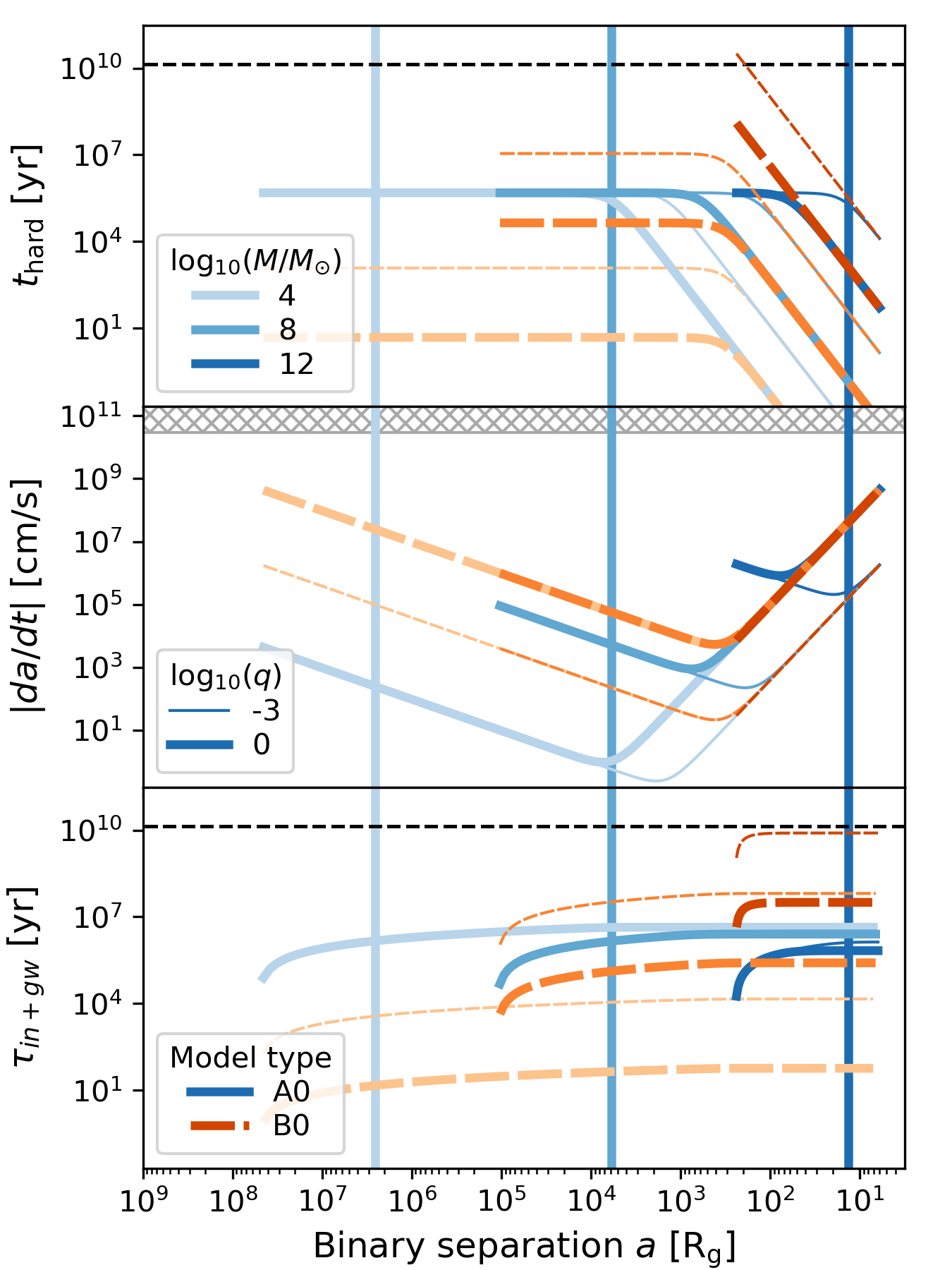}
    \includegraphics[width=0.48\textwidth]{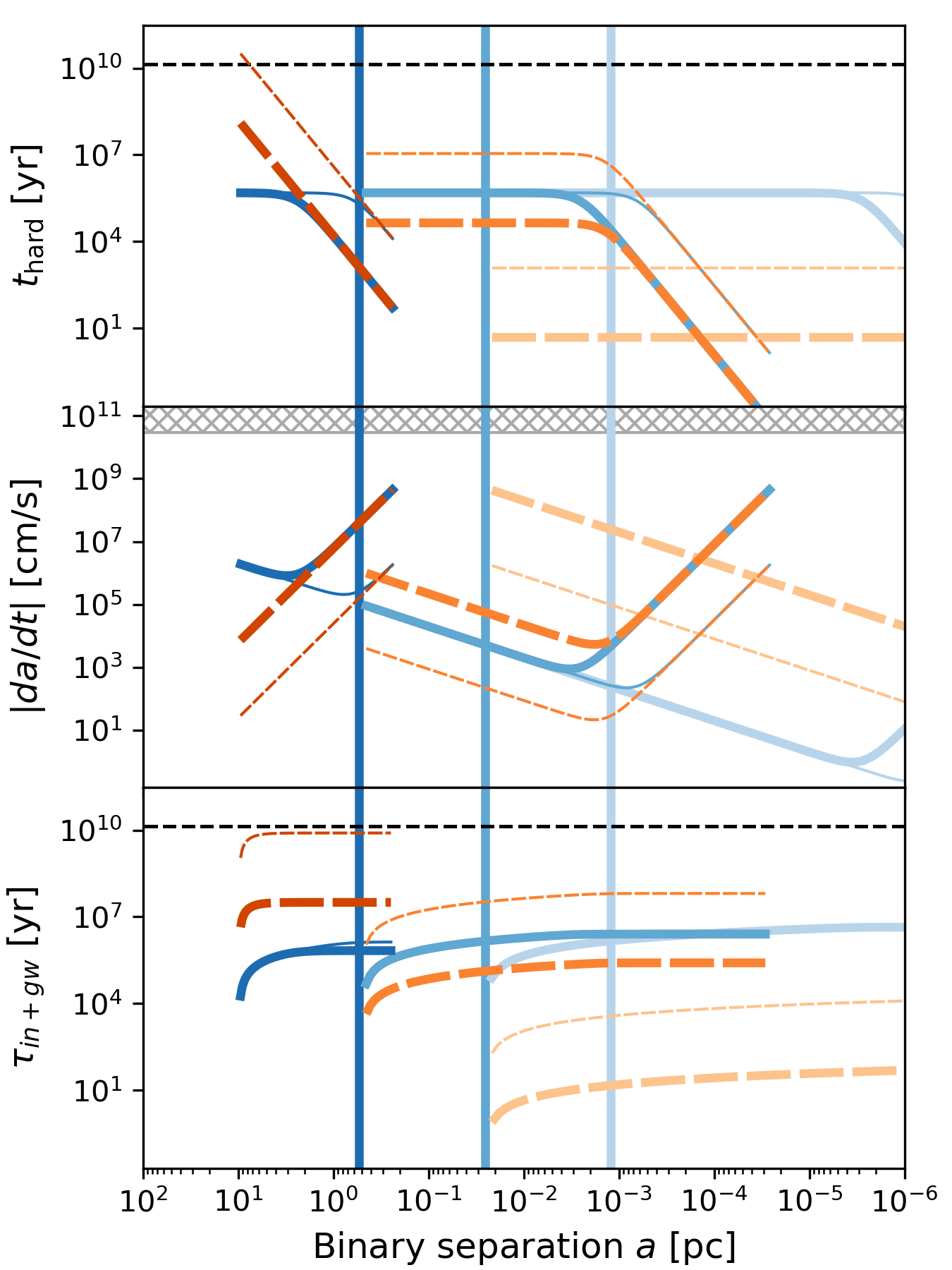}
    \caption{SMBHB hardening timescales and rates versus separation are shown for the fiducial hardening models \ctin\ (blue solid curves) and \cagw\ (orange dashed curves). The $x$-axis in the {\em left panels} is the binary separation $a$ in units of $\rg$ (which scales linearly with $M$), while the {\em right panels} use units of pc. Note that $a$ decreases from left to right on the plot, in the direction of forward time. All curves start at $\rchar$ and stop at $\risco$; the values of both quantities vary with $M$. In all panels, line color scales with $M$ and line thickness scales with $q$, as indicated in the legends. The {\em top panels} show the binary hardening timescale $t=a/(-\dadt)$, the {\em middle panels} show the hardening rate $|\dadt|$, and the {\em bottom panels} show the cumulative inspiral time from $\rchar$ to $\risco$), $\tauingw$. Curves are shown for four values of mass $M$ and mass ratio $q$ spanning the range used in the GWB calculations. Note that in the \cagw\ model, $\dadt$ versus $a [\rg]$ has no mass dependence, such that the orange dashed lines for different masses overlap in the middle left panel. The vertical solid lines mark the low-frequency edge of the PTA regime, defined here as the orbital separation at which a circular binary of mass $M$ has a GW frequency of $(20\, {\rm yr})\inv$. The horizontal dashed black line in the top and bottom panels marks $\thub$, and the gray hatched region in the middle panels denotes hardening rates that are disallowed by the $\dadtmax$ criterion. PTA measurements of the GWB are sensitive to the transition of SMBHBs from the astrophysical to the GW regime.}
    \label{fig:fid_dadt}
\end{figure*}

\begin{figure}
    \centering
    \includegraphics[width=\linewidth]{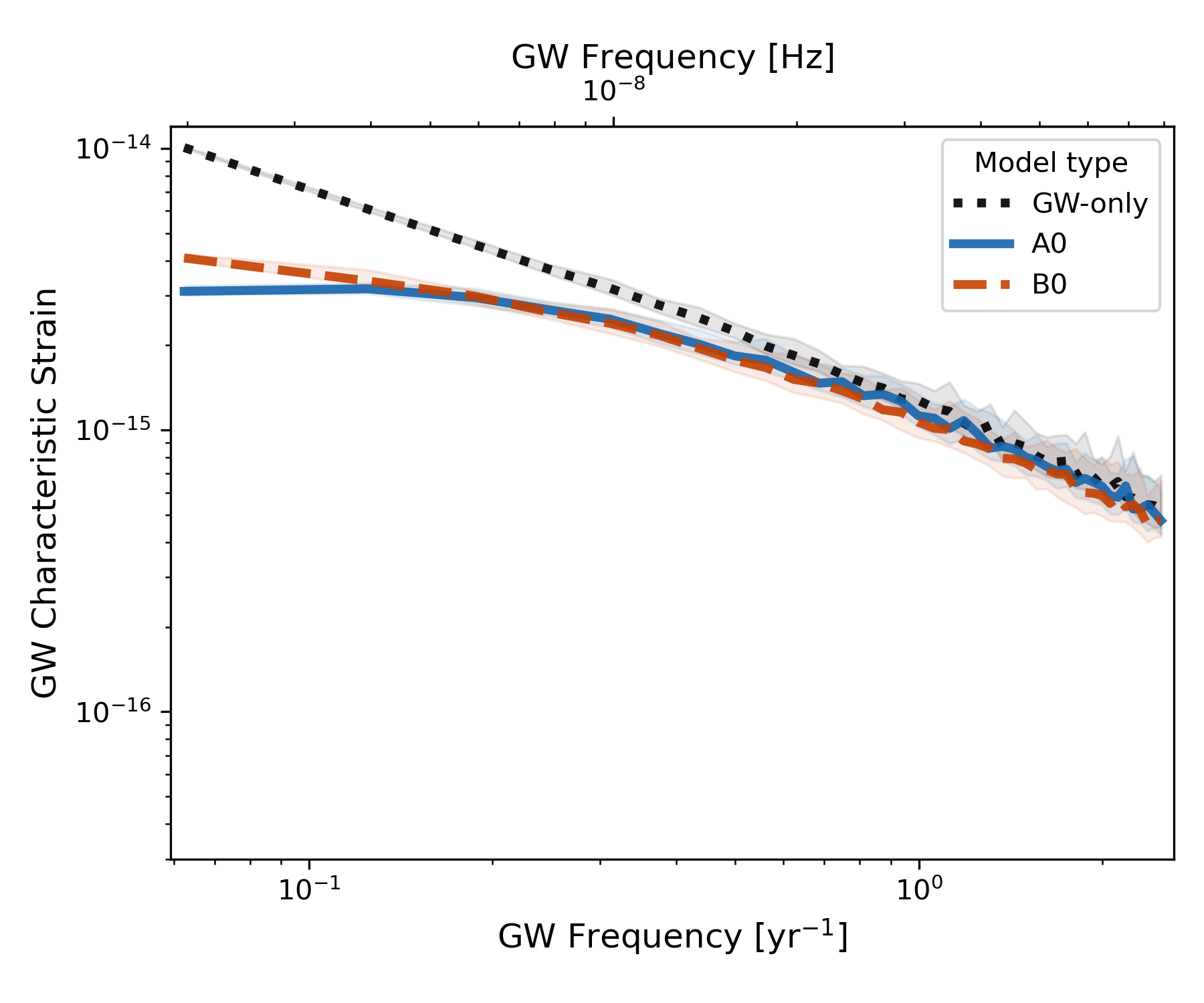}
    \caption{GWB spectra are shown for the same fiducial hardening models as in Figure~\ref{fig:fid_dadt}: \ctin\ (blue solid curves) and \cagw\ (orange dashed curves). For comparison, the black dotted line shows a GW-only hardening model as described in Section~\ref{ssec:gwonly}. The GWB for GW-only hardening deviates slightly from a pure power-law owing to discrete source effects at high frequencies. Shaded regions denote the 50\% interval over 100 realizations of the GWB for each model. Both the \ctin\ and the \cagw\ models have low-frequency attenuation due to astrophysical hardening, though only the \ctin\ model has an actual low-frequency turnover.}
    \label{fig:fid_gwb}
\end{figure}

The right-hand panels of Figure~\ref{fig:allowedparams} show how the allowed parameter space varies with binary mass and mass ratio. The minimum $\nuin$ values vary little over three orders of magnitude in mass ratio (dark vs light lines)
but varies strongly with binary mass. In both the \ctin\ and the \cagw\ models, the lowest-$M$, lowest-$q$ binaries place the strongest constraints on $\nuin$ for most values of $\agw$.

In all panels of Figure~\ref{fig:allowedparams}, the yellow-green dotted line denotes the orbital separation at which circular binaries of a given mass enter the PTA regime (defined here as $f_{\rm obs,min}=(20{\rm yr})^{-1})$. $\rchar>a_{\rm orb}(f_{\rm obs,min})$ is required to ensure that all GW emission at relevant frequencies is included. For the fiducial value $\alphach=-2/3$, $\rchar$ and $a(f_{\rm obs,min})$ have the same mass scaling, and all three plotted values of $\rchnine=0.1, 1, 10$pc satisfy this criterion. The fiducial value $\rchnine=1$pc corresponds to $10^{4.3}\rg$ for $M=10^9\msun$, such that $\rchar$ ranges from $0.02{\rm pc}=10^{7.7}\rg$ at $M=10^4\msun$ to $\rchar=10{\rm pc}=10^{2.3}\rg$  at $M=10^{12}\msun$.

\subsection{Semi-analytic SMBHB population models}
\label{ssec:sams}
I use the semi-analytic / semi-empirical SMBHB population synthesis code \texttt{holodeck} to create cosmological populations of SMBHBs, apply analytic hardening models, and calculate the resulting GWB spectra. A detailed description of the procedure and the elements of the SMBHB population models can be found in NG15Astro, with some recent updates as outlined in \citet{kelley24} and \citet{matt26a}. I briefly review the most salient points here.

The desired input for GWB calculations is the number density of SMBHBs as a function of $M$, $q$, $z$, and $a$. This depends on the number density of galaxies (i.e., the GSMF), the relation between SMBH and galaxy masses, and the galaxy merger rate. Combining these elements yields a number density of binaries forming (at a given $M$, $q$, \& $z$). The SMBHB hardening models then specify the orbital evolution of $a$ from formation to merger. 

I use the $z=0$ $M_{\rm BH}-M_{\rm bulge}$ relation of  \citet{kormendy13} as in NG15Astro. Following \citet{matt26a}, I use the double-Schechter GSMF of \citet{leja20}  and the \citet{rodriguez15} galaxy merger rates derived from the Illustris cosmological simulation \citep[e.g.,][]{nelson15b}. I do not consider redshift evolution of the GSMF or $M_{\rm BH}-M_{\rm bulge}$ relation in this work. The rest of this paper focuses on the impact of SMBHB hardening parameters on GWB spectra, rather than the impact of these galactic-scale parameter choices.

\section{Results}
\label{sec:results}

Figure~\ref{fig:fid_dadt} shows $\thard$, $\dadt$, \& $\tauingw$ versus separation for the fiducial \ctin\ and \cagw\ model parameters, for a range of binary masses and mass ratios, and Figure~\ref{fig:fid_gwb} shows the corresponding GWB spectra. I discuss the \ctin\ model in Section~\ref{ssec:fiducial} and the \cagw\ model in Section~\ref{ssec:cagw}.

\subsection{The constant-timescale \ctin\ model}
\label{ssec:fiducial}

The binary hardening timescales and rates for the \ctin\ model are shown via the blue curves in Figure~\ref{fig:fid_dadt}. The middle panels show the hardening rate $|\dadt|$ versus separation, illustrating two key features of this hardening model: the transition radius $\agw$ between the inner and GW-dominated hardening, and the constant power-law index $\nuin$ for inner astrophysical hardening  at $\rchar>a>\agw$. 
The vertical solid lines in Figure~\ref{fig:fid_dadt} show the binary separation at which an equal-mass, circular BH binary of mass $M$ at $z\sim 0$ would enter the low-frequency edge of the PTA regime, defined here as $f_{\rm obs} = (20\,{\rm yr})\inv=1.6\times\rev{10^{-9}}$ Hz. $10^9\msun$ binaries evolve from $\rchar=1$ pc to $a=0.057$ pc before reaching this frequency and then shrink by another factor of 3.8 before transitioning to the GW regime at $\aninerg=10^{2.5}\rg$ ($\aninecm=0.015$ pc). Because $t_{\rm hard,in}$ is constant in the \ctin\ model, the cumulative inspiral time $\tauingw$ (lower panels) varies by less than a factor of 8, from $\tauingw=0.64$-$5.1$ Myr, across the masses and mass ratios considered.

For all binaries, $\rchar$ is $\sim 18\times$ larger than the separation at which $z\sim0$ sources enter the PTA regime,
and it is a factor of $\sim 6$ larger than the corresponding separation for $f_{\rm em}=(100 \,{\rm yr})\inv$. 
$\rchar>\agw$ for most binaries, but $M>2.4\times10^{10}\msun$ binaries are already GW-dominated when they enter the PTA band. Within the inner astrophysical hardening phase, $t_{\rm hard,in}=4.9\times10^5$ yr for all binaries. 
$\agwrg$ ranges from $10^{3.75}\rg=2.7\times10^{-6} $pc for equal-mass $M=10^4\msun$ binaries to $10\rg=0.48$ pc for binaries with $M=10^{12}\msun, q=10^{-3}$. 

The middle-left panel of Figure~\ref{fig:fid_dadt} illustrates that $\dadt_{\rm GW}$ is exactly the same for all binaries of a given mass ratio. In other words, the GW-driven hardening rate ($\propto M^3/a^3$) is independent of binary mass when $a$ is expressed in $\rg$ units. The parameter $\agw$ simply specifies where binaries begin to follow this GW-driven curve. 
For the \cagw\ model, this transition has no $M$ or $q$ dependence at all (i.e., all the orange curves in the middle-left panel exactly overlap). This further motivates the inside-out hardening model: modeling PTA-observable GWs from SMBHBs requires only specifying the point at which binaries transition to the GW regime and the (often small) portion of  astrophysical evolution that falls within the PTA frequency range.

The blue solid curve in Figure~\ref{fig:fid_gwb} shows the GWB spectrum for the \ctin\ model. It is significantly attenuated relative to a pure power-law, with a low-frequency turnover. The median characteristic strain amplitude is $10^{-14.5}$ at $(10\, {\rm yr})\inv$ and $10^{-14.9}$ at yr$\inv$. \rev{This is about a factor of two lower than the GWB amplitude inferred in \citet{agazie23a}, as discussed above.} Note that because these SMBHB populations are not fit to GW data, the absolute GWB amplitudes are less important than the relative differences between hardening models, which I focus on in the remainder of this paper. \rev{In particular, as is described in more detail below, variation in $\aninerg$ can produce changes in GWB amplitude that are much larger than this factor of two.} 

\subsection{The constant-$\agw$ \cagw\ model}
\label{ssec:cagw}

The orange curves in Figure~\ref{fig:fid_dadt} show the binary hardening rates and timescales for the \cagw\ model. All binaries follow the same $|\dadt|$ versus $a$ curve when $a$ is plotted in $\rg$ units, modulo the scaling with mass ratio. $\thard$ and $\tauingw$ both scale linearly with $M$, such that massive, low-mass-ratio binaries take longer to merge. Because $\agwrg=10^{2.5}\rg$ is independent of $M$, $\agwcm$ varies linearly with $M$. As in the \ctin\ model, the most massive binaries are already in the GW phase at $\rchar$; in the \cagw\ model this occurs for $M>10^{10}\msun$. $q=1,\: M=10^9\msun$ binaries in the \cagw\ model have a total inspiral time from 1pc of $\sim 10^{6.5}$ yr, which is very similar to the \ctin\ model. Unlike the \ctin\ model, though, $\tauingw$ varies by about eight orders of magnitude across the masses and mass ratios considered, from $< 100$ yr to almost $10^{10}$ yr.

The GWB spectrum above $f_{\rm obs}\sim6$ nHz is largely identical for the \ctin\ and \cagw\ models (Figure~\ref{fig:fid_gwb}). At lower frequencies, the \cagw\ GWB is shallower than a pure power-law, but it has less attenuation than the \ctin\ GWB. Its amplitude at 2 nHz is $\sim 30\%$ higher, and its spectrum does not actually ``turn over."

\subsection{Dependence on $\aninerg$}
\label{ssec:a9}

\begin{figure*}
    \centering
    \begin{subfigure}{0.495\textwidth}
    \centering
    \includegraphics[width=0.8\textwidth]{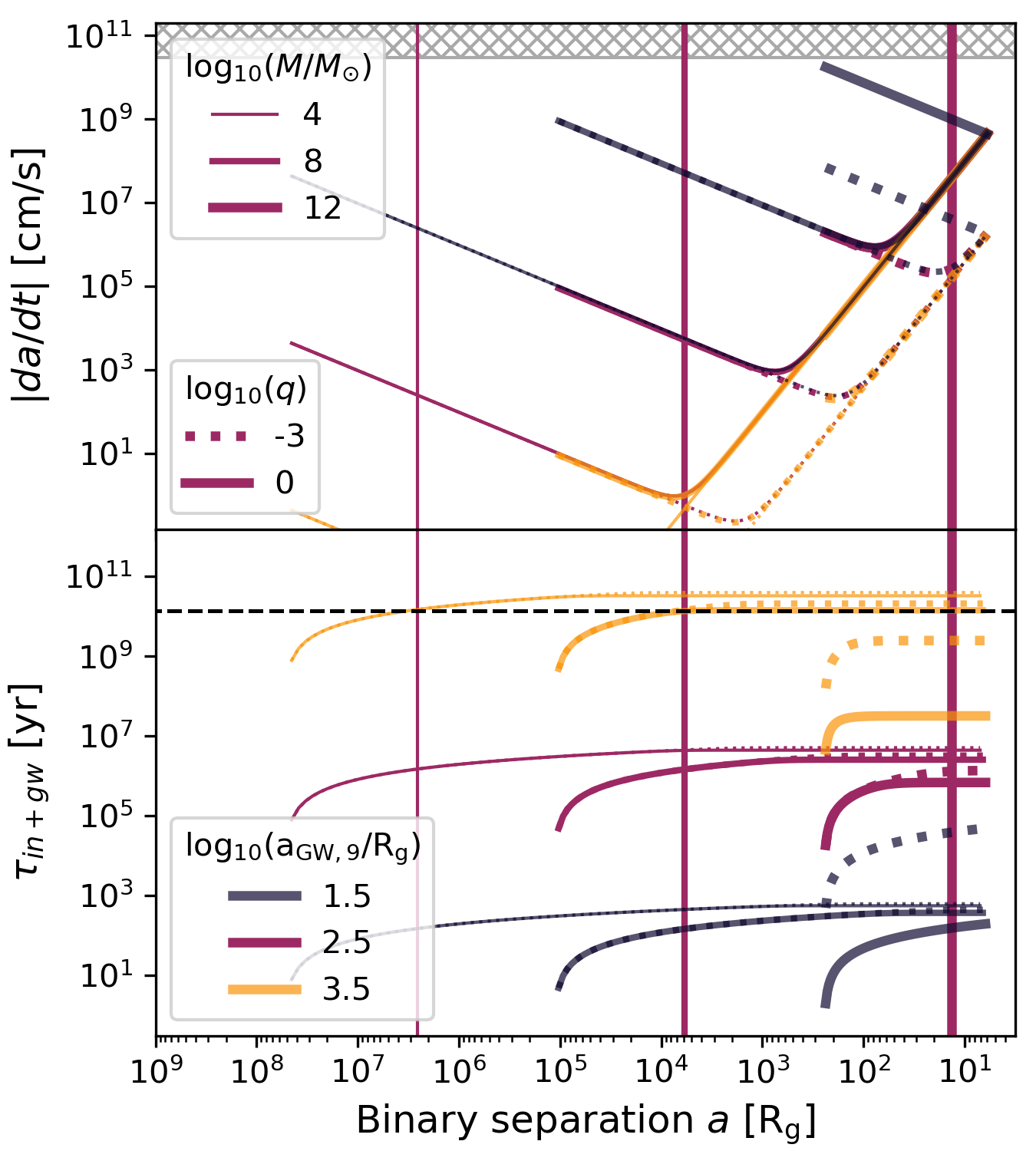}
    \vspace{-5pt} 
    \caption{\label{subfig:ctin_r9var_dadt}}
    \end{subfigure}
    \begin{subfigure}{0.495\textwidth}
    \centering
    \includegraphics[width=0.72\textwidth]{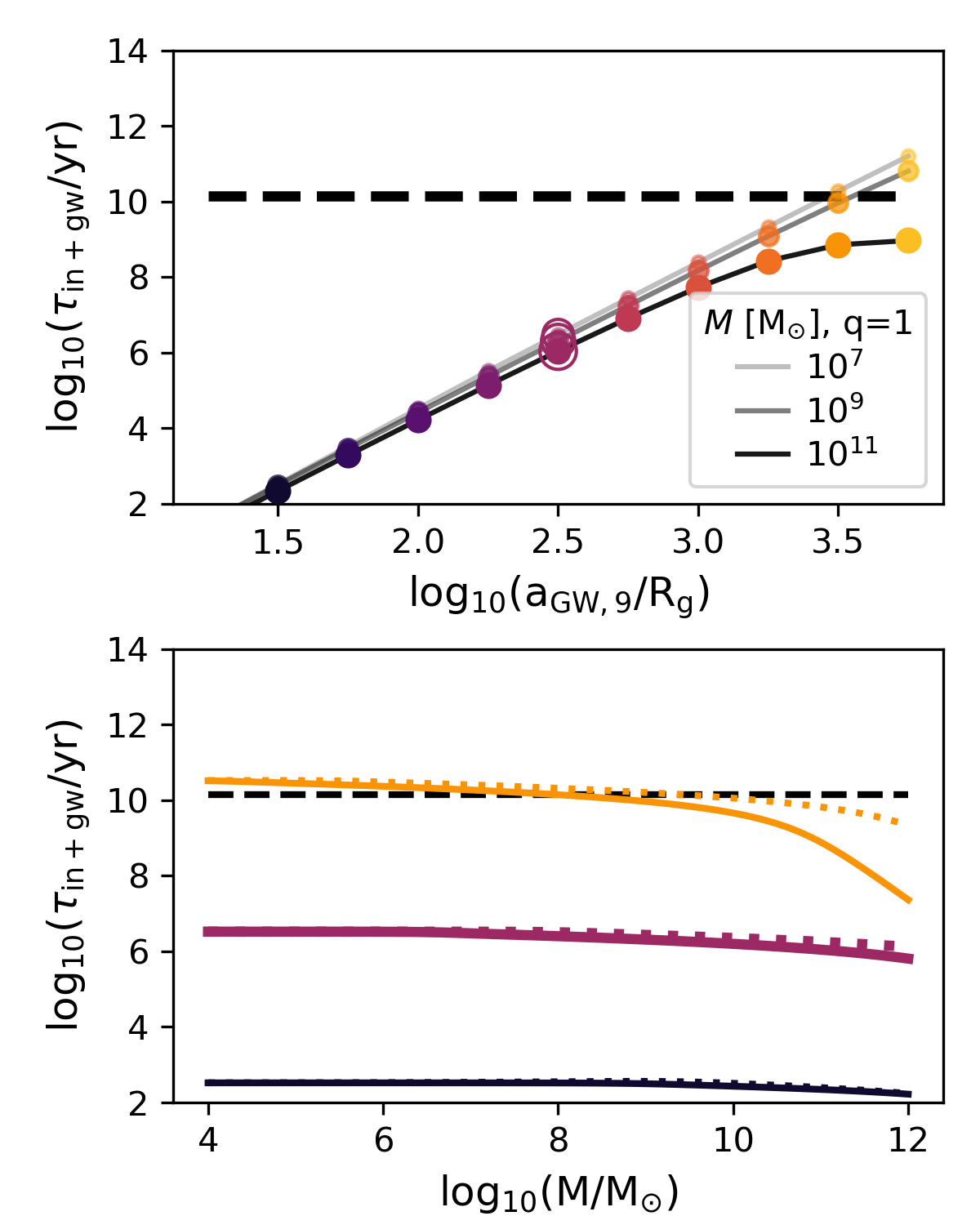}
    \vspace{-5pt} 
    \caption{\label{subfig:ctin_r9var_tau}}
    \end{subfigure}
    \begin{subfigure}{0.54\textwidth}
    \centering
    \includegraphics[width=\textwidth, trim=0 0.6cm 0 0.5cm,clip]{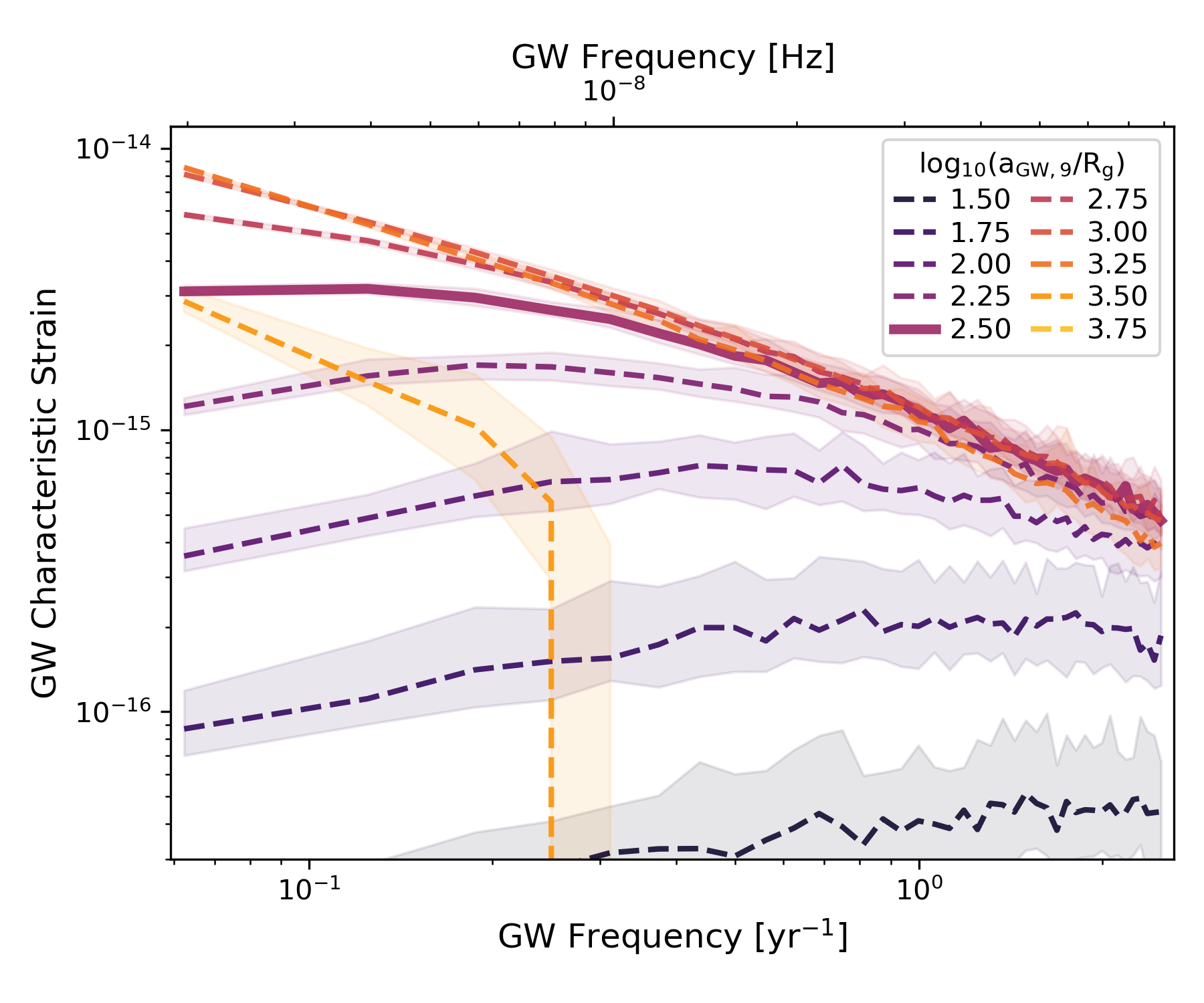}
    \vspace{-10pt} 
    \caption{\label{subfig:ctin_r9var_gwb}}
    \end{subfigure}
    \begin{subfigure}{0.42\textwidth}
    \centering
    \includegraphics[width=0.85\textwidth,trim=0 0cm 0 0.35cm,clip]{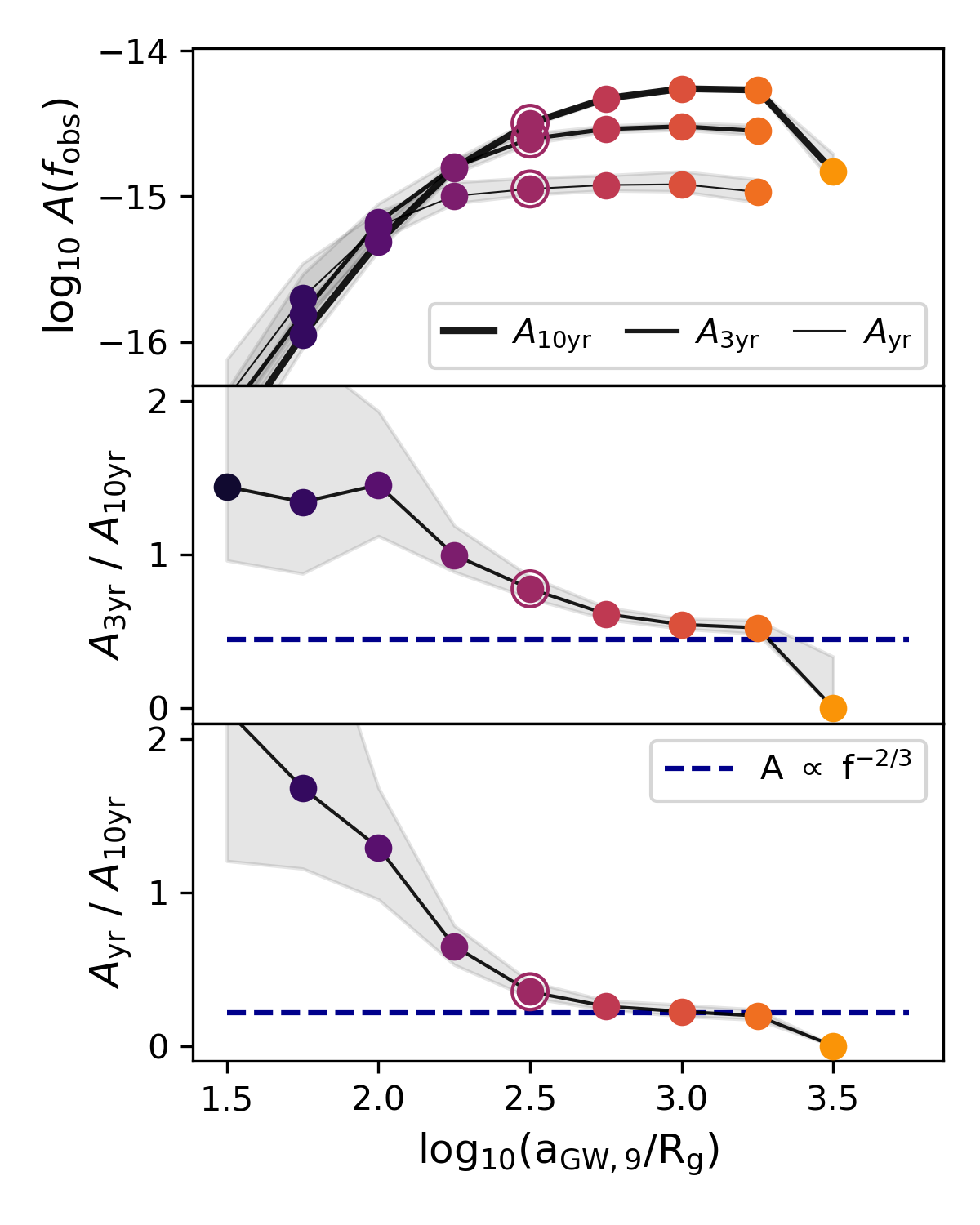}
    \vspace{-12pt} 
    \caption{\label{subfig:ctin_r9var_amps}}
    \end{subfigure}
    \vspace{-10pt}     
    \caption{Key results for the \ctin\aninevar\ model that varies $\aninerg$ with other parameters at fiducial \ctin\ values. The color scale indicates the value of $\aninerg$. {\bf(a)} {\em Top panel:} binary hardening rate versus separation (in $\rg$ units)  for select masses, mass ratios, and $\aninerg$ values, in a similar manner as the middle panels of Figure~\ref{fig:fid_dadt}. Here, thicker curves denote more massive binaries, solid curves denote $q=1$, and dotted curves denote $q=10^{-3}$. {\em Bottom panel:} $\tauingw$ versus separation.  {\bf(b)} {\em Top panel:} $\tauingw$ versus $\aninerg$ for equal-mass binaries with $M=10^7, 10^9, \& 10^{11}$, as indicated in the legend. Filled$+$open circles denote the fiducial model. {\em Bottom panel:} $\tauingw$ versus $M$ for the same three $\aninerg$ values as in (a). Solid (dotted) lines denote $q=1$ ($q=10^{-3}$). In both panels, the black dashed line indicates $\thub$. {\bf(c)} GWB spectra for $\aninerg$ varied between $10^{1.5}\rg$ and $10^{3.75}\rg$. (Note that $\aninerg=10^{3.75}\rg$ does not produce an observable GWB and does not appear on the plot.) The thick solid line denotes the fiducial model. Shaded regions denote the 50\% range over 100 GWB realizations. {\bf(d)} {\em Top panel:} The GWB characteristic strain amplitude at three frequencies versus $\aninerg$. {\em Middle panel:} Ratio of GWB amplitude at $f_{\rm obs}=(3\, {\rm yr})\inv$ \& $(10\, {\rm yr})\inv$, versus $\aninerg$. {\em Bottom panel:} GWB amplitude ratio at $f_{\rm obs}={\rm yr}\inv$ \& $(10\, {\rm yr})\inv$, versus $\aninerg$. The  dashed lines indicate the ratios for $A\propto f^{-2/3}$. In all panels, shaded regions denote the 50\% range over 100 realizations. Peak GWB amplitudes occur for $\aninerg=1000-2000\rg$, with no observable GWB for $\aninerg<10^{1.5}\rg$ or $>10^{3.5}\rg$. Between these limits, the GWB is quite sensitive to $\aninerg$, with large variations in spectral shape, and amplitudes varying by factors of up to $\sim10^3$.}
    \label{fig:ctin_r9var}
\end{figure*}

\begin{figure*}
    \centering
    \begin{subfigure}{0.45\textwidth}
    \includegraphics[width=\textwidth]{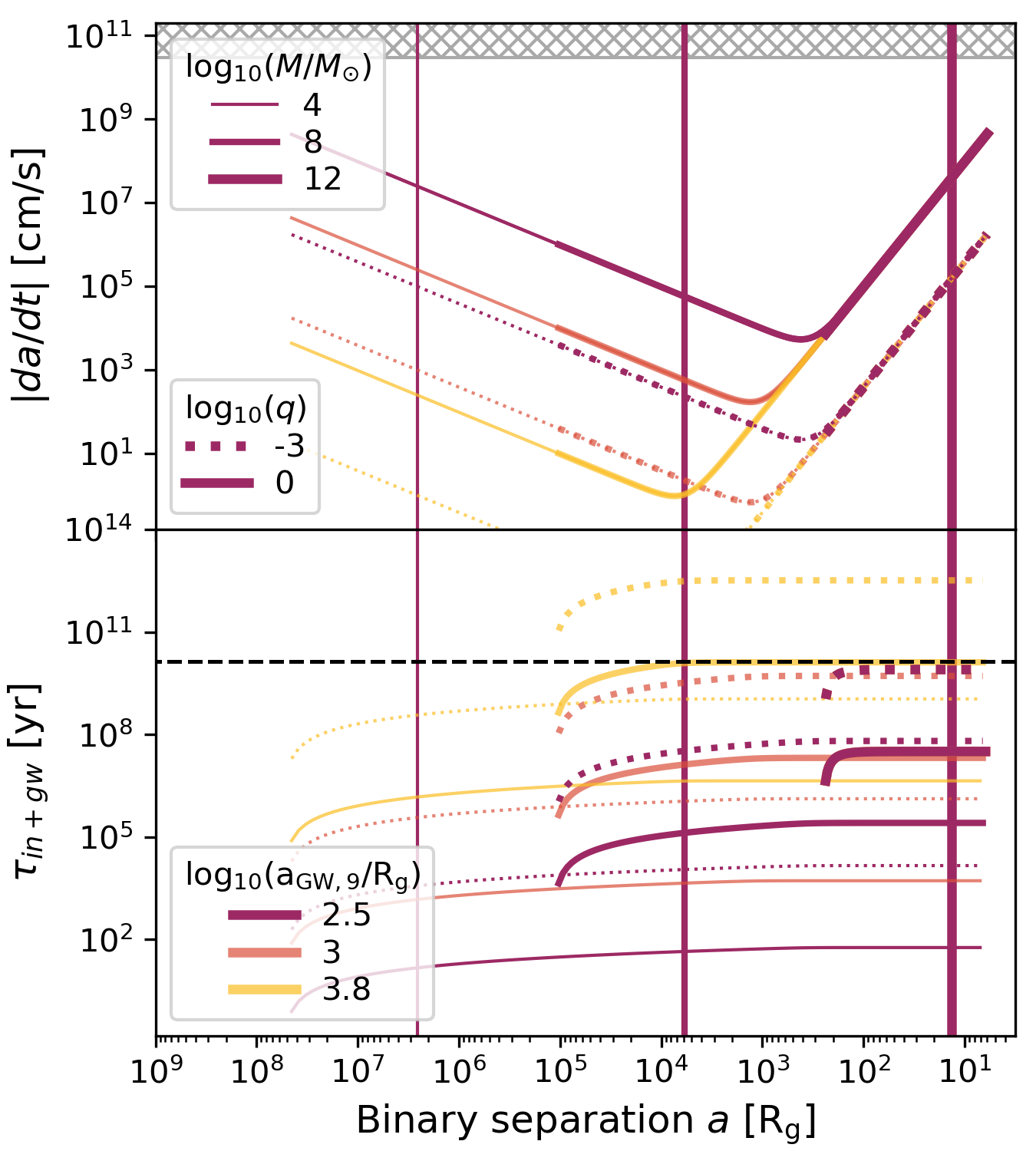}
    \caption{\label{subfig:cagw_r9var_dadt}}
    \end{subfigure}
    \begin{subfigure}{0.41\textwidth}
    \includegraphics[width=\textwidth]{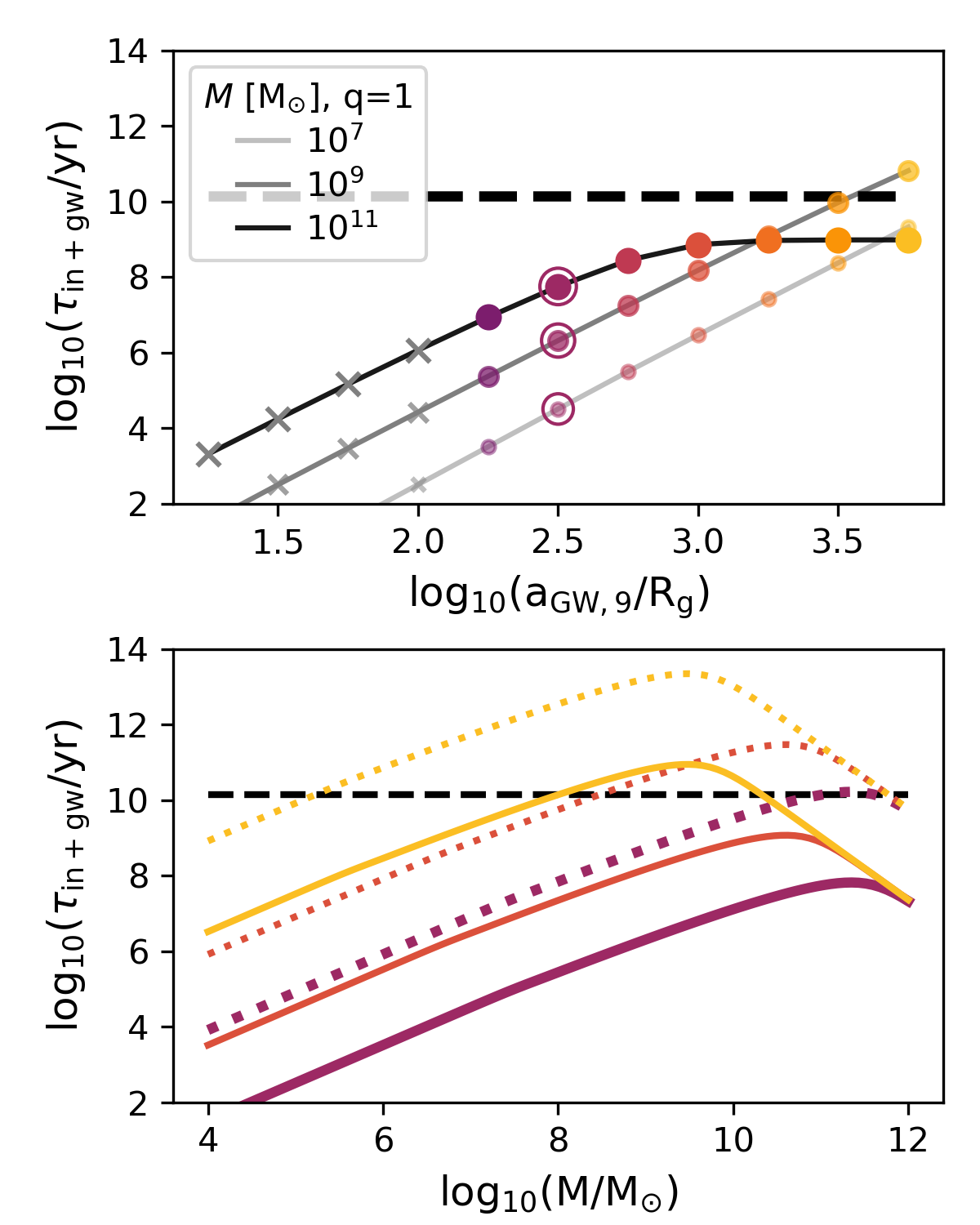}
    \caption{\label{subfig:cagw_r9var_tau}}
    \end{subfigure}
    \begin{subfigure}{0.52\textwidth}
    \includegraphics[width=\textwidth]{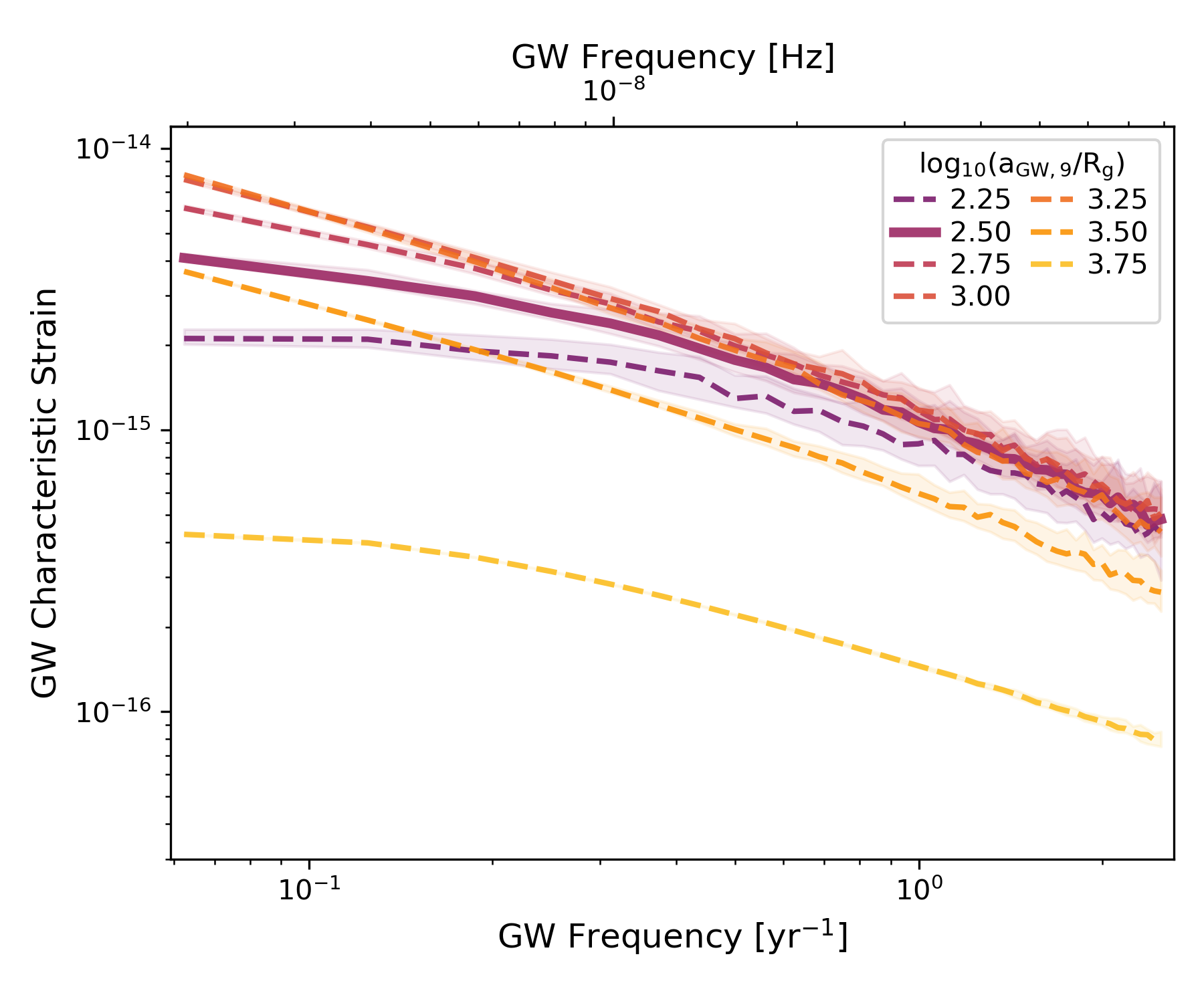}
    \caption{\label{subfig:cagw_r9var_gwb}}
    \end{subfigure}
    \begin{subfigure}{0.34\textwidth}
    \includegraphics[width=\textwidth]{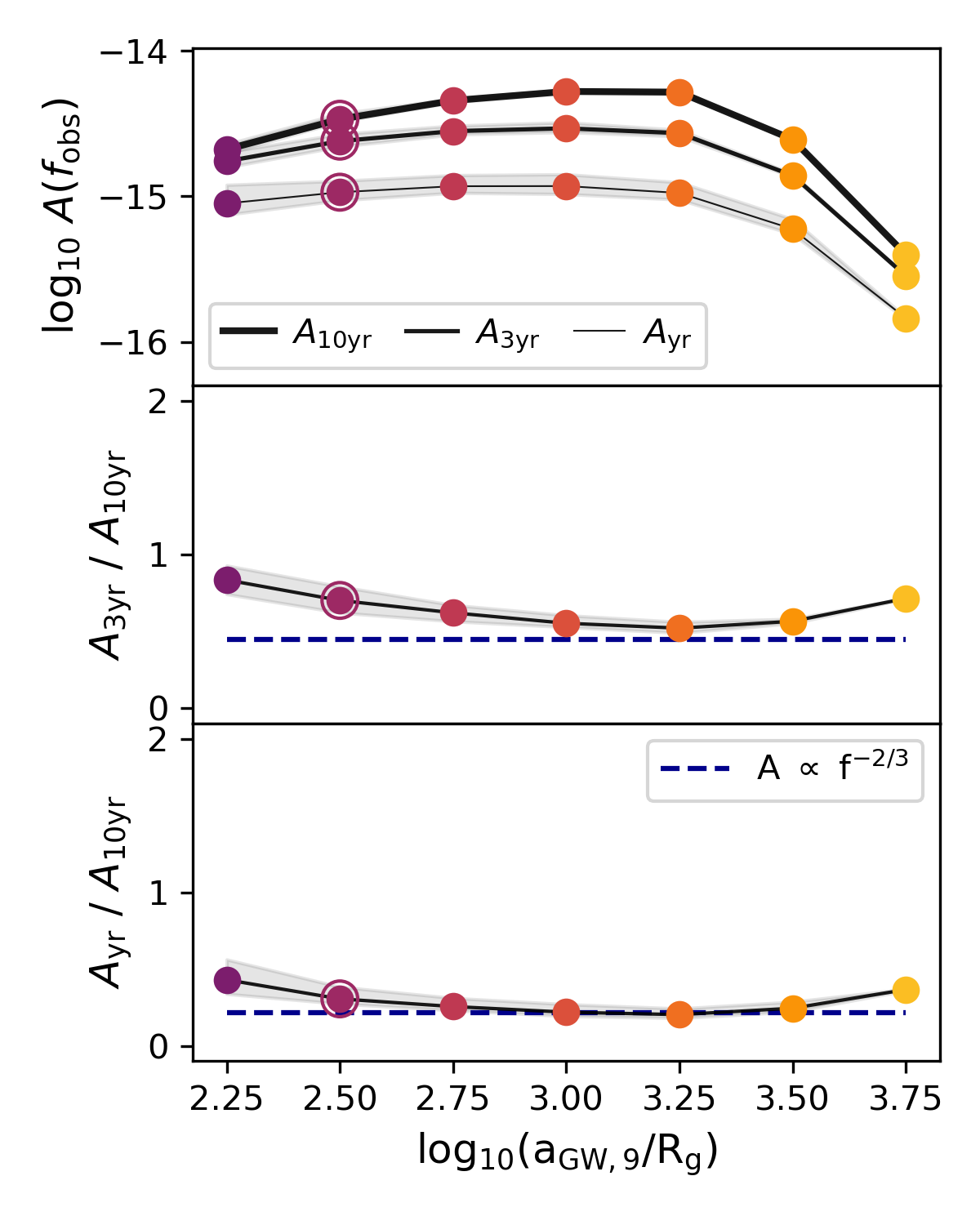}
    \caption{\label{subfig:cagw_r9var_amps}}
    \end{subfigure}
    \caption{Key results for the \cagw\aninevar\ model that varies $\aninerg$ with other parameters at fiducial \cagw, in the same manner as in Figure~\ref{subfig:ctin_r9var_dadt}-\ref{subfig:ctin_r9var_amps}. The upper panel of  Figure~\ref{subfig:cagw_r9var_tau} includes small $\aninerg$ values that do not satisfy $\dadt<\dadtmax$ (gray $\times$ symbols). As in the \ctin\aninevar\ models, the GWB amplitudes peak at $\aninerg=1000-2000\rg$, but the high-frequency amplitudes do not drop as sharply for larger $\aninerg$. Small $\aninerg$ ($\leq100\rg$) are disallowed by the $\dadtmax$ criterion.}
    \label{fig:cagw}
\end{figure*}

Figures~\ref{subfig:ctin_r9var_dadt} \& Figure~\ref{subfig:ctin_r9var_tau} show binary inspiral for \ctin\aninevar\ model variants with $\aninerg$ ranging from $10^{1.5}$-$10^{3.75}\rg$.
Massive binaries with small $\agw$ approach relativistic $|\dadt|$ at $\rchar$, and smaller values of $\aninerg$ for which $|\dadtrc| > c$ are not shown. $\tauingw$ increases sharply from a few hundred yr for $\aninerg=10^{1.5}\rg$ to a few $\times 10^{10}$ yr for $\aninerg=10^{3.75}\rg$. 

The resulting GWB spectra  are shown in Figure~\ref{subfig:ctin_r9var_gwb}, and Figure~\ref{subfig:ctin_r9var_amps} shows the GWB amplitude ratios $A_{\rm 3yr}/A_{\rm 10yr}$ and $A_{\rm 1yr}/A_{\rm 10yr}$, with a dashed line marking ratios for an $A\propto f^{-2/3}$ power law.  Both the GWB amplitude and spectral shape vary dramatically across the $\aninerg$ values considered. For $\aninerg\lesssim100 \rg$, the GWB spectrum is essentially flat with very low amplitudes, due to strong astrophysical hardening. As $\aninerg$ increases, so does the GWB amplitude, until it reaches a peak at $\aninerg\sim1000-2000 \rg$. At this peak there is  little to no low-frequency attenuation, and the spectra closely resemble an $f^{-2/3}$ power-law model for idealized GW-only inspiral. These cases represent a near-perfect balance in which massive binaries are efficiently brought to the edge of the PTA regime by astrophysics and then driven through it predominantly by GWs.   

At larger values of $\aninerg$, most binaries are unable to reach the PTA band within $\thub$, and the GWB amplitude drops precipitously from near maximal at $\aninerg=10^{3.25}\rg$, to zero above $f=0.4\, {\rm yr}\inv$ at $\aninerg=10^{3.5}\rg$ (where low-mass binaries stall), to zero overall  at $\aninerg=10^{3.75}\rg$ (where virtually all binaries stall). The amplitude ratios in Figure~\ref{subfig:ctin_r9var_amps} show a clear trend from very flat spectral shapes (and low amplitudes) at small $\aninerg$, to a nearly power-law shape at $\aninerg=1000\rg$, to a steep spectrum at the largest $\aninerg$, before the GWB disappears entirely. 

We see that only $\aninerg$ in the range $10^{1.5}$-$10^{3.5}\rg$ are consistent with an observable GWB in this model, and the extremes of this range are only marginally observable. (The low-frequency amplitude is suppressed by a factor of $\sim1000$ at $\aninerg=10^{1.5}\rg$ relative to the maximum at $\sim10^3\rg$.) In other words, simply by requiring that binary evolution obeys $\dadtmax<c$ and produces a detectable GWB, $\aninerg$ is constrained to within a factor of $\sim 30-100$. And within that range, the low-frequency amplitude varies by an even larger factor, with huge variation in spectral shapes. Similar results are found for the other models in Table~\ref{tab:models}, discussed in more detail below. The range of allowed $\aninerg$ values that produce an observable GWB varies between the models but in all cases span $\lesssim$ two orders of magnitude and have significant spectral variation within this range. The GWB is therefore quite sensitive to $\aninerg$, motivating its use as a free parameter to be constrained by PTAs. 

The default GSMF, galaxy merger rate, and $M_{\rm BH}-M_{\rm bulge}$ parameters used in this work produce GWB amplitudes slightly below the published value of $2.4\times10^{-15}$ at $f={\rm yr}\inv$ from \citep{agazie23a}, even for the hardening models that maximize GW emission in the PTA band. The tension between the observed GWB and that produced by SMBHB population models that match EM constraints has been discussed extensively in recent literature \citep[e.g.,][]{agazie23e,sato23,liepold24,sato25a,sato25b,matt26a,matt26b,harris26,buttigieg26}. Because the SMBHB models in this work are not fit to PTA data, the relative GWB amplitudes and shapes are more informative than their exact values. Clearly, however, SMBHB hardening models with $\aninerg\lesssim100\rg$ will be inconsistent with observed GWB spectra that do not feature a very strong low-frequency turnover. Matching such a strongly attenuated GWB would require SMBHB masses and/or number densities well above those inferred from the $z\sim0$ EM constraints.

Figure~\ref{fig:cagw} explores variation of $\aninerg$ in the \cagw\aninevar\ models.
In Figure~\ref{subfig:cagw_r9var_tau}, models with $\agwrg<10^{2.25}\rg$ exceed $|\dadt|_{\rm max}=c$ and are indicated with gray ``$\times$'s."
$\tauingw$ increases with $\aninerg$ (corresponding to more astrophysical stalling), but unlike the \ctin\aninevar\ models, here $\tauingw$ also increases with $M$ and $q$, up to the highest masses where strong GW emission shortens the merger time. The $\tauingw$ versus $M$ curves thus have a peak at $M\sim10^{9}-10^{11}\msun$. This effect is most dramatic for the largest value of $\aninerg=10^{3.75}\rg$, for which equal-mass binaries in the mass range $\sim 10^8-10^{10}\msun$ have $\tauingw>\thub$ and therefore do not contribute to the GWB. (For comparison, no binaries above $\sim10^{10}\msun$ merge within $\thub$ for the \ctin\aninevar\ model with $\aninerg=10^{3.75}\rg$.)

Figures~\ref{subfig:cagw_r9var_gwb} \& \ref{subfig:cagw_r9var_amps} show the corresponding GWB amplitudes and amplitude ratios. 
As in the \ctin\aninevar\ models, the amplitudes peak at $\aninerg\sim 10^3-10^{3.25}\rg$ with a nearly $f^{-2/3}$ spectrum. However, the spectral shape varies less with $\aninerg$ than in the \ctin\aninevar\ models, and the trend in amplitude ratios is not monotonic. 
Rather than dramatically steepening for $\agwrg\geq10^{3.25}\rg$, the spectrum again becomes shallower than a power-law, and the amplitudes for $\aninerg=10^{3.75}\rg$ are only $\sim8\times$ lower than the fiducial model (rather than disappearing entirely). This difference between the \ctin\aninevar\ and \cagw\aninevar\ models arises from their different trends in $\tauingw$ with mass. The inspiral timescales in the \cagw\aninevar\ model vary strongly and non-monotonically with $M$, such that highest- and lowest-$M$ SMBHBs merge efficiently while $M\sim10^8-10^{10}\msun$ SMBHBs stall for large $\aninerg$. GWB amplitudes are suppressed for large $\aninerg$ in the \cagw\aninevar\ models, but less severely, and with little change in the GWB spectral shape.

The effects of varying $\aninerg$ can be summarized as follows. Minimum $\aninerg$ values are firmly limited by $\dadtmax<c$ ($\aninerg\gtrsim10^{1.5}\rg$ and $\gtrsim10^{2}\rg$ for \ctin\aninevar\ and \cagw\aninevar, respectively). Maximum $\aninerg$ values are firmly limited by the fact that only binaries with $\tau_{\rm tot}<\thub$ ($\aninerg\lesssim10^{3.75}\rg$ for \ctin\aninevar\ and \cagw\aninevar) can produce an observable GWB. $\aninerg$ is therefore constrained to an allowed range of $\lesssim$ two orders of magnitude, and the GWB shape and amplitude vary widely within that range. 

For small $\aninerg$ values, astrophysical hardening causes strong GWB attenuation. Intermediate values of $\aninerg$ ($\sim 1000-2000\rg$ for these models) optimize the balance between efficient astrophysical hardening that brings binaries to the PTA regime but allows GWs to dominate thereafter, yielding a maximal GWB amplitude with little low-frequency attenuation. 
For large $\aninerg$, binary hardening begins to stall. The GWB shape and amplitude depend very sensitively on {\em which} binaries stall outside the PTA regime versus those that have $\tauingw\lesssim\thub$ and spend maximal time in the PTA band. The variation in $\tauingw$ with $M$ dominates these effects in the  \ctin\aninevar\ and \cagw\aninevar\ models. I explore these mass trends further in the next subsection.

\begin{figure*}
    \centering
    \subfloat[\label{subfig:ctin_alphgwvar_dadt}]{\includegraphics[width=0.45\textwidth]{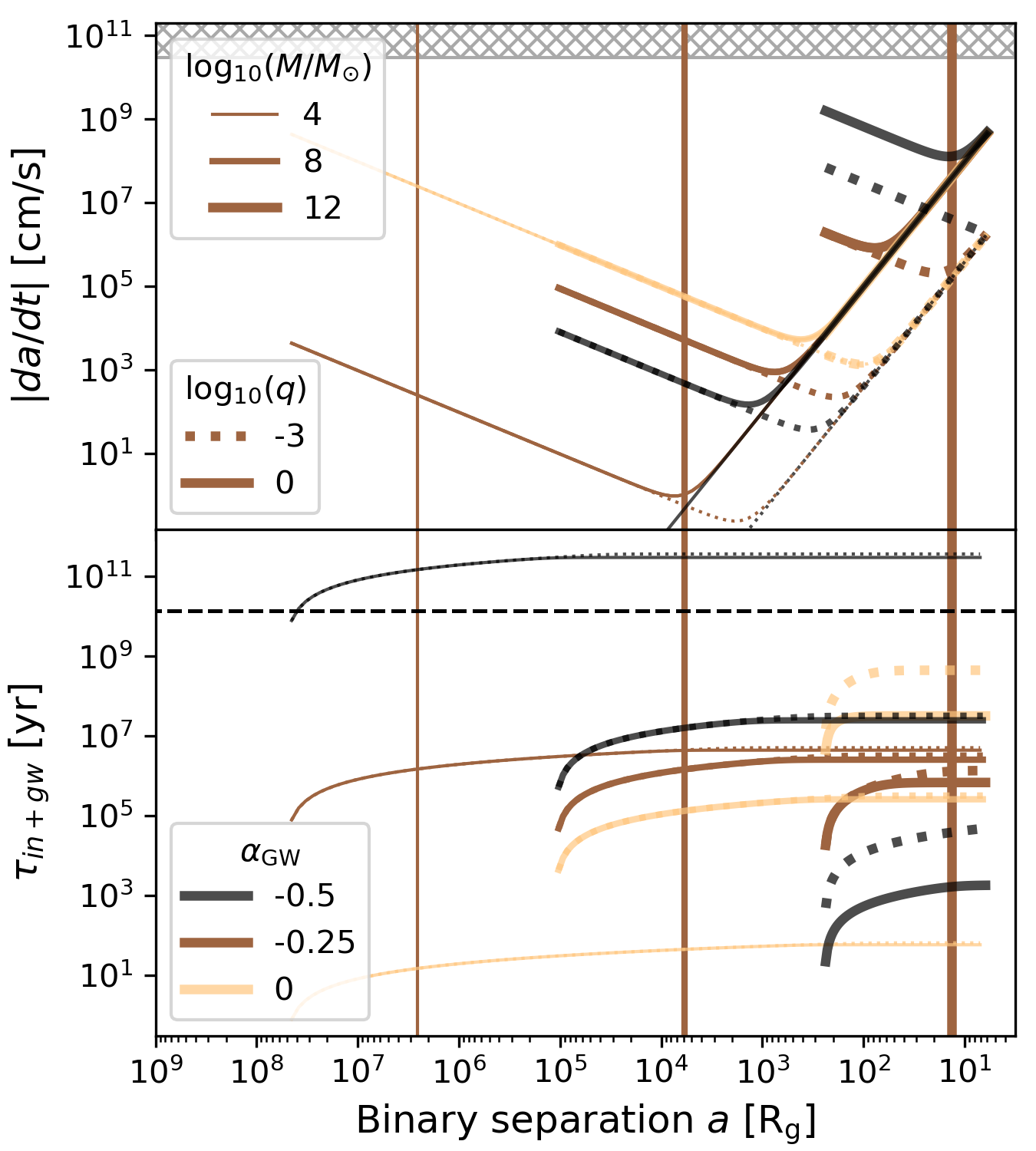}}
    \subfloat[\label{subfig:ctin_alphgwvar_tau}]{\includegraphics[width=0.41\textwidth]{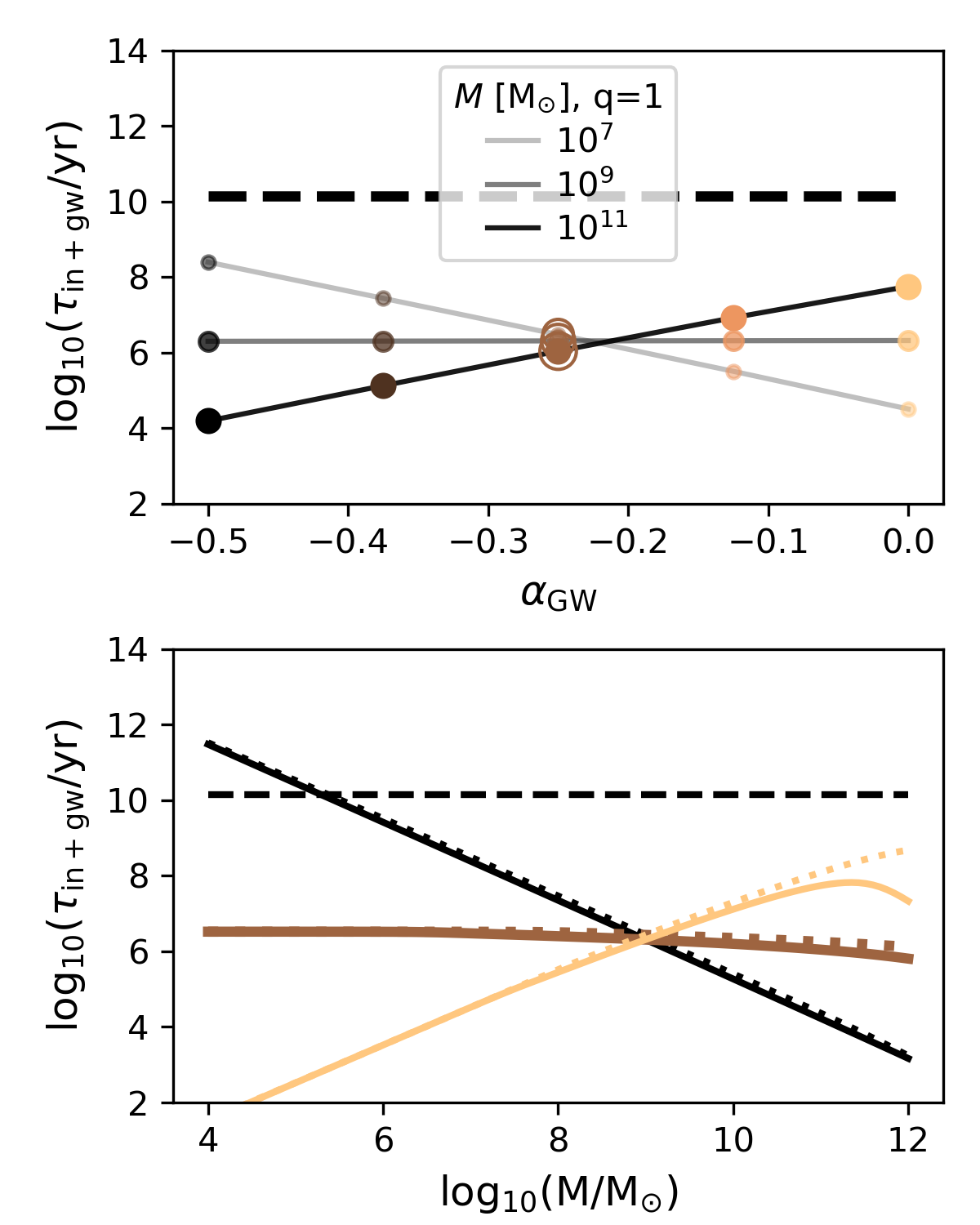}}\\
    \subfloat[\label{subfig:ctin_alphgwvar_gwb}]
    {\includegraphics[width=0.52\linewidth]{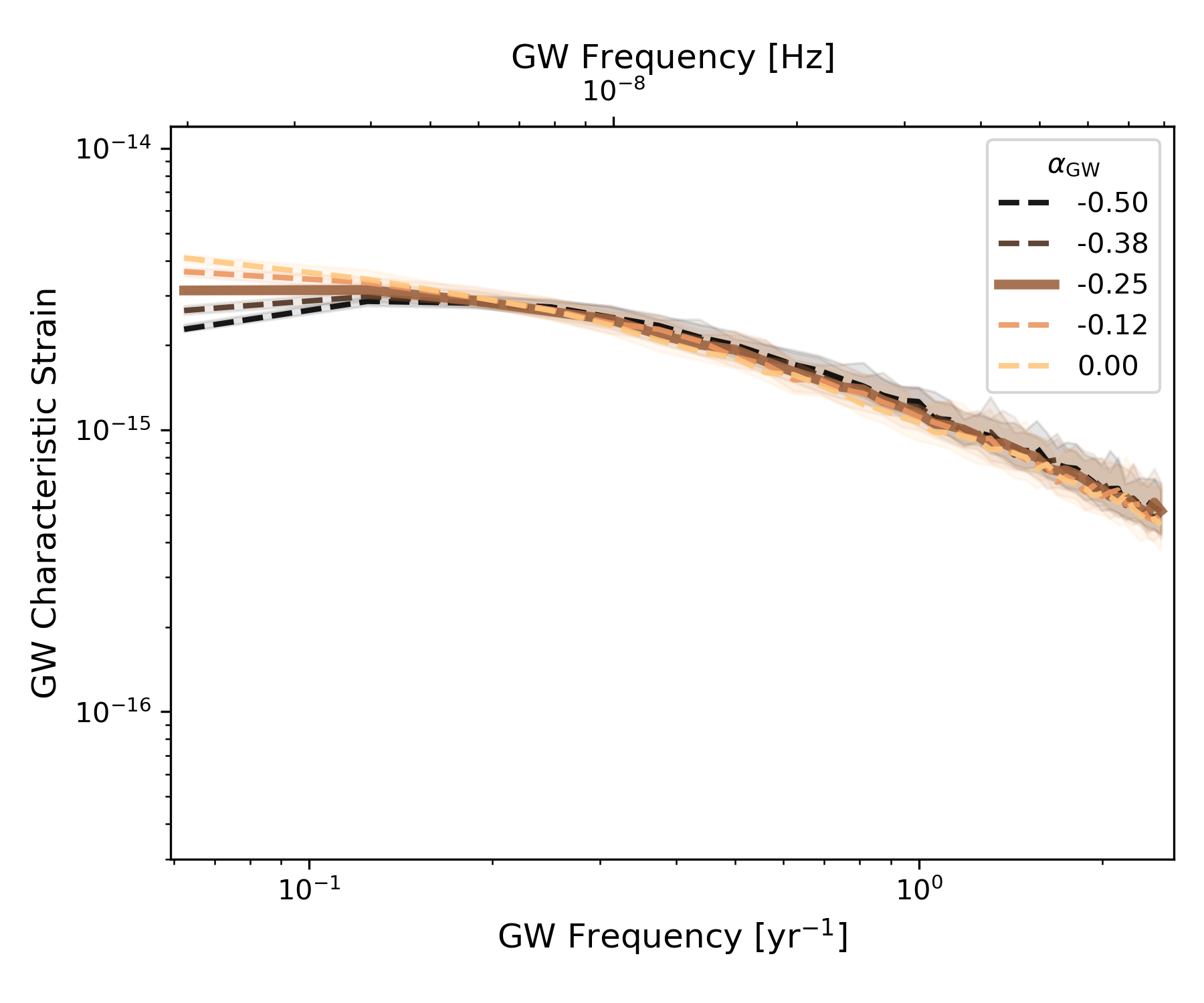}}
    \subfloat[\label{subfig:ctin_alphgwvar_amps}]{\includegraphics[width=0.34\linewidth]{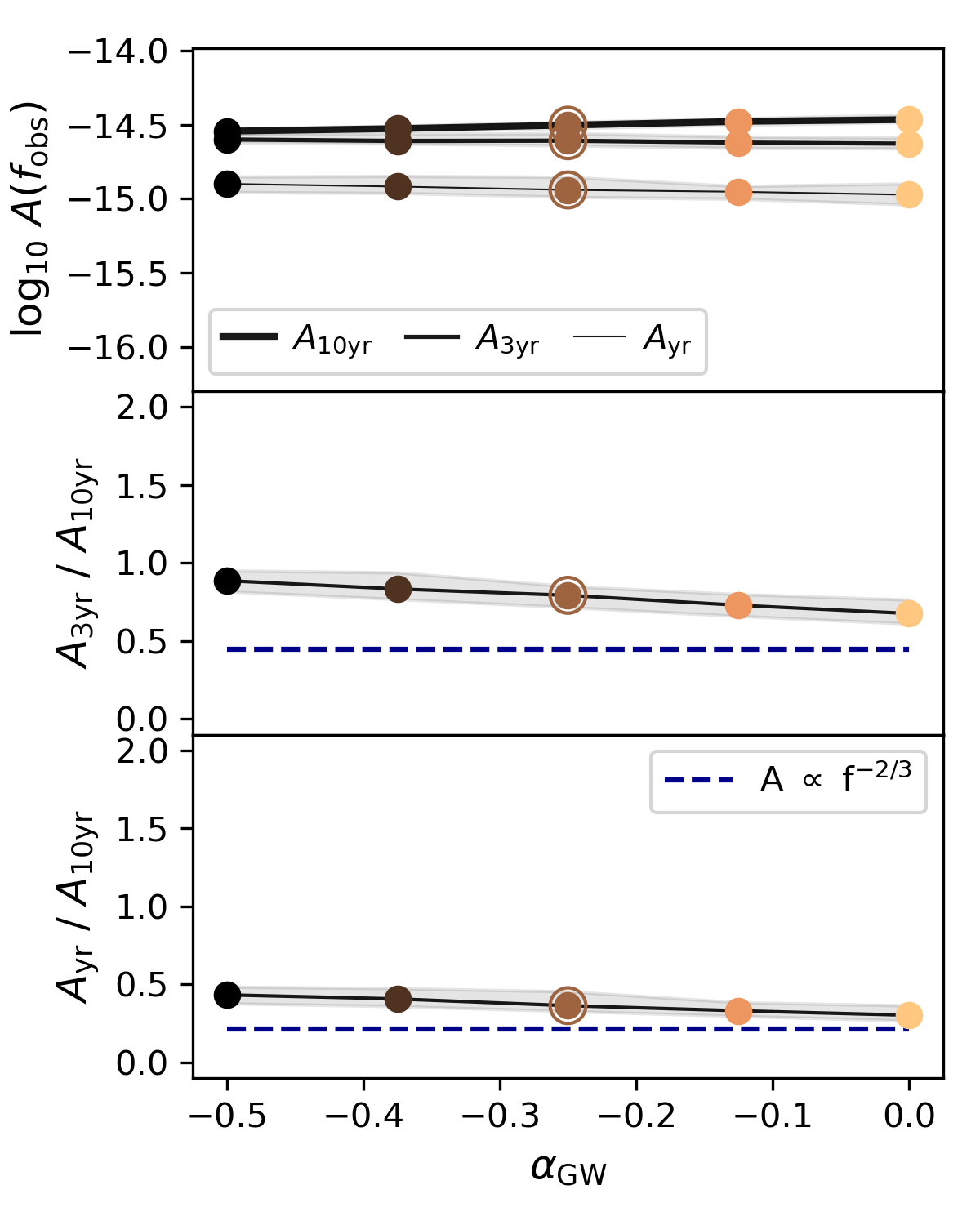}}
    \caption{Key results are shown for the \ctin\aninevar\ model with varying $\alphagw$ (which controls the mass scaling of $\aninerg$), and with fiducial values of all other parameters, in the same manner as in Figure~\ref{subfig:ctin_r9var_dadt}-\ref{subfig:ctin_r9var_amps}. Filled$+$open circles in (b) \& (d), and the thick solid line in (c), denote the fiducial value $\alphagw=-1/4$. Despite large differences in $\tauingw$ for $\alphagw=-0.5$ - 0, GWB amplitudes vary by less than a factor of 2, with negligible difference at $f_{\rm obs}\gtrsim6$ nHz.}
    \label{fig:fid_alphavar}
\end{figure*}

\begin{figure*}
    \centering
    \begin{subfigure}{0.45\textwidth}
    \includegraphics[width=\textwidth]{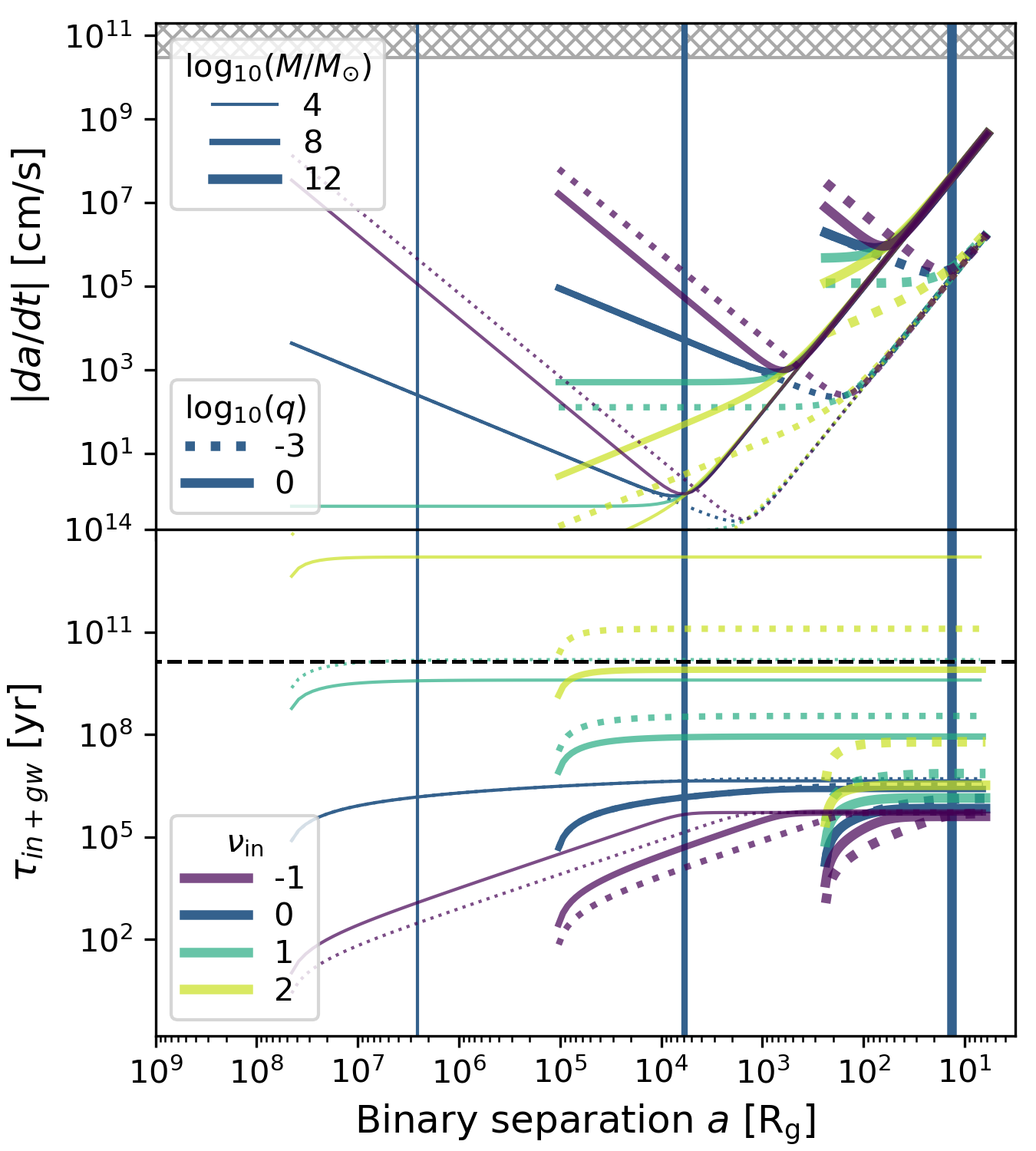}
    \caption{\label{subfig:ctin_nuivar_dadt}}
    \end{subfigure}
    \begin{subfigure}{0.41\textwidth}
    \includegraphics[width=\textwidth]{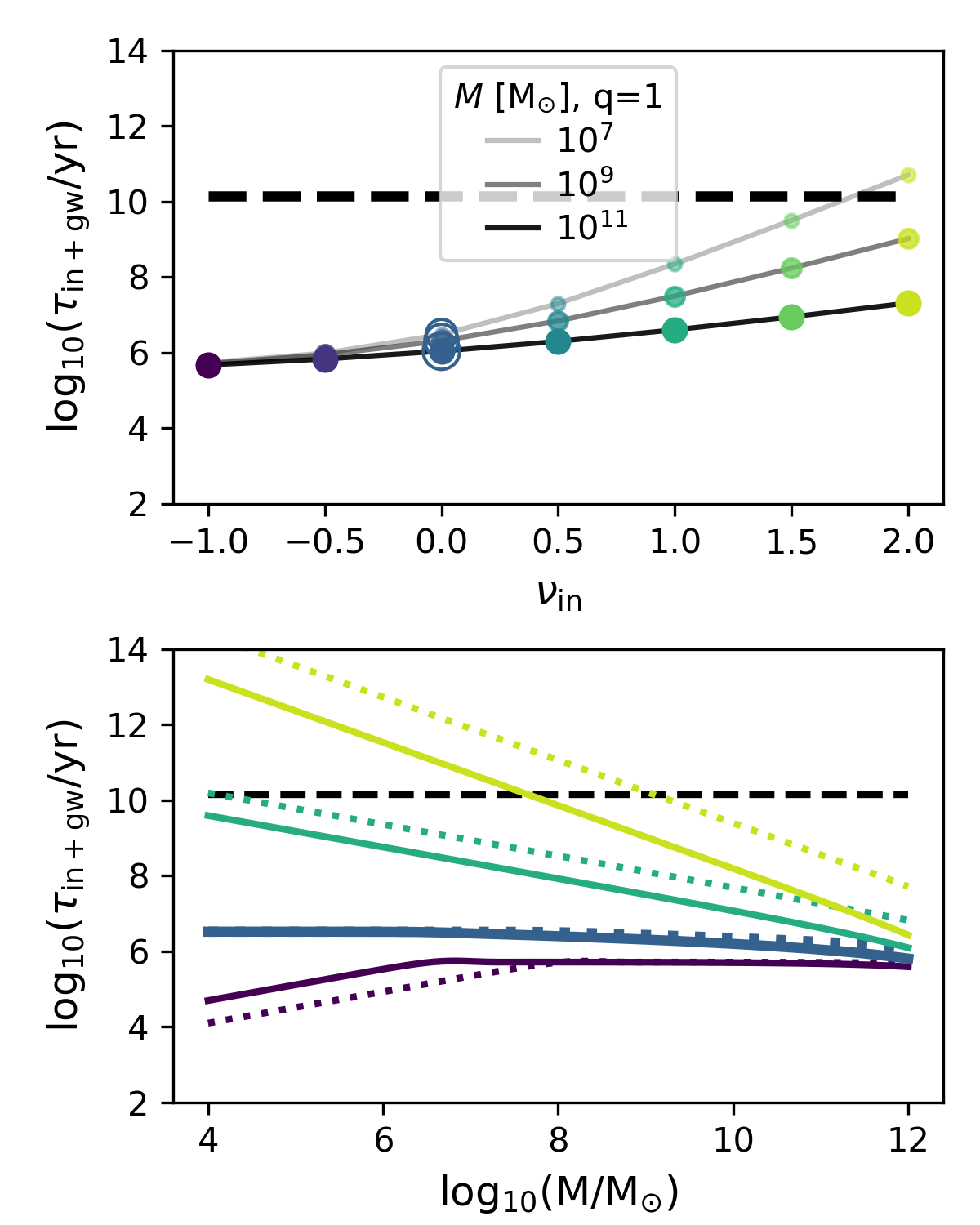}
    \caption{\label{subfig:ctin_nuivar_tau}}
    \end{subfigure}
    \begin{subfigure}{0.52\textwidth}
    \includegraphics[width=\textwidth]{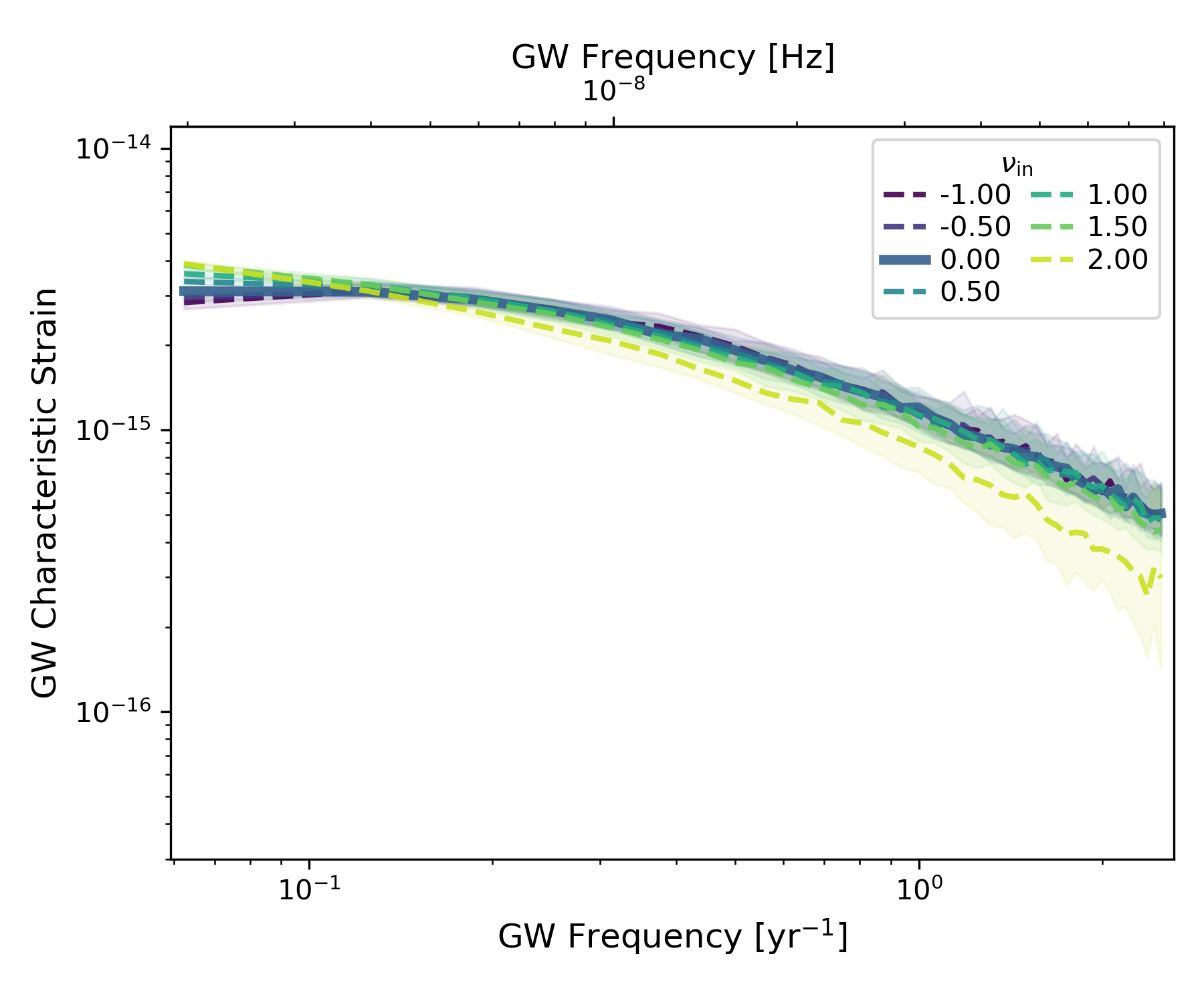}
    \caption{\label{subfig:ctin_nuivar_gwb}}
    \end{subfigure}
    \begin{subfigure}{0.34\textwidth}
    \includegraphics[width=\textwidth]{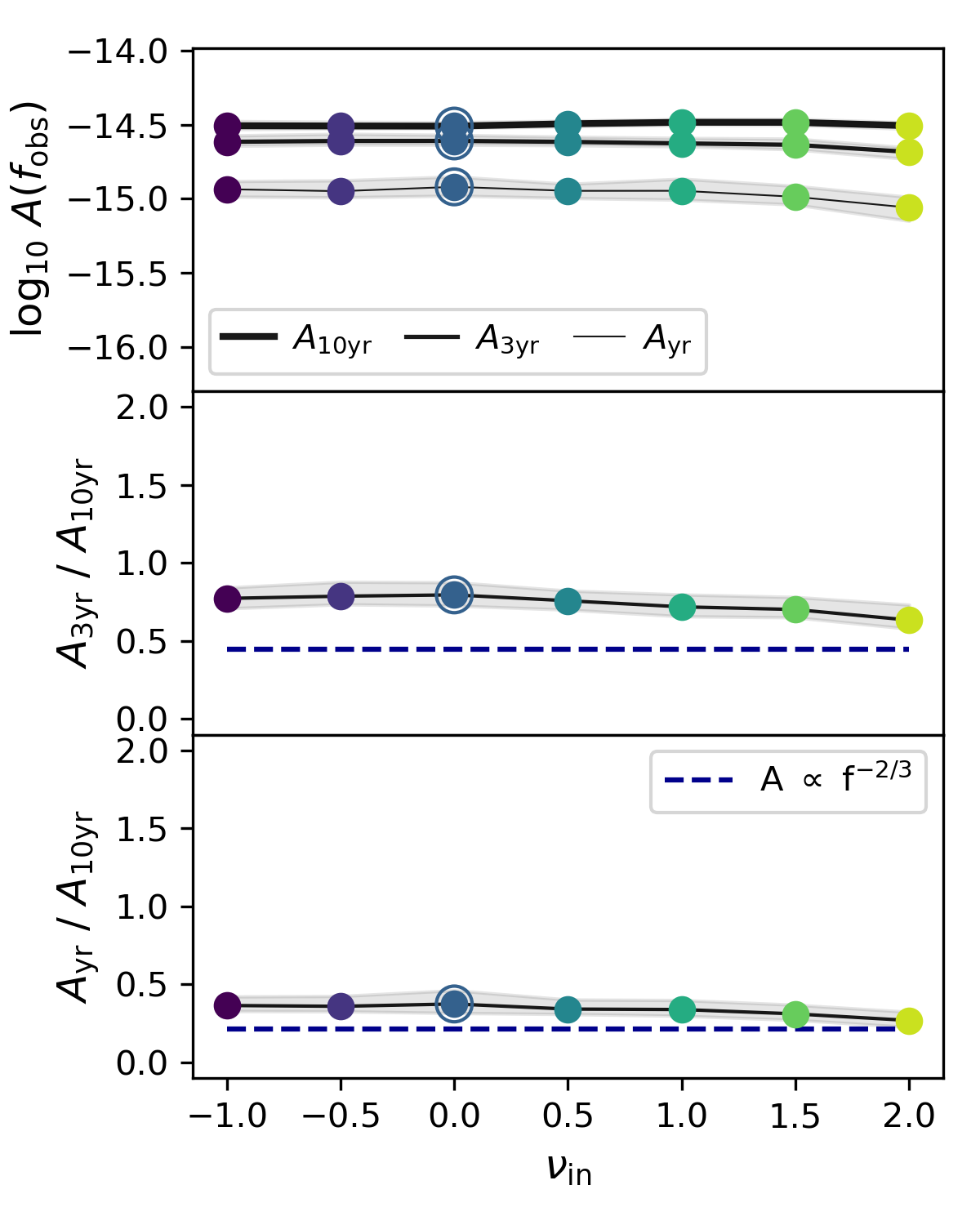}
    \caption{\label{subfig:ctin_nuivar_amps}}
    \end{subfigure}    
    \caption{Key results are shown for the \ctin\nuvar\ models with varying $\nuin$ (which controls the power-law index of $\thard$ in the inner astrophysical regime) and fiducial \ctin\ values of all other parameters, in the same manner as in Figure~\ref{subfig:ctin_r9var_dadt}-\ref{subfig:ctin_r9var_amps}. Filled$+$open circles in (b) \& (d), and the thick solid line in (c), denote the fiducial value $\nuin=0$.  Despite large differences in $\tauingw$ for $\nuin=-1$-$2$, GWB amplitudes are nearly identical at $f_{\rm obs}\sim({\rm 10 \,yr})\inv$, with only slight spectral steepening for larger $\nuin$.}
    \label{fig:fid_nuinvar}
\end{figure*}
\subsection{The mass and mass-ratio scaling of $\agwrg$}
\label{ssec:alphabetavar}

The above comparison between the \ctin\ model with $\alphagw=-1/4,\: \betagw=+1/4$ and the \cagw\ model with $\alphagw=\betagw=0$ provides insight into how SMBHB evolution depends on the mass and mass-ratio scaling of $\agwrg$. Here I systematically vary $\alphagw$ \& $\betagw$  to better understand these trends.

Figure~\ref{fig:fid_alphavar} shows \ctin\alphvar\ model variants.  At SMBHB masses $\ll$ or $\gg 10^9\msun,$ the inspiral time varies by many orders of magnitude between $\alphagw=-1/2$ and $\alphagw=0$. For $\alphagw=-1/2$, $10^4\msun$ binaries have $\tauingw>10^{11}$ yr while $10^{12}\msun$ binaries have $\tauingw\lesssim10^{3}$ yr. For $\alphagw=0$, the trend is reversed: $10^4\msun$  binaries have $\tauingw<100$ yr and approach relativistic $\dadt$ at $\rchar$, while $10^{12}\msun$ binaries 
have $\tauingw\sim10^7-10^8$ yr. \ctin\alphvar\ model variants with $\alphagw>0$ exceed $\dadtmax$. Note that the curves cross slightly above $\alphagw=-1/4$ because GW emission contributes to the inspiral rate.

Despite the wide range in $\tauingw$, the GWB spectra are remarkably consistent from $\alphagw=-1/2$ to $0$. Lower $\alphagw$ yields slightly more low-frequency turnover (below $f_{\rm obs}\sim6$ nHz), because these models drive massive SMBHBs together very quickly. The GWB \rev{amplitude} at 2 nHz decreases by a factor of $\sim 2$ from $\alphagw=0$ to $-1/2$. The \cagw\alphvar\ models have similarly weak dependence on $\alphagw$, as do the gas-like and stellar-like model variants discussed in Section~\ref{ssec:litcomparison}.

Values of $\alphagw$ outside the range $[-1/2,\,0]$ could also be considered if they satisfy the $\dadtmax$ criterion, but this tends to create a situation in which SMBHBs at one end of the mass spectrum cannot merge within $\thub$, while those at the other end approach relativistic speeds. As a result, the GWB is entirely comprised of SMBHBs in a fairly narrow mass range---a  finely tuned scenario that seems unlikely to arise in nature. As a cautionary note, any GWB calculation that creates realizations of SMBHB populations from a discrete grid of mass bins (such as the semi-empirical framework in \texttt{holodeck}) is susceptible to artifacts if the relevant mass range is coarsely sampled. When the mass range producing the GWB is narrow, a more finely spaced mass grid is required to produce well-converged GWB predictions.

I find that varying $\betagw$, which controls the mass-ratio dependence of $\agw$, has essentially no effect on the GWB spectral shape or amplitude. In the \ctin\betavar\ models, the GWB amplitudes vary by $<1\%$ as $\betagw$ is varied from $-1/2$ to $+1/2$. This is also true in the \cagw\betavar\ model, as well as the stellar-like and gas-like model variants discussed in Section~\ref{ssec:litcomparison}. The insensitivity of the GWB to $\betagw$ reflects the fact that SMBHBs with high mass ratios (near $q=1$) dominate the GWB. 

However, the total inspiral timescales for low-mass-ratio binaries {\em do} vary significantly with $\betagw$. For $q\sim0.01$, $\tauingw$ varies from $\lesssim 10^5$ yr for $\betagw=+1/2$ to $\gtrsim10^{11}$ yr for $\betagw=-1/2$ in the \ctin\betavar\ model. $\tauingw$ for $q\sim0.1$ varies by about two orders of magnitude over the same range in $\betagw$. Likewise, $\tauingw$ varies strongly with $\alphagw$ for low-mass and ultramassive SMBHBs. This has the important consequence that, even though the GWB is not sensitive to $\alphagw$ and $\betagw$, SMBHB merger rates (and LISA event rates) {\em are} sensitive to these scalings. Similarly, the detection likelihood of individual continuous-wave (CW) sources above the background is sensitive to the mass distribution of SMBHBs \citep{gardiner24,gardiner25,gardiner26}. CW occurrence therefore may be more sensitive to $\alphagw$ and $\betagw$ than the GWB itself.

\begin{figure*}
    \centering
    \begin{subfigure}{0.495\textwidth}
    \includegraphics[width=\textwidth]{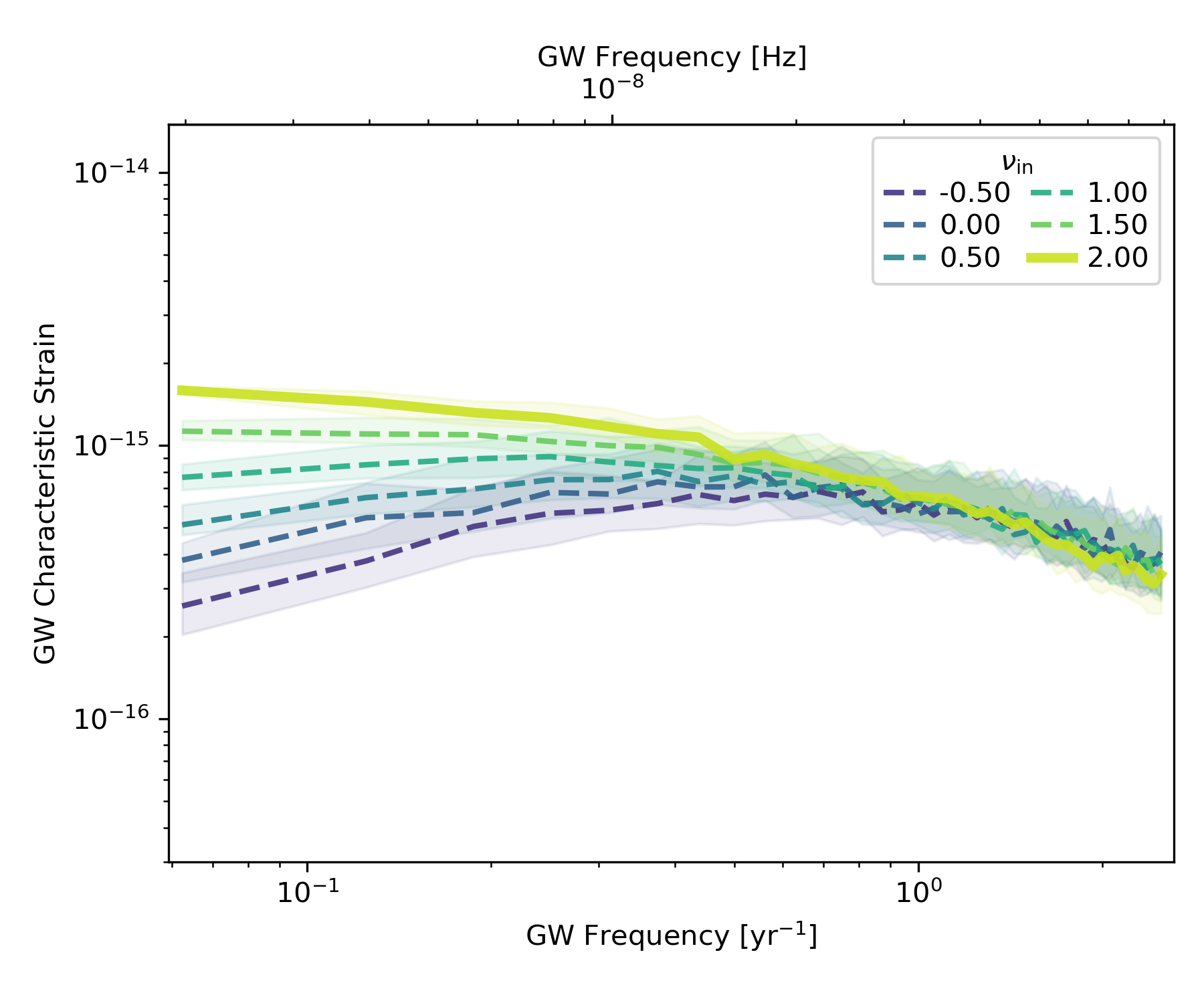}
    \caption{\label{subfig:ctin_gas_nuivar_gwb}}
    \end{subfigure}
    \begin{subfigure}{0.495\textwidth}
    \includegraphics[width=\textwidth]{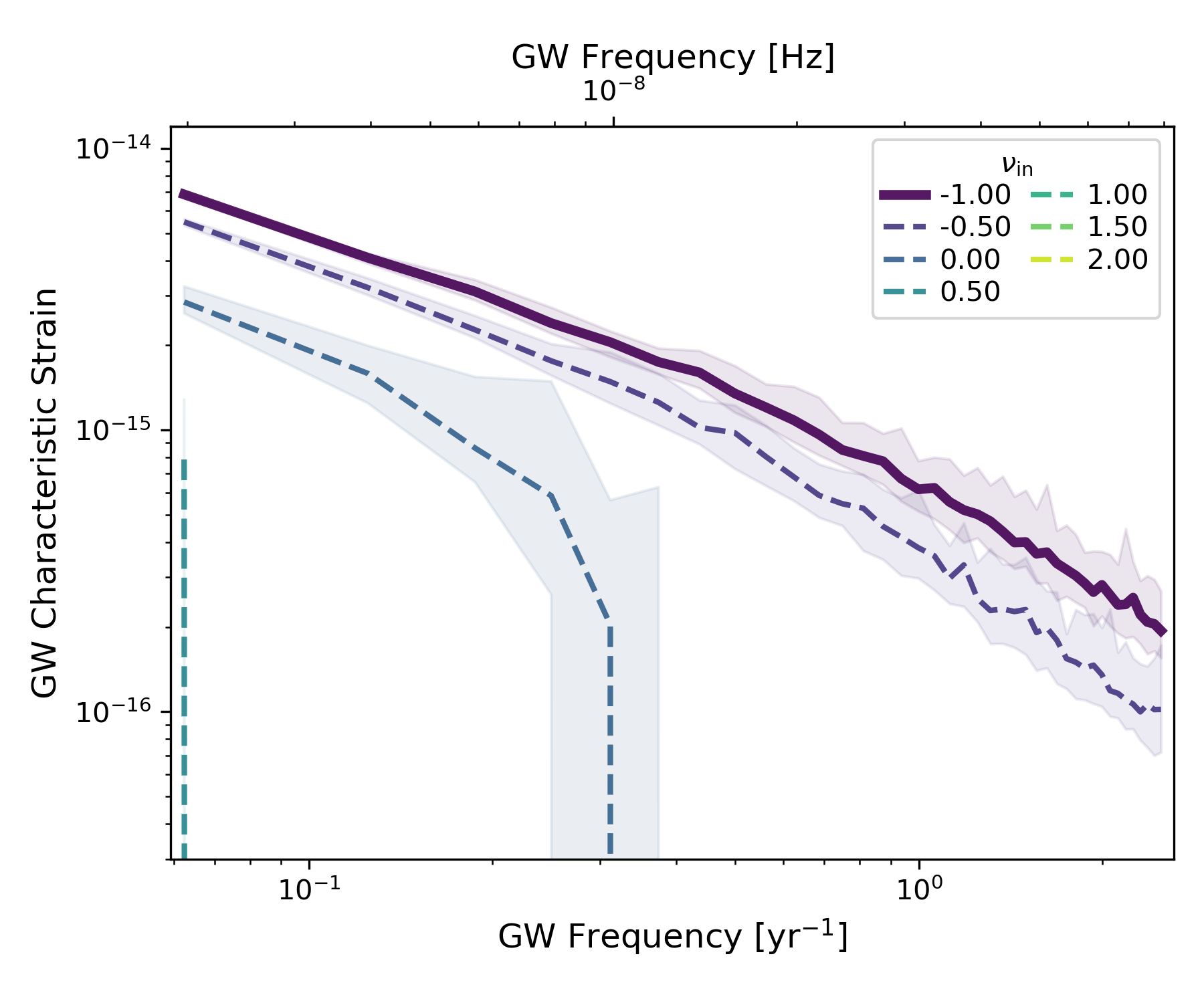}
    \caption{\label{subfig:ctin_star_nuivar_gwb}}
    \end{subfigure}
    \caption{{\bf (a)} GWB spectra are shown for the \ctingas\ model with $\aninerg=100\rg$, for various values of $\nuin$. The thick solid line indicates the fiducial value of $\nuin=+2$. Lower $\nuin$ creates stronger low-frequency GWB attenuation. {\bf(b)} GWB spectra are shown for the \ctinstar\ model with $\aninerg=10^{3.5}\rg$, for varying $\nuin$. The thick solid line indicates the fiducial value $\nuin=-1$. Larger $\nuin$ creates high-frequency spectral steepening, with no observable GWB for $\nuin\geq0.5$. In both the \ctingas\ and \ctinstar\ models, the GWB dependence on $\nuin$ is much larger than in the \ctin\ model, but with opposite trends for the gas-like versus stellar-like models.}
    \label{fig:fid_nuinvar_stargas}
\end{figure*}

\subsection{Dependence on inner power-law index $\nuin$}
\label{ssec:nuinvar}
In addition to $\agw$, $\nuin$ is the other key hardening parameter in the inner astrophysical regime. Figure~\ref{fig:fid_nuinvar} shows \ctin\nuvar\ model variants with $\nuin=[-1, \,+2]$. 
For $\nuin<1$, the hardening rate decreases from $\rchar$ to $\agw$ before increasing sharply in the GW regime. 

The  GWB spectra are very similar for $\nuin=-1$ to $+1.5$, with  slight low-frequency attenuation for lower $\nuin$  due to faster astrophysical hardening. For $\nuin=+2$, the GWB spectrum is also slightly steeper across the relevant frequency range, because binaries $\lesssim 10^8\msun$ cannot merge within $\thub$. Notably, this is the {\em opposite} trend seen with $\nuin$ in the 2PL hardening model (Section~\ref{ssec:2PL}). In NG15Astro, $\nuin$ is the only parameter varied that directly controls hardening through the PTA band. The large GWB variation with $\aninerg$ shown here is thus entirely folded into $\nuin$ and other hardening parameters in the 2PL model. In my approach, these effects can be disentangled. Larger $\nuin$ means more time spent emitting GWs in the PTA band, but only if the SMBHBs can reach the PTA regime before $z=0$. 

Results for the \cagw\nuvar\ models are similar to the \ctin\nuvar\ models, except that $\nuin<0$ is disallowed by $\dadtmax$. Thus, in both of these models with $\aninerg$ fixed at $10^{2.5}\rg$, the GWB varies little with $\nuin$, much less than for variations in $\aninerg$ with fixed $\nuin=0$.
However, if $\aninerg$ is either larger or smaller than the fiducial value of $10^{2.5}\rg$, the GWB {\em does} vary significantly with $\nuin$. Figures~\ref{fig:fid_nuinvar_stargas}a \& b show GWB spectra for varying $\nuin$ in the \ctingas\nuvar\ and \ctinstar\nuvar\ models. In the former, with $\aninerg=10^2\rg$, the GWB spectra are attenuated at all frequencies relative to Figure~\ref{subfig:ctin_nuivar_gwb}. $\nuin<-0.5$ is disallowed by the $\dadtmax$ criterion. All $\nuin$ values yield similar GWB amplitudes at $f_{\rm obs}>{\rm yr}\inv$, but faster astrophysical hardening (lower $\nuin$) yields increasingly strong low-frequency turnovers. At $f_{\rm obs}>({\rm\, 10 yr})\inv$, the amplitude for $\nuin=-0.5$ is 3.8 times lower than for $\nuin=+2$. This is again the opposite trend seen with $\nuin$\ in the 2PL hardening model, but here it is much more pronounced than for \ctin\nuvar\ and more closely resembles the low-frequency behavior with $\nuin$ shown in Figure 4 of NG15Astro.  Overall, we see that $\agwrg$ is still the primary determinant of the overall GWB spectral shape and amplitude, but when astrophysical hardening dominates through most of the PTA band, $\nuin$ significantly modulates the GWB as well.

For the stellar-like models with $\aninerg=10^{3.5}\rg$ (Figure~\ref{fig:fid_nuinvar_stargas}b), efficient astrophysical hardening at $a>\agw$ is required to avoid binary stalling. This is achieved for $\nuin\leq-0.5$, but above this value the inspiral timescales start to exceed $\thub$, and there is essentially no observable GWB for $\nuin>0$. The amplitude drops off starting at the highest frequencies. Once again this behavior differs from the 2PL model used in NG15Astro, where the amplitude decreases with lower $\nuin$, but starting from {\em low} frequencies instead of the high-frequency steepening shown in Figure~\ref{fig:fid_nuinvar_stargas}b.
Together, Figures~\ref{fig:fid_nuinvar}, \ref{fig:fid_nuinvar_stargas}a \& \ref{fig:fid_nuinvar_stargas}b indicate that $\aninerg$ generally has a larger effect on the GWB spectrum, but that $\nuin$ significantly modulates the GWB spectrum for some values of $\aninerg$. Variations in both parameters can produce very different GWB spectral shapes and amplitudes.

\begin{figure*}
    \centering
    \begin{subfigure}{0.52\textwidth}
    \includegraphics[width=\linewidth]{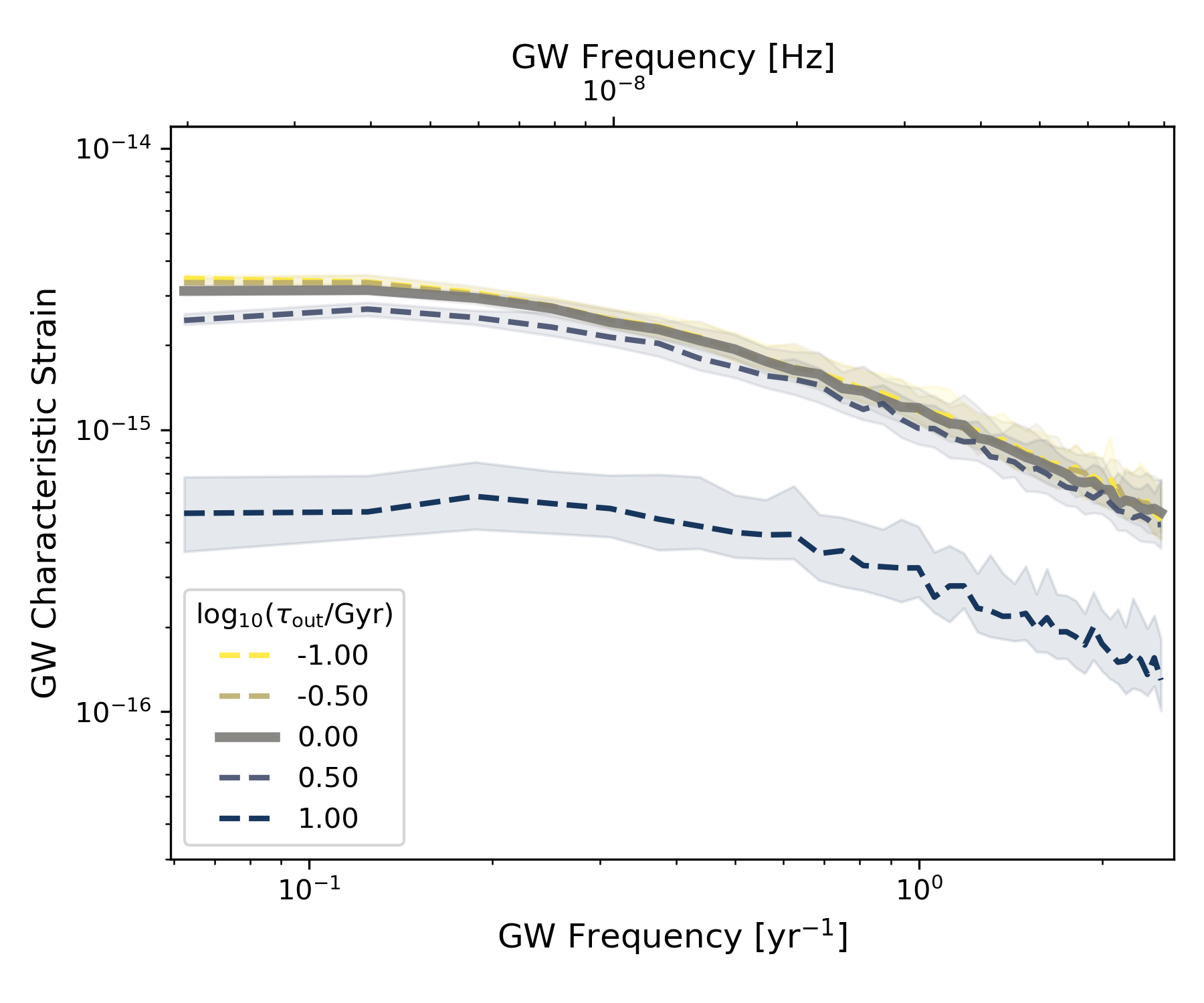}
    \caption{\label{subfig:fid_tout_gwb}}    
    \end{subfigure}
    \begin{subfigure}{0.34\textwidth}    
    \includegraphics[width=\linewidth]{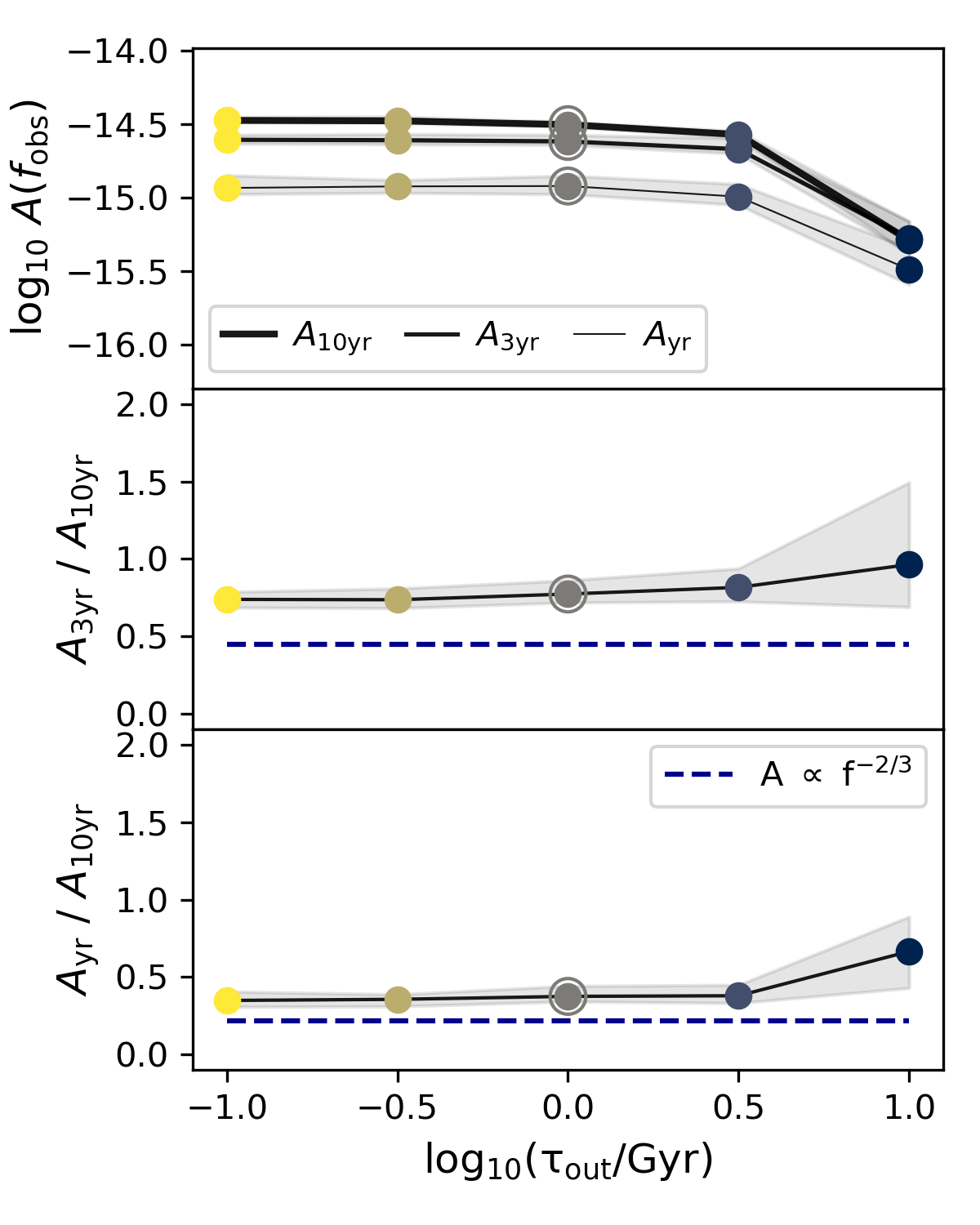}   
    \caption{\label{subfig:fid_tout_amps}}    
    \end{subfigure}
    \caption{GWB characteristic strain amplitudes and amplitude ratios are shown for the \ctin\toutvar\ model variants with varying $\tau_{\rm out}$ (the inspiral timescale in the outer astrophysical phase where $a>\rchar$). {\bf(a)}: the GWB spectra for each model variant are shown in the same manner as in Figure~\ref{subfig:ctin_r9var_gwb}. {\bf(b)}: the GWB amplitudes and amplitude ratios are shown in the same manner as in Figure~\ref{subfig:ctin_r9var_amps}. The filled$+$open circles denote the fiducial value $\tau_{\rm out}=1$ Gyr. The GWB spectra are nearly identical for $\tau_{\rm out}\leq1$ Gyr and only slightly attenuated at $10^{0.5}$ Gyr. Longer inspiral times attenuate the GWB more significantly, as many binaries are unable to reach the PTA regime by $z\sim0$.}
    \label{fig:fid_toutvar}
\end{figure*}

\subsection{Connecting to the outer hardening regime}
\label{ssec:outer}
Continuing analysis of the hardening model parameters in an inside-out fashion, I now examine the \ctin\rchninevar\ model variants with $\rchnine$ ranging from 0.1 pc to 100 pc. (Recall that $\rchar$ defines the boundary between the inner and outer hardening phases.) 
Across this factor of $10^3$ variation in $\rchnine$, the GWB is virtually indistinguishable from the fiducial case $\rchnine=1$ pc;  the low-frequency amplitudes vary by $0.1\%$ and the high-frequency amplitudes vary by $2\%$. The \cagw\rchninevar\ models are similarly insensitive to $\rchnine$ (except that in that case, $\rchnine=100$pc is disallowed by the $\dadtmax$ criterion). The insensitivity of the GWB to $\rchnine$ simply reflects that $\rchar$ is outside the PTA regime. Its primary impact is to limit the range of other model parameters for which $|\dadtrc| < \dadtmax$. Models with steep $\nuin\lesssim-1$, large $\rchar$, and/or small $\agw$ quickly reach unphysical hardening rates. For example, all of the hardening rates in Figure~\ref{fig:fid_nuinvar} are well below $\dadtmax$, but if $\rchar$ were larger by a factor of $\sim10-100$, the steepest $\nuin=-1$ models would exceed $\dadtmax$ at $\rchar$. 
This motivates a choice of $\rchar$ that is larger than $a(f_{\rm obs,min})$ but not by a large factor.

Finally, I consider the outer astrophysical hardening regime, parameterized by $\tau_{\rm out}$. Figure~\ref{fig:fid_toutvar} shows the GWB spectra and amplitudes for $\tau_{\rm out}=$ 0.1 to 10 Gyr in the \ctin\toutvar\ models. $\dadt$ and $\tauingw$ are not shown, because these model variations have identical binary evolution at $a<\rchar$. $\tau_{\rm out}$ modulates only the redshift at which each binary enters the inner hardening phase and thus has minimal impact on  the GWB spectral shape. Recall that $\tauingw\sim10^6$ yr in the fiducial \ctin\ model, such that $\tau_{\rm out}$ dominates the total binary inspiral time. 

The GWB amplitude is largely unaffected until $\tau_{\rm out}$ exceeds several Gyr. For $\tau_{\rm out}=10$ Gyr, binaries forming at $z<1.8$ cannot merge by $z=0$. This yields an overall drop by a factor of $\sim 4-6$ in the GWB amplitude across the spectrum, relative to the fiducial value of 1 Gyr. 
Because $\tau_{\rm out}$ primarily impacts the amplitude rather than the shape of the GWB, it is subject to degeneracies with galaxy population parameters that control the number and mass of SMBHBs formed (i.e., the GSMF and the $M_{\rm BH}-M_{\rm bulge}$ parameters). 
These galactic-scale parameters are also degenerate among themselves to some extent, and care should always be taken when interpreting the results of SMBHB population parameter inference.

\begin{figure*}
    \centering
    \begin{subfigure}{0.435\textwidth}
    \centering
    \includegraphics[width=\textwidth]{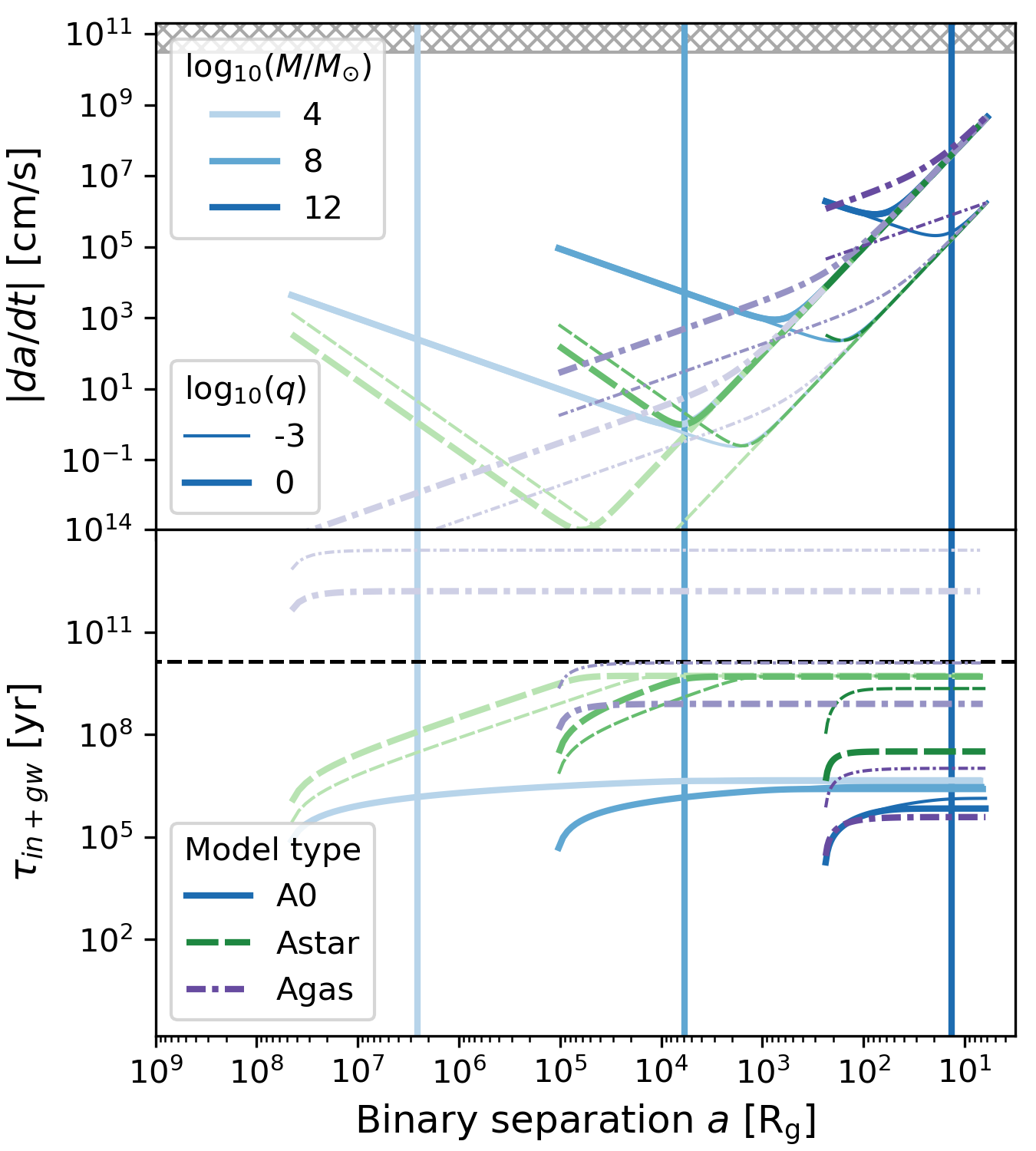}
    \vspace{-10pt} 
    \caption{\label{subfig:stargas_dadt}}
    \end{subfigure}
    \begin{subfigure}{0.555\textwidth}
    \centering
    \includegraphics[width=\textwidth]{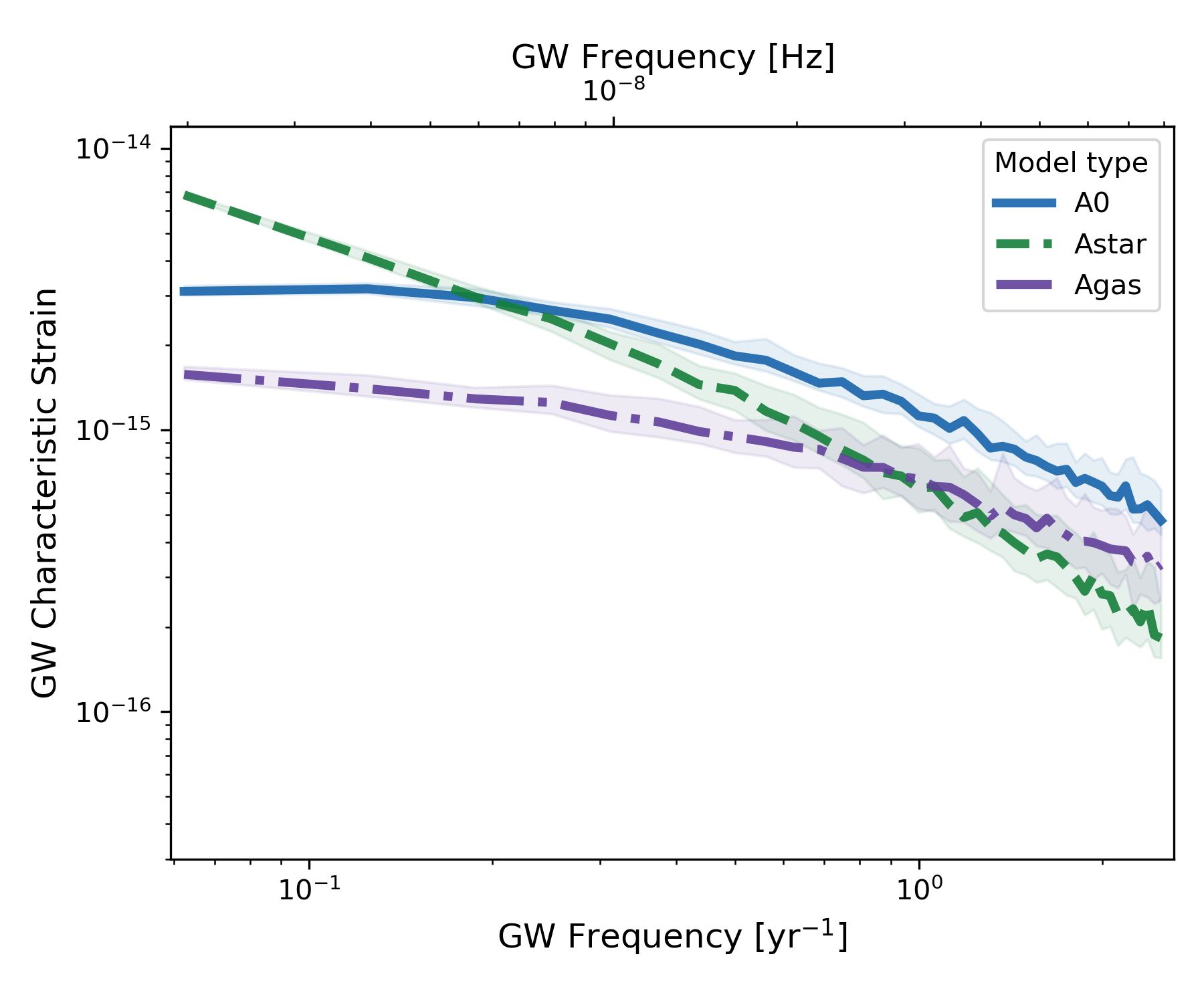}
    \vspace{-10pt} 
    \caption{\label{subfig:stargas_gwb}}
    \end{subfigure}
    \caption{{\bf(a)} SMBHB hardening rates (top panel) and $\tauingw$ (bottom panel) are shown versus binary separation for three fiducial model variants: \ctin\ (solid blue curves), \ctinstar\ (dashed green curves), and \ctingas\ (dash-dotted purple curves). As in Figure~\ref{fig:fid_dadt}, curves for different binary $M$ \& $q$ are as indicated in the legends, and vertical lines indicate where binaries of a given mass enter the PTA band. {\bf(b)} GWB spectra are shown for the same three hardening models, which differ in their $\aninerg$ and $\nuin$ values. These parameter variations between the primary fiducial model and its stellar-like and gas-like variants impact both the shape and the amplitude of the resulting GWB spectra. \label{fig:stargasnu0}}
\end{figure*}

\subsection{Comparison to stellar- and gas-driven inspiral}
\label{ssec:litcomparison}

I now consider how this analytic hardening model relates to stellar-driven and gas-driven hardening mechanisms in the literature. For simplicity and clarity, I vary $\nuin$ and $\aninerg$ while keeping other parameters at their fiducial values. These model variants are intended to be illustrative rather than a precise match of any stellar- or gas-driven hardening models.
Of course, reality is likely to lie somewhere between the stellar-like and gas-like limits, either because both stars and gas commonly contribute to binary hardening at PTA frequencies, or because the binaries that dominate the GWB include a mixture of stellar-driven and gas-driven systems. Any estimates of $\aninerg$ and $\nuin$ obtained from the GWB using this hardening model will be strain-weighted, population-averaged values. $\nuin$ also represents the average power-law index of the hardening timescale through the PTA regime, if the true timescale is not described by a single power law. However, the tantalizing possibility remains that future PTA data sets could constrain the dominant astrophysical hardening channel for SMBHBs in the nHz regime.

\subsubsection{Stellar-like hardening models}
\label{sssec:stellar}

Binary inspiral rates are sometimes expressed in terms of the hardening parameter $s\equiv d(a^{-1})/dt$.
Numerical stellar scattering experiments show that once a hard binary forms, the hardening parameter is roughly constant in the stellar scattering regime \citep[e.g.,][]{quinlan96,sesana15,khan18a}. $s$ is also strongly correlated with the density of the stellar loss cone \citep[e.g.,][]{quinlan97,khan12b,holley25}. If the loss cone is depleted by the binary faster than it can be replenished, $s$ will decrease (i.e., binary hardening will slow or stall). In the power-law parameterization of my hardening model, the constant-$s$ case corresponds to $\nuin=-1$.

\cite{holley25} recently used a large, observationally informed suite of $N$-body simulations to quantify the dependence of binary merger timescales on host galaxy properties, focusing on binary evolution driven by stellar scattering. 
Using linear regression, \citet{holley25} 
derived a best-fit relation that predicts coalescence time to within a factor of two.

For the \ctinstar\ model in this work, the inner hardening parameter ranges from $s\sim10^{-3}$ (pc Myr)$^{-1}$ for $\sim10^{10}\msun$ binaries to $s\sim 10$ (pc Myr)$^{-1}$ for $\sim10^5\msun$ binaries. ($s$ varies from $\sim 10^{-5}$-$10^2$ (pc Myr)$^{-1}$ across the same mass range in the \cagwstar\ model.) This fairly closely traces the range in $s$ measured from the simulations of \citet{holley25}. 
\citet{harris26} find a similar range in $s$ for their stellar-scattering models based on IllustisTNG simulations and tuned to observed core elliptical galaxy properties. I leave a more detailed comparison to these and other stellar hardening models for future work.

Figure~\ref{fig:stargasnu0} shows the hardening rates and GWB spectra for the \ctinstar\ model with $\nuin=-1$ and $\aninerg=10^{3.5}\rg$ (green dashed line), compared to the fiducial \ctin\ model with $\nuin=0$ and $\aninerg=10^{2.5}\rg$ (blue solid line). (The \ctingas\ model with $\nuin=+2$ and $\aninerg=100\rg$ is also shown and discussed in Section~\ref{sssec:gas}.) The GWB amplitude is higher with less low-frequency turnover in the stellar model where $\tauingw\sim$ few$\times10^9$ yr, as opposed to $\tauingw\sim10^6$-$10^8$ yr in the fiducial model. Also, more of the inspiral is GW-driven in the \ctinstar\ model. This emphasizes a point made above, that the GWB will be highest and closest to a power-law shape when SMBHBs are brought efficiently to the edge of the PTA band and then driven to merger at low redshift by GW emission. The \cagwstar\ model behaves similarly. \citet{harris26} also find minimal GWB attenuation in their best-fit stellar-scattering model for SMBHB inspiral.

I now consider parameter variations for the \ctinstar\ and \cagwstar\ models. Relative to \ctin\ and \cagw, these models with $\nuin=-1$ impose stricter limits on the allowed $\aninerg$ for which $\dadt<\dadtmax$, permitting only $\aninerg\geq10^{2.25}\rg$ and $\geq10^{3.25}\rg$, respectively, for  \ctinstar\aninevar\ and \cagwstar\aninevar. In both stellar models, GWB amplitudes sharply decline for $\aninerg\gtrsim10^{3.5}\rg$. 

As discussed in Section~\ref{ssec:nuinvar}, Figure~\ref{fig:fid_nuinvar_stargas}b shows very strong GWB spectral steepening for larger $\nuin$ in the \ctinstar\nuvar\ models, with  similarly large variation in the \cagwstar\nuvar\ models. This is because  $\aninerg=10^{3.5}\rg$ for the stellar models, causing binaries with slower astrophysical inspiral (larger $\nuin$) to stall outside the PTA regime. 

As with the fiducial \ctin\alphvar\ and \cagw\alphvar\ models, the stellar-like models are largely insensitive to variations in $\alphagw$. The small impact that $\alphagw$ does have on the GWB spectra is reversed, however, owing to the longer timescales for all binaries in the stellar models. For low $\alphagw$, stellar models exhibit high-frequency spectral steepening because $\tauingw\gg\thub$ for low $M$, versus the low-frequency attenuation in the \ctin\alphvar\ because $\tauingw\lesssim\thub$ for low $M$.
Overall, the dependence on $\alphagw$ is secondary to the dependence on $\aninerg$ and $\nuin$.

The \ctinstar\rchninevar\ models are insensitive to $\rchnine$, although here the GWB amplitude variation from $\rchnine=0.1$ to 100pc is at the $\sim 30\%$ level instead of the $\lesssim1\%$ level as for \ctin\rchninevar\ and \cagw\rchninevar. For the \cagwstar\rchninevar\ models, $\rchnine\geq10$ pc is disallowed by the $\dadtmax$ criterion.

\subsubsection{Gas-like hardening models}
\label{sssec:gas}

\citet{haiman09} presented analytic formulae for the evolution of a SMBHB in a circumbinary gas disk, delineating (1) an inner region where radiation pressure and electron-scattering opacity dominate, (2) an middle region where gas pressure and electron-scattering opacity dominate, and (3) an outer region where gas pressure and free-free opacity dominate. This prescription has been used in numerous SMBHB inspiral models \citep[e.g.,][]{kocsis11, kelley17a, kelley17b}. Assuming that binary  evolution is determined by the viscous time, they found $\thard \propto a^{7/2}$ in region 1, $\propto a^{7/5}$ in region 2, and $\propto a^{5/4}$ in region 3. These scalings correspond to the power-law index $\nuin$ in my generalized hardening model. The \ctingas\ and \cagwgas\ models use a single intermediate value of $\nuin=2$ as a simple representation of the average evolution of SMBHBs embedded in a circumbinary gas disk. I also use  $\aninerg=100\rg$ for these gas-like models as a rough estimate of the point at which the viscous disk timescale and GW inspiral timescale are equal. 

The purple dot-dashed lines in Figure~\ref{fig:stargasnu0}a\&b show the  \ctingas\ model. The gas-like GWB has a lower amplitude and flatter spectrum than the fiducial and stellar models, largely because GW emission does not dominate until the final $100 \rg$. Another difference is the sharpness of the transition to the GW regime. In the stellar limit, the transition from $\thard \propto a^{-1}$ to $\thard \propto a^{+4}$ is dramatic, and the GWB shape is very sensitive to the separation $\agw$ at which this transition happens---more so than in the gas limit where $\thard$ transitions from  $\propto a^{+2}$ to $ \propto a^{+4}$.
Nonetheless, the \ctingas\aninevar\ and \cagwgas\aninevar\ models still depend sensitively on $\aninerg$. The GWB amplitudes increase by a factor of $\sim7$-$14$ as $\aninerg$ is varied from $10^{1.25}\rg$ to $10^{2.5}\rg$. For larger $\aninerg$, the GWB amplitude rapidly drops off at high frequencies as more binaries exceed $\thub$. There is no observable GWB for $\aninerg>1000\rg$.

Because binaries in the gas-like models  are driven by astrophysics until $100\rg$, fast astrophysical hardening (lower $\nuin$) causes strong GWB attenuation (Figure~\ref{fig:fid_nuinvar_stargas}a). The \cagwgas\nuvar\ models show similar attenuation for larger $\nuin$, but in that case $\nuin\leq0$ is disallowed by $\dadtmax$. 
In both the stellar-like and the gas-like model variants, the variation with $\nuin$ is significantly stronger than in the \ctin\nuvar\ and \cagw\nuvar\ models, which represent a ``middle-ground" in ($\agwrg$, $\nuin$) space between these stellar and gas limits.
The gas-like models are largely insensitive to variations in $\alphagw$. When $\rchnine$ is increased above the fiducial value of 1 pc, $\tauingw$ exceeds $\thub$ starting from low binary masses. This causes a sharp drop in the GWB at high frequencies, and no GWB is observable for $\rchnine>10$ pc.

\rev{SMBHBs are expected to dynamically decouple from their circumbinary gas disks once the GW inspiral time becomes shorter than the viscous timescale in the disk. Because the total binary hardening rate is the sum of the inner and GW-driven hardening rates, the sharpness of the transition at decoupling is determined by $\nuin$. For the \ctingas\ and \cagwgas\ models with $\nuin=2$, the gas-driven hardening rate contributes $1\%$ to the total hardening rate when the binary has shrunk by a factor of 10 past the point of decoupling. Sharper decouplings could be considered (such as an exponential cutoff of the inner hardening rate below $\agw$), but at the expense of introducing additional model parameters. For the sake of model simplicity, such scenarios are not considered here.}

This simple parameterization of binary hardening, in which it is assumed that $\dadt$ is always negative, cannot capture the possibility of binary orbital expansion \citep[e.g.,][]{munoz19, moody19}. 
Because I assume a single power-law for $\dadt_{\rm in}$ versus $a$, the case of positive torques from a circumbinary gas disk would effectively yield a slower average inner hardening rate, corresponding to larger values of $\agw$ and/or $\nuin$.

\section{Discussion}
\label{sec:discuss}

Here I briefly outline possibilities for the extension of this inside-out hardening model to include nonzero orbital eccentricity. I also discuss the applicability of this method to LISA predictions and implications for multi-messenger sources.

\subsection{Eccentric binaries}
\label{ssec:ecc}
Although only circular binaries are considered in this work, numerous studies of stellar- and gas-driven eccentricity indicate that SMBH binaries may have \rev{significant} eccentricities when they enter the PTA regime \rev{\citep[e.g.,][]{sesana10, holley15, khan18a, zrake21, dorazio21, gualandris22, siwek23a, fastidio24, fastidio25}}. Triple SMBH interactions can also drive binaries to high eccentricities  \rev{\citep[][]{bonetti16, fastidio25}}. The inside-out hardening model can be readily extended to include eccentricity evolution with a few considerations. 

$\nuin$ could be defined as the value for circular binaries, and $\aninerg$ should be defined more precisely as the separation at which a {\em circular} $M = 10^9 M_{\odot}$ binary enters the GW-dominated regime. Since eccentric binaries harden faster via GW emission, they would transition to the GW regime earlier (at separations larger than $\agw$). \rev{In addition, some degeneracy exists in the impact on the GWB of a population of eccentric SMBHBs driven purely by GWs versus a population of circular SMBHBs driven by both astrophysical and GW hardening in the PTA band. Both can produce low-frequency attenuation of the GWB. Unless SMBHB eccentricities are extreme in the PTA band ($e\gtrsim0.95$), only modest attenuation of the low-frequency GWB spectrum should result from the redistribution of GW energy to higher-frequency harmonics \citep[e.g.,][]{kelley17b, chen2017}. However, this remains an important caveat for models of circular-only SMBHB evolution, and the sensitivity of the GWB spectrum to $\agw$ in the case of eccentric binaries deserves further study.}

\rev{Another} consideration for eccentric binaries is the choice of $\rchar$. Here I have chosen $\rchar$ such that the inner hardening regime begins prior to when (circular) binaries reach PTA frequencies, even for a future 100-yr data set. However, eccentric binaries will emit not only at twice the orbital frequency, but also at higher harmonics. \rev{Eccentric binaries will therefore} enter the PTA band at larger semi-major axes than circular binaries. Accounting for this effect would require choosing larger values of $\rchar$ (thereby restricting the space for other hardening parameters to satisfy $|\dadtrc|<\dadtmax$), ignoring the contribution of higher harmonics from wide, eccentric binaries to the low-frequency GWB (which may be small but cannot {\em a priori} be assumed negligible), or abandoning the premise of a single value of $\nuin$ for all binaries. Further exploration of this issue is left for future work. 

\subsection{Applications to the LISA regime}

This model can be applied to calculations of LISA event rates, although there are important differences to bear in mind. The huge difference in relevant masses and frequencies for PTAs versus LISA challenges the assumption that a single value for each binary hardening parameter can represent all binaries. The nHz GWB from SMBHBs is dominated by the most massive binaries ($\gtrsim 10^9\msun$) at relatively low redshifts. Thus, it is not unreasonable to assume that the binaries dominating the GWB have similarities in their hardening parameters. It is much less clear whether $\sim 10^4-10^7\msun$ binaries merging at higher redshifts should be expected to share the same hardening parameters. Caution should therefore be used when applying PTA-constrained hardening parameters to the LISA regime. 

Binaries enter the mHz frequency regime late in their inspiral, and only lower-mass SMBHBs will reach the LISA band at all before merging. Thus, the details of inner SMBHB hardening are important mainly for determining the merger timescale. As discussed above, the merger rate is much more sensitive to binary inspiral timescales than is the nHz GWB amplitude. Nonetheless, because LISA event rates remain highly uncertain, any constraints from PTA data on SMBH merger rates for LISA (even weak ones) will be valuable.

\subsection{Implications for multi-messenger astrophysics}

A SMBHB that is simultaneously accreting gas and emitting GWs could appear as a multi-messenger source, which would provide a wealth of information about SMBH accretion, environments, and binary inspiral, in addition to opening  a new regime of precision cosmology.

The question of whether gaseous or stellar processes dominate SMBHB environments is important for their EM signatures in addition to their inspiral rates.  It is possible that the massive SMBHBs dominating the nHz GWB are preferentially hosted in gas-poor, early-type galaxies, greatly lowering the likelihood of observing EM accretion signatures. In systems with plentiful cold nuclear gas that drives SMBHB inspiral, EM counterparts are more likely. However, observing distinct variability signatures of the binary orbital motion presents numerous challenges including host contamination, obscuration, and years-long required observation times. 

A concern for multi-messenger observations of LISA sources is that even if gas drives SMBHBs nearly all the way to merger, if the gas cannot follow the SMBHBs to fuel accretion after they enter the GW regime, the merger event may be EM-dark \citep[e.g.,][]{milphi05}. Recent simulations indicate that SMBHBs can accrete from their individual ``mini-disks" well after their inspiral starts to outpace the viscous disk evolution \citep[e.g.,][]{noble12, farris15, bowen17, tang18}, and EM signatures may also arise from shocks in radiatively inefficient flows \citep[e.g.,][]{bode10, farris10} or jets launched during the merger \citep[e.g.,][]{gold14, paschalidis21}. Thus, an inspiral model with $\agw>\risco$ does not necessarily indicate EM-dark mergers. 

For PTAs, multi-messenger studies have the advantage that SMBHB sources are essentially stationary on human timescales ($\tauingw \gg \tau_{\rm orb}$). If individual loud CW sources can be identified above the GWB \citep[e.g.,][]{ravi12, kelley18, agazie23_cw, gardiner24}, targeted searches for EM counterparts and/or likely host galaxies are a promising avenue for multi-messenger science \citep[][]{petrov24, gardiner25, agarwal26_targeted}. If PTA data are able to constrain the dominant astrophysical hardening mechanism even weakly, this will inform the prospects for EM counterparts. But given the large uncertainties and complexity of the relevant physical processes, a result that points to stellar-driven hardening should not be taken as strong evidence that EM counterparts to SMBHBs will not be found.

\section{Summary and conclusions}
\label{sec:conclude}

I present an analytic model for SMBHB inspiral designed to optimize physical constraints obtained from PTA observations of the nHz GWB. Because PTAs probe the innermost phase of SMBHB inspiral, I model this regime in detail while using a simple delay timescale to connect the inner regime to a cosmological SMBHB population at large scales.
Central to my approach is recognition that the GWB amplitude and spectral shape are quite sensitive to the binary semi-major axis $\agwrg$ at which inspiral transitions from the astrophysical to GW-driven regime, and that this quantity should therefore be treated as a free parameter to be constrained with PTA data. I refer to this  approach as an ``inside-out" model for SMBHB inspiral.

The sensitivity of the GWB to $\agwrg$ arises from the strong dependence of GW-driven inspiral on binary separation ($\dadt_{\rm GW}\propto a^{-3}$) and from the requirement that if a GWB is observed, $\agwrg$ must be neither far outside the PTA regime nor close to $\risco$ for most binaries. Because GW-driven inspiral also depends strongly on binary mass and mass ratio ($\dadt_{\rm GW} \propto \eta M^3$), I parameterize $\agwrg$ (in gravitational units) as a function of both quantities, normalized to $10^9\msun$, equal-mass binaries: $\agwrg = \aninerg M_9^{\alphagw} (4\eta)^{\betagw}$. 
 
The inspiral rate at $a>\agwrg$ is modeled as a single power law, $\dadt \propto a^{1-\nuin}$. This is referred to as the ``inner" astrophysical hardening regime, and if multiple processes (i.e., gas and stars) contribute to binary inspiral in this phase, $\nuin$ will capture the average hardening rate driven by these processes. The boundary of the inner regime is set by $\rchar$, which should be outside the PTA regime, but not by a large factor. A binary separation of $\rchnine=1$ pc is comfortably larger than the point at which circular, $10^9\msun$ SMBHBs enter the PTA regime, even with a hypothetical 100-yr PTA data set.

I do not explicitly model SMBHB hardening rates far outside the PTA regime.  The outer phase of binary evolution is  modeled as a delay timescale $\tau_{\rm out}$ between the time of binary formation (or the time of galaxy merger) and the time at which it reaches $\rchar$. $\rchar$ also scales with mass, and I choose a scaling that guarantees all binaries obey $\rchar > a(f_{\rm obs,min})$, where $f_{\rm obs,min}=1/T_{\rm obs}$ is the minimum GW frequency accessible to a PTA with an observation time of $T_{\rm obs}$. 

The allowed parameter values are limited such that the binary hardening rate never exceeds $\dadtmax$. I use $\dadtmax=c$ for simplicity, but a more realistic limit such as $\dadtmax=v_{\rm orb}(a)$ could also be used. The results are not very sensitive to this choice, as models with large $|\dadt|$ emit little GW radiation through the PTA regime. 

This model has similar complexity and form to the 2PL model used in NG15Astro. However, the fixed total timescale in that model imposes a normalization on hardening rates at all scales, which artificially suppresses variation in the inner regime to which PTAs are most sensitive. I also compare another simple ``inside-out" approach that is widely used: the \citet{phinney01} calculation for a GWB arising from  SMBHBs inpsiraling via GW emission alone, working backwards from a merger remnant density. This approach cannot account for how SMBHBs reach the  PTA regime, nor whether astrophysical hardening attenuates the low-frequency GWB. However, there is a minimum binary mass for which this approach is self-consistent even {\em within} the PTA regime. For current PTA data, this mass threshold is $M\gtrsim$ \rev{few $\times 10^7 \msun$ for equal-mass binaries at $z\sim0.4-0.5$, with a higher threshold for lower redshift or $q<1$. L}ower-mass SMBHBs cannot merge by $z=0$ via GWs alone, starting from the lowest frequencies accessible to PTAs. NG15Astro find that such binaries contribute \rev{$\gtrsim 0.1\%$} to the GWB, but the mass threshold increases as $T_{\rm obs}^{8/5}$. This further motivates an SMBHB hardening model that accounts for astrophysical processes but retains enough simplicity to be used in parameter inference of GW source populations.

My central findings can be summarized as follows.
\begin{itemize}
    \item Only values of $\aninerg\sim 20$-$6000\rg$ are consistent with the existence of an observable GWB produced by SMBHBs, across a wide range in other model parameters. Moreover, for a given set of other parameters, viable $\aninerg$ values are constrained to within a factor of $\sim30$-$100$. Within this narrow range the GWB amplitude and spectral shape varies significantly. The GWB amplitude peaks, and the spectrum most closely resembles a power-law, for $\aninerg\sim500$-$2000\rg$. 
    
    \item When $\aninerg$ is near the values that maximize the GWB amplitude, the GWB spectrum is relatively insensitive to $\nuin$. However, models with large $\aninerg$ and large positive $\nuin$ cause binaries to stall  outside the PTA regime. Conversely, for small $\aninerg$ and $\nuin$, rapid inspiral causes strong attenuation of the low-frequency GWB, and allowed models are strongly constrained by $\dadtmax$. 

    \item The GWB is insensitive to $\betagw$, the mass-ratio scaling of $\agwrg$, and it is only weakly sensitive to the mass scaling $\alphagw$. Steep scalings of either parameter (e.g., $|\alphagw|,|\betagw|\gtrsim1/2$) yield a GWB that arises entirely from a narrow range of SMBHB masses or mass ratios, or more often, such models are disallowed by the  $\dadtmax$ criterion.

    \item Total inspiral timescales are much more sensitive to $\alphagw$ and $\betagw$ than the GWB is. These parameters are thus more important for LISA  predictions. CW detection probabilities for individual SMBHBs may also be more sensitive to the mass and mass-ratio scaling of $\agwrg$, relative to the unresolved stochastic background. 

    \item The GWB is largely insensitive to the boundary between the outer and inner hardening phases, $\rchar$. This is true as long as $\rchar$ is not so large that satisfying $|\dadt|<\dadtmax$ at $\rchar$ imposes strong limits on the allowed ranges of other parameters. The outer phase timescale $\tau_{\rm out}$ does not strongly impact the GWB unless $\tau_{\rm out}\gtrsim 3$ Gyr, at which point the amplitude drops as more and more low-redshift binaries stall. 

    \item I examine combinations of the key model parameters $\agwrg$ and $\nuin$ that are tuned to resemble literature models for stellar- and gas-driven evolution, finding reasonable agreement between my stellar-like models and $N$-body simulation results \citep{holley25,harris26}. Although more study is needed, PTA constraints on $\agwrg$ and/or $\nuin$ would point towards the SMBHB hardening mechanism that dominates through the nHz regime.
    
\end{itemize}

This simple, computationally inexpensive framework isolates the parameters that are critical to late-stage SMBHB inspiral, where SMBHBs are emitting at PTA frequencies and where very few EM constraints exist. Future work will extend the model to include nonzero binary eccentricities and to predictions for low-mass binary mergers detectable with LISA. This inside-out SMBHB hardening model can be readily applied to current and future PTA data sets to infer the nature of the cosmic SMBHB population that is the likely source of the observed GWB.

\begin{acknowledgments}
I thank Kayhan Gültekin and Cayenne Matt for helpful discussions, and I thank Kayhan Gültekin, Joseph Lazio, Cayenne Matt, Alexander Criswell, \rev{and the anonymous referee} for thoughtful comments that significantly improved this manuscript. I gratefully acknowledge support from NSF awards AST-2307171, AST-2509457, and PHY-2607948.
\end{acknowledgments}

\appendix
\section{An alternate parameterization of the inside-out hardening model}\label{sec:model1}

Here I present an alternate parameterization of the inside-out hardening model that uses $\dadtrc$ (the hardening rate of equal-mass binaries at $\rchar$) as an input model parameter instead of $\nuin$. This could be useful if one wishes to consider binary models that have a common inspiral rate at the start of the inner regime, rather than a common power-law index within the inner regime. It also has the advantage that the $\dadtmax$ criterion is automatically satisfied as long as $|\dadtrc|<\dadtmax$. In this case, $\nuin$ is defined via
\begin{equation}
     \nuin = 1 - \frac{\log_{\rm 10}|\dadtrc| - \log_{\rm 10} |\dadt_{\rm GW}|}{\log_{\rm 10} \rchar-\log_{\rm 10} \rev{\agwcm}}
\end{equation} 
I assume that $\dadtrc$ for unequal-mass binaries has the same scaling with $\eta$ as does $\dadt(\agw)$, such that the value of $\nuin$ is constant for binaries of a given primary mass. 

Because $\nuin$ is derived from $\dadtrc$, the $\nuinmax$ criterion must be expressed in terms of limits on allowed values of $\dadtrc$ as follows:
    \begin{align}
        \log_{10}\left|\dadtrc\right| &> -\log_{10}(4\eta) + (1-\nuinmax)\log_{10}\left (\frac{\rchar}{\rev{\agwcm}} \right ) + \log_{10}\dadt(\rev{\agwcm}) \:\:\:\:\:\:{\rm and}\nonumber\\
        \log_{10}\left|\,\dadtrc\right| &< {\rm min} \left [ -\log_{10}(4\eta) + (1+\nuinmax)\log_{10}\left (\frac{\rchar}{\rev{\agwcm}} \right ) + \log_{10}\dot a(\rev{\agwcm}),\:\:\:\: \log_{10}\dadtmax  \,\right ].
    \end{align}
This criterion is especially important for the $\dadtrc$ parameterization to prevent numerical instabilities when $\rchar\sim\agw$, or $|\dadtrc|\gg$ or $\ll |\dadt_{\rm GW}|$.

To summarize, the inputs for this $\dadtrc$ hardening model parameterization are $\dadtrc$, $\agw$, $\rchar$, and $\tau_{\rm out}$. As in the standard parameterization presented in the main text, $\agw$ is specified by three parameters that control its scaling with mass and mass ratio ($\anine$, $\alphagw$, \& $\betagw$), while two parameters control the mass scaling of $\rchar$.

\bibliography{refs_new_hardening}{}

@String( prd = "Phys. Rev. D" )

@ARTICLE{arzoumanian2018,
       author = {{Arzoumanian}, Z. and {Baker}, P.~T. and {Brazier}, A. and {Burke-Spolaor}, S. and {Chamberlin}, S.~J. and {Chatterjee}, S. and {Christy}, B. and {Cordes}, J.~M. and {Cornish}, N.~J. and {Crawford}, F. and {Thankful Cromartie}, H. and {Crowter}, K. and {DeCesar}, M. and {Demorest}, P.~B. and {Dolch}, T. and {Ellis}, J.~A. and {Ferdman}, R.~D. and {Ferrara}, E. and {Folkner}, W.~M. and {Fonseca}, E. and {Garver-Daniels}, N. and {Gentile}, P.~A. and {Haas}, R. and {Hazboun}, J.~S. and {Huerta}, E.~A. and {Islo}, K. and {Jones}, G. and {Jones}, M.~L. and {Kaplan}, D.~L. and {Kaspi}, V.~M. and {Lam}, M.~T. and {Lazio}, T.~J.~W. and {Levin}, L. and {Lommen}, A.~N. and {Lorimer}, D.~R. and {Luo}, J. and {Lynch}, R.~S. and {Madison}, D.~R. and {McLaughlin}, M.~A. and {McWilliams}, S.~T. and {Mingarelli}, C.~M.~F. and {Ng}, C. and {Nice}, D.~J. and {Park}, R.~S. and {Pennucci}, T.~T. and {Pol}, N.~S. and {Ransom}, S.~M. and {Ray}, P.~S. and {Rasskazov}, A. and {Siemens}, X. and {Simon}, J. and {Spiewak}, R. and {Stairs}, I.~H. and {Stinebring}, D.~R. and {Stovall}, K. and {Swiggum}, J. and {Taylor}, S.~R. and {Vallisneri}, M. and {van Haasteren}, R. and {Vigeland}, S. and {Zhu}, W.~W. and {NANOGrav Collaboration}},
        title = "{The NANOGrav 11 Year Data Set: Pulsar-timing Constraints on the Stochastic Gravitational-wave Background}",
      journal = {\apj},
         year = 2018,
        month = may,
       volume = {859},
       number = {1},
          eid = {47},
        pages = {47},
          doi = {10.3847/1538-4357/aabd3b},
archivePrefix = {arXiv},
       eprint = {1801.02617},
 primaryClass = {astro-ph.HE},
       adsurl = {https://ui.adsabs.harvard.edu/abs/2018ApJ...859...47A}
}

@ARTICLE{sesana06,
       author = {{Sesana}, Alberto and {Haardt}, Francesco and {Madau}, Piero},
        title = "{Interaction of Massive Black Hole Binaries with Their Stellar Environment. I. Ejection of Hypervelocity Stars}",
      journal = {\apj},
         year = 2006,
        month = nov,
       volume = {651},
       number = {1},
        pages = {392-400},
          doi = {10.1086/507596},
archivePrefix = {arXiv},
       eprint = {astro-ph/0604299},
 primaryClass = {astro-ph},
       adsurl = {https://ui.adsabs.harvard.edu/abs/2006ApJ...651..392S}
}

@ARTICLE{afzal23,
       author = {{Afzal}, Adeela and {Agazie}, Gabriella and {Anumarlapudi}, Akash and {Archibald}, Anne M. and {Arzoumanian}, Zaven and {Baker}, Paul T. and {B{\'e}csy}, Bence and {Blanco-Pillado}, Jose Juan and {Blecha}, Laura and {Boddy}, Kimberly K. and {Brazier}, Adam and {Brook}, Paul R. and {Burke-Spolaor}, Sarah and {Burnette}, Rand and {Case}, Robin and {Charisi}, Maria and {Chatterjee}, Shami and {Chatziioannou}, Katerina and {Cheeseboro}, Belinda D. and {Chen}, Siyuan and {Cohen}, Tyler and {Cordes}, James M. and {Cornish}, Neil J. and {Crawford}, Fronefield and {Cromartie}, H. Thankful and {Crowter}, Kathryn and {Cutler}, Curt J. and {Decesar}, Megan E. and {Degan}, Dallas and {Demorest}, Paul B. and {Deng}, Heling and {Dolch}, Timothy and {Drachler}, Brendan and {von Eckardstein}, Richard and {Ferrara}, Elizabeth C. and {Fiore}, William and {Fonseca}, Emmanuel and {Freedman}, Gabriel E. and {Garver-Daniels}, Nate and {Gentile}, Peter A. and {Gersbach}, Kyle A. and {Glaser}, Joseph and {Good}, Deborah C. and {Guertin}, Lydia and {G{\"u}ltekin}, Kayhan and {Hazboun}, Jeffrey S. and {Hourihane}, Sophie and {Islo}, Kristina and {Jennings}, Ross J. and {Johnson}, Aaron D. and {Jones}, Megan L. and {Kaiser}, Andrew R. and {Kaplan}, David L. and {Kelley}, Luke Zoltan and {Kerr}, Matthew and {Key}, Joey S. and {Laal}, Nima and {Lam}, Michael T. and {Lamb}, William G. and {Lazio}, T. Joseph W. and {Lee}, Vincent S.~H. and {Lewandowska}, Natalia and {Lino Dos Santos}, Rafael R. and {Littenberg}, Tyson B. and {Liu}, Tingting and {Lorimer}, Duncan R. and {Luo}, Jing and {Lynch}, Ryan S. and {Ma}, Chung-Pei and {Madison}, Dustin R. and {McEwen}, Alexander and {McKee}, James W. and {McLaughlin}, Maura A. and {McMann}, Natasha and {Meyers}, Bradley W. and {Meyers}, Patrick M. and {Mingarelli}, Chiara M.~F. and {Mitridate}, Andrea and {Nay}, Jonathan and {Natarajan}, Priyamvada and {Ng}, Cherry and {Nice}, David J. and {Ocker}, Stella Koch and {Olum}, Ken D. and {Pennucci}, Timothy T. and {Perera}, Benetge B.~P. and {Petrov}, Polina and {Pol}, Nihan S. and {Radovan}, Henri A. and {Ransom}, Scott M. and {Ray}, Paul S. and {Romano}, Joseph D. and {Sardesai}, Shashwat C. and {Schmiedekamp}, Ann and {Schmiedekamp}, Carl and {Schmitz}, Kai and {Schr{\"o}der}, Tobias and {Schult}, Levi and {Shapiro-Albert}, Brent J. and {Siemens}, Xavier and {Simon}, Joseph and {Siwek}, Magdalena S. and {Stairs}, Ingrid H. and {Stinebring}, Daniel R. and {Stovall}, Kevin and {Stratmann}, Peter and {Sun}, Jerry P. and {Susobhanan}, Abhimanyu and {Swiggum}, Joseph K. and {Taylor}, Jacob and {Taylor}, Stephen R. and {Trickle}, Tanner and {Turner}, Jacob E. and {Unal}, Caner and {Vallisneri}, Michele and {Verma}, Sonali and {Vigeland}, Sarah J. and {Wahl}, Haley M. and {Wang}, Qiaohong and {Witt}, Caitlin A. and {Wright}, David and {Young}, Olivia and {Zurek}, Kathryn M. and {NANOGrav Collaboration}},
        title = "{The NANOGrav 15 yr Data Set: Search for Signals from New Physics}",
      journal = {\apjl},
         year = 2023,
        month = jul,
       volume = {951},
       number = {1},
          eid = {L11},
        pages = {L11},
          doi = {10.3847/2041-8213/acdc91},
archivePrefix = {arXiv},
       eprint = {2306.16219},
 primaryClass = {astro-ph.HE},
       adsurl = {https://ui.adsabs.harvard.edu/abs/2023ApJ...951L..11A}
}

@ARTICLE{agarwal26_targeted,
       author = {{Agarwal}, Nikita and {Agazie}, Gabriella and {Anumarlapudi}, Akash and {Archibald}, Anne M. and {Arzoumanian}, Zaven and {Baier}, Jeremy G. and {Baker}, Paul T. and {B{\'e}csy}, Bence and {Blecha}, Laura and {Brazier}, Adam and {Brook}, Paul R. and {Burke-Spolaor}, Sarah and {Burnette}, Rand and {Case}, Robin and {Casey-Clyde}, J. Andrew and {Chang}, Yu-Ting and {Charisi}, Maria and {Chatterjee}, Shami and {Cohen}, Tyler and {Coppi}, Paolo and {Cordes}, James M. and {Cornish}, Neil J. and {Crawford}, Fronefield and {Cromartie}, H. Thankful and {Crowter}, Kathryn and {Decesar}, Megan E. and {Demorest}, Paul B. and {Deng}, Heling and {Dey}, Lankeswar and {Dolch}, Timothy and {D'Orazio}, Daniel J. and {Eisenberg}, Ellis and {Ferrara}, Elizabeth C. and {Doskoch}, Graham and {Fiore}, William and {Fonseca}, Emmanuel and {Freedman}, Gabriel E. and {Gardiner}, Emiko C. and {Garver-Daniels}, Nate and {Gentile}, Peter A. and {Gersbach}, Kyle A. and {Glaser}, Joseph and {Graham}, Matthew J. and {Good}, Deborah C. and {G{\"u}ltekin}, Kayhan and {Harris}, C.~J. and {Hazboun}, Jeffrey S. and {Hutchison}, Forrest and {Jennings}, Ross J. and {Johnson}, Aaron D. and {Jones}, Megan L. and {Kaplan}, David L. and {Kelley}, Luke Zoltan and {Kerr}, Matthew and {Key}, Joey S. and {Laal}, Nima and {Lam}, Michael T. and {Lamb}, William G. and {Larsen}, Bjorn and {Lazio}, T. Joseph W. and {Lewandowska}, Natalia and {Liu}, Tingting and {Lorimer}, Duncan R. and {Luo}, Jing and {Lynch}, Ryan S. and {Ma}, Chung-Pei and {Madison}, Dustin R. and {Matt}, Cayenne and {McEwen}, Alexander and {McKee}, James W. and {McLaughlin}, Maura A. and {McMann}, Natasha and {Meyers}, Bradley W. and {Meyers}, Patrick M. and {Mingarelli}, Chiara M.~F. and {Mitridate}, Andrea and {Natarajan}, Priyamvada and {Ng}, Cherry and {Nice}, David J. and {Nichols}, Shania and {Ocker}, Stella Koch and {Olum}, Ken D. and {Pennucci}, Timothy T. and {Perera}, Benetge B.~P. and {Petrov}, Polina and {Pol}, Nihan S. and {Radovan}, Henri A. and {Ransom}, Scott M. and {Ray}, Paul S. and {Romano}, Joseph D. and {Runnoe}, Jessie C. and {Saffer}, Alexander and {Sardesai}, Shashwat C. and {Schmiedekamp}, Ann and {Schmiedekamp}, Carl and {Schmitz}, Kai and {Semenzato}, Federico and {Shapiro-Albert}, Brent J. and {Shivakumar}, Rohan and {Siemens}, Xavier and {Simon}, Joseph and {Sosa Fiscella}, Sophia V. and {Stairs}, Ingrid H. and {Stinebring}, Daniel R. and {Stovall}, Kevin and {Susobhanan}, Abhimanyu and {Swiggum}, Joseph K. and {Taylor}, Jacob A. and {Taylor}, Stephen R. and {Thompson}, Mercedes S. and {Turner}, Jacob E. and {Vallisneri}, Michele and {van Haasteren}, Rutger and {Vigeland}, Sarah J. and {Wahl}, Haley M. and {Willson}, London and {Wilson}, Kevin P. and {Witt}, Caitlin A. and {Wright}, David and {Young}, Olivia and {Zheng}, Qinyuan and {Nanograv Collaboration}},
        title = "{The NANOGrav 15 yr Dataset: Targeted Searches for Supermassive Black Hole Binaries}",
      journal = {\apjl},
         year = 2026,
        month = feb,
       volume = {998},
       number = {1},
          eid = {L11},
        pages = {L11},
          doi = {10.3847/2041-8213/ae3719},
archivePrefix = {arXiv},
       eprint = {2508.16534},
 primaryClass = {astro-ph.HE},
       adsurl = {https://ui.adsabs.harvard.edu/abs/2026ApJ...998L..11A}
}

@ARTICLE{agazie23a,
       author = {{Agazie}, Gabriella and {Anumarlapudi}, Akash and {Archibald}, Anne M. and {Arzoumanian}, Zaven and {Baker}, Paul T. and {B{\'e}csy}, Bence and {Blecha}, Laura and {Brazier}, Adam and {Brook}, Paul R. and {Burke-Spolaor}, Sarah and {Burnette}, Rand and {Case}, Robin and {Charisi}, Maria and {Chatterjee}, Shami and {Chatziioannou}, Katerina and {Cheeseboro}, Belinda D. and {Chen}, Siyuan and {Cohen}, Tyler and {Cordes}, James M. and {Cornish}, Neil J. and {Crawford}, Fronefield and {Cromartie}, H. Thankful and {Crowter}, Kathryn and {Cutler}, Curt J. and {Decesar}, Megan E. and {Degan}, Dallas and {Demorest}, Paul B. and {Deng}, Heling and {Dolch}, Timothy and {Drachler}, Brendan and {Ellis}, Justin A. and {Ferrara}, Elizabeth C. and {Fiore}, William and {Fonseca}, Emmanuel and {Freedman}, Gabriel E. and {Garver-Daniels}, Nate and {Gentile}, Peter A. and {Gersbach}, Kyle A. and {Glaser}, Joseph and {Good}, Deborah C. and {G{\"u}ltekin}, Kayhan and {Hazboun}, Jeffrey S. and {Hourihane}, Sophie and {Islo}, Kristina and {Jennings}, Ross J. and {Johnson}, Aaron D. and {Jones}, Megan L. and {Kaiser}, Andrew R. and {Kaplan}, David L. and {Kelley}, Luke Zoltan and {Kerr}, Matthew and {Key}, Joey S. and {Klein}, Tonia C. and {Laal}, Nima and {Lam}, Michael T. and {Lamb}, William G. and {Lazio}, T. Joseph W. and {Lewandowska}, Natalia and {Littenberg}, Tyson B. and {Liu}, Tingting and {Lommen}, Andrea and {Lorimer}, Duncan R. and {Luo}, Jing and {Lynch}, Ryan S. and {Ma}, Chung-Pei and {Madison}, Dustin R. and {Mattson}, Margaret A. and {McEwen}, Alexander and {McKee}, James W. and {McLaughlin}, Maura A. and {McMann}, Natasha and {Meyers}, Bradley W. and {Meyers}, Patrick M. and {Mingarelli}, Chiara M.~F. and {Mitridate}, Andrea and {Natarajan}, Priyamvada and {Ng}, Cherry and {Nice}, David J. and {Ocker}, Stella Koch and {Olum}, Ken D. and {Pennucci}, Timothy T. and {Perera}, Benetge B.~P. and {Petrov}, Polina and {Pol}, Nihan S. and {Radovan}, Henri A. and {Ransom}, Scott M. and {Ray}, Paul S. and {Romano}, Joseph D. and {Sardesai}, Shashwat C. and {Schmiedekamp}, Ann and {Schmiedekamp}, Carl and {Schmitz}, Kai and {Schult}, Levi and {Shapiro-Albert}, Brent J. and {Siemens}, Xavier and {Simon}, Joseph and {Siwek}, Magdalena S. and {Stairs}, Ingrid H. and {Stinebring}, Daniel R. and {Stovall}, Kevin and {Sun}, Jerry P. and {Susobhanan}, Abhimanyu and {Swiggum}, Joseph K. and {Taylor}, Jacob and {Taylor}, Stephen R. and {Turner}, Jacob E. and {Unal}, Caner and {Vallisneri}, Michele and {van Haasteren}, Rutger and {Vigeland}, Sarah J. and {Wahl}, Haley M. and {Wang}, Qiaohong and {Witt}, Caitlin A. and {Young}, Olivia and {Nanograv Collaboration}},
        title = "{The NANOGrav 15 yr Data Set: Evidence for a Gravitational-wave Background}",
      journal = {\apjl},
         year = 2023,
        month = jul,
       volume = {951},
       number = {1},
          eid = {L8},
        pages = {L8},
          doi = {10.3847/2041-8213/acdac6},
archivePrefix = {arXiv},
       eprint = {2306.16213},
 primaryClass = {astro-ph.HE},
       adsurl = {https://ui.adsabs.harvard.edu/abs/2023ApJ...951L...8A}
}

@ARTICLE{agazie23e,
       author = {{Agazie}, Gabriella and {Anumarlapudi}, Akash and {Archibald}, Anne M. and {Baker}, Paul T. and {B{\'e}csy}, Bence and {Blecha}, Laura and {Bonilla}, Alexander and {Brazier}, Adam and {Brook}, Paul R. and {Burke-Spolaor}, Sarah and {Burnette}, Rand and {Case}, Robin and {Casey-Clyde}, J. Andrew and {Charisi}, Maria and {Chatterjee}, Shami and {Chatziioannou}, Katerina and {Cheeseboro}, Belinda D. and {Chen}, Siyuan and {Cohen}, Tyler and {Cordes}, James M. and {Cornish}, Neil J. and {Crawford}, Fronefield and {Cromartie}, H. Thankful and {Crowter}, Kathryn and {Cutler}, Curt J. and {D'Orazio}, Daniel J. and {Decesar}, Megan E. and {Degan}, Dallas and {Demorest}, Paul B. and {Deng}, Heling and {Dolch}, Timothy and {Drachler}, Brendan and {Ferrara}, Elizabeth C. and {Fiore}, William and {Fonseca}, Emmanuel and {Freedman}, Gabriel E. and {Gardiner}, Emiko and {Garver-Daniels}, Nate and {Gentile}, Peter A. and {Gersbach}, Kyle A. and {Glaser}, Joseph and {Good}, Deborah C. and {G{\"u}ltekin}, Kayhan and {Hazboun}, Jeffrey S. and {Hourihane}, Sophie and {Islo}, Kristina and {Jennings}, Ross J. and {Johnson}, Aaron and {Jones}, Megan L. and {Kaiser}, Andrew R. and {Kaplan}, David L. and {Kelley}, Luke Zoltan and {Kerr}, Matthew and {Key}, Joey S. and {Laal}, Nima and {Lam}, Michael T. and {Lamb}, William G. and {Lazio}, T. Joseph W. and {Lewandowska}, Natalia and {Littenberg}, Tyson B. and {Liu}, Tingting and {Luo}, Jing and {Lynch}, Ryan S. and {Ma}, Chung-Pei and {Madison}, Dustin R. and {McEwen}, Alexander and {McKee}, James W. and {McLaughlin}, Maura A. and {McMann}, Natasha and {Meyers}, Bradley W. and {Meyers}, Patrick M. and {Mingarelli}, Chiara M.~F. and {Mitridate}, Andrea and {Natarajan}, Priyamvada and {Ng}, Cherry and {Nice}, David J. and {Ocker}, Stella Koch and {Olum}, Ken D. and {Pennucci}, Timothy T. and {Perera}, Benetge B.~P. and {Petrov}, Polina and {Pol}, Nihan S. and {Radovan}, Henri A. and {Ransom}, Scott M. and {Ray}, Paul S. and {Romano}, Joseph D. and {Runnoe}, Jessie C. and {Sardesai}, Shashwat C. and {Schmiedekamp}, Ann and {Schmiedekamp}, Carl and {Schmitz}, Kai and {Schult}, Levi and {Shapiro-Albert}, Brent J. and {Siemens}, Xavier and {Simon}, Joseph and {Siwek}, Magdalena S. and {Stairs}, Ingrid H. and {Stinebring}, Daniel R. and {Stovall}, Kevin and {Sun}, Jerry P. and {Susobhanan}, Abhimanyu and {Swiggum}, Joseph K. and {Taylor}, Jacob and {Taylor}, Stephen R. and {Turner}, Jacob E. and {Unal}, Caner and {Vallisneri}, Michele and {Vigeland}, Sarah J. and {Wachter}, Jeremy M. and {Wahl}, Haley M. and {Wang}, Qiaohong and {Witt}, Caitlin A. and {Wright}, David and {Young}, Olivia and {Nanograv Collaboration}},
        title = "{The NANOGrav 15 yr Data Set: Constraints on Supermassive Black Hole Binaries from the Gravitational-wave Background}",
      journal = {\apjl},
         year = 2023,
        month = aug,
       volume = {952},
       number = {2},
          eid = {L37},
        pages = {L37},
          doi = {10.3847/2041-8213/ace18b},
archivePrefix = {arXiv},
       eprint = {2306.16220},
 primaryClass = {astro-ph.HE},
       adsurl = {https://ui.adsabs.harvard.edu/abs/2023ApJ...952L..37A}
}

@ARTICLE{agazie23_cw,
       author = {{Agazie}, Gabriella and {Anumarlapudi}, Akash and {Archibald}, Anne M. and {Arzoumanian}, Zaven and {Baker}, Paul T. and {B{\'e}csy}, Bence and {Blecha}, Laura and {Brazier}, Adam and {Brook}, Paul R. and {Burke-Spolaor}, Sarah and {Case}, Robin and {Casey-Clyde}, J. Andrew and {Charisi}, Maria and {Chatterjee}, Shami and {Cohen}, Tyler and {Cordes}, James M. and {Cornish}, Neil J. and {Crawford}, Fronefield and {Cromartie}, H. Thankful and {Crowter}, Kathryn and {Decesar}, Megan E. and {Demorest}, Paul B. and {Digman}, Matthew C. and {Dolch}, Timothy and {Drachler}, Brendan and {Ferrara}, Elizabeth C. and {Fiore}, William and {Fonseca}, Emmanuel and {Freedman}, Gabriel E. and {Garver-Daniels}, Nate and {Gentile}, Peter A. and {Glaser}, Joseph and {Good}, Deborah C. and {G{\"u}ltekin}, Kayhan and {Hazboun}, Jeffrey S. and {Hourihane}, Sophie and {Jennings}, Ross J. and {Johnson}, Aaron D. and {Jones}, Megan L. and {Kaiser}, Andrew R. and {Kaplan}, David L. and {Kelley}, Luke Zoltan and {Kerr}, Matthew and {Key}, Joey S. and {Laal}, Nima and {Lam}, Michael T. and {Lamb}, William G. and {Lazio}, T. Joseph W. and {Lewandowska}, Natalia and {Liu}, Tingting and {Lorimer}, Duncan R. and {Luo}, Jing and {Lynch}, Ryan S. and {Ma}, Chung-Pei and {Madison}, Dustin R. and {McEwen}, Alexander and {McKee}, James W. and {McLaughlin}, Maura A. and {McMann}, Natasha and {Meyers}, Bradley W. and {Meyers}, Patrick M. and {Mingarelli}, Chiara M.~F. and {Mitridate}, Andrea and {Ng}, Cherry and {Nice}, David J. and {Ocker}, Stella Koch and {Olum}, Ken D. and {Pennucci}, Timothy T. and {Perera}, Benetge B.~P. and {Petrov}, Polina and {Pol}, Nihan S. and {Radovan}, Henri A. and {Ransom}, Scott M. and {Ray}, Paul S. and {Romano}, Joseph D. and {Sardesai}, Shashwat C. and {Schmiedekamp}, Ann and {Schmiedekamp}, Carl and {Schmitz}, Kai and {Shapiro-Albert}, Brent J. and {Siemens}, Xavier and {Simon}, Joseph and {Siwek}, Magdalena S. and {Stairs}, Ingrid H. and {Stinebring}, Daniel R. and {Stovall}, Kevin and {Susobhanan}, Abhimanyu and {Swiggum}, Joseph K. and {Taylor}, Jacob and {Taylor}, Stephen R. and {Turner}, Jacob E. and {Unal}, Caner and {Vallisneri}, Michele and {van Haasteren}, Rutger and {Vigeland}, Sarah J. and {Wahl}, Haley M. and {Witt}, Caitlin A. and {Young}, Olivia and {Nanograv Collaboration}},
        title = "{The NANOGrav 15 yr Data Set: Bayesian Limits on Gravitational Waves from Individual Supermassive Black Hole Binaries}",
      journal = {\apjl},
         year = 2023,
        month = jul,
       volume = {951},
       number = {2},
          eid = {L50},
        pages = {L50},
          doi = {10.3847/2041-8213/ace18a},
archivePrefix = {arXiv},
       eprint = {2306.16222},
 primaryClass = {astro-ph.HE},
       adsurl = {https://ui.adsabs.harvard.edu/abs/2023ApJ...951L..50A}
}

@ARTICLE{amaro17,
       author = {{Amaro-Seoane}, Pau and {Audley}, Heather and {Babak}, Stanislav and
         {Baker}, John and {Barausse}, Enrico and {Bender}, Peter and
         {Berti}, Emanuele and {Binetruy}, Pierre and {Born}, Michael and
         {Bortoluzzi}, Daniele and {Camp}, Jordan and {Caprini}, Chiara and
         {Cardoso}, Vitor and {Colpi}, Monica and {Conklin}, John and
         {Cornish}, Neil and {Cutler}, Curt and {Danzmann}, Karsten and
         {Dolesi}, Rita and {Ferraioli}, Luigi and {Ferroni}, Valerio and
         {Fitzsimons}, Ewan and {Gair}, Jonathan and {Gesa Bote}, Lluis and
         {Giardini}, Domenico and {Gibert}, Ferran and {Grimani}, Catia and
         {Halloin}, Hubert and {Heinzel}, Gerhard and {Hertog}, Thomas and
         {Hewitson}, Martin and {Holley-Bockelmann}, Kelly and
         {Hollington}, Daniel and {Hueller}, Mauro and {Inchauspe}, Henri and
         {Jetzer}, Philippe and {Karnesis}, Nikos and {Killow}, Christian and
         {Klein}, Antoine and {Klipstein}, Bill and {Korsakova}, Natalia and
         {Larson}, Shane L and {Livas}, Jeffrey and {Lloro}, Ivan and
         {Man}, Nary and {Mance}, Davor and {Martino}, Joseph and
         {Mateos}, Ignacio and {McKenzie}, Kirk and {McWilliams}, Sean T and
         {Miller}, Cole and {Mueller}, Guido and {Nardini}, Germano and
         {Nelemans}, Gijs and {Nofrarias}, Miquel and {Petiteau}, Antoine and
         {Pivato}, Paolo and {Plagnol}, Eric and {Porter}, Ed and
         {Reiche}, Jens and {Robertson}, David and {Robertson}, Norna and
         {Rossi}, Elena and {Russano}, Giuliana and {Schutz}, Bernard and
         {Sesana}, Alberto and {Shoemaker}, David and {Slutsky}, Jacob and
         {Sopuerta}, Carlos F. and {Sumner}, Tim and {Tamanini}, Nicola and
         {Thorpe}, Ira and {Troebs}, Michael and {Vallisneri}, Michele and
         {Vecchio}, Alberto and {Vetrugno}, Daniele and {Vitale}, Stefano and
         {Volonteri}, Marta and {Wanner}, Gudrun and {Ward}, Harry and
         {Wass}, Peter and {Weber}, William and {Ziemer}, John and
         {Zweifel}, Peter},
        title = "{Laser Interferometer Space Antenna}",
      journal = {arXiv e-prints},
         year = 2017,
        month = feb,
          eid = {arXiv:1702.00786},
        pages = {arXiv:1702.00786},
archivePrefix = {arXiv},
       eprint = {1702.00786},
 primaryClass = {astro-ph.IM},
       adsurl = {https://ui.adsabs.harvard.edu/abs/2017arXiv170200786A}
}

@ARTICLE{amaro23,
       author = {{Amaro-Seoane}, Pau and {Andrews}, Jeff and {Arca Sedda}, Manuel and {Askar}, Abbas and {Baghi}, Quentin and {Balasov}, Razvan and {Bartos}, Imre and {Bavera}, Simone S. and {Bellovary}, Jillian and {Berry}, Christopher P.~L. and {Berti}, Emanuele and {Bianchi}, Stefano and {Blecha}, Laura and {Blondin}, St{\'e}phane and {Bogdanovi{\'c}}, Tamara and {Boissier}, Samuel and {Bonetti}, Matteo and {Bonoli}, Silvia and {Bortolas}, Elisa and {Breivik}, Katelyn and {Capelo}, Pedro R. and {Caramete}, Laurentiu and {Cattorini}, Federico and {Charisi}, Maria and {Chaty}, Sylvain and {Chen}, Xian and {Chru{\'s}li{\'n}ska}, Martyna and {Chua}, Alvin J.~K. and {Church}, Ross and {Colpi}, Monica and {D'Orazio}, Daniel and {Danielski}, Camilla and {Davies}, Melvyn B. and {Dayal}, Pratika and {De Rosa}, Alessandra and {Derdzinski}, Andrea and {Destounis}, Kyriakos and {Dotti}, Massimo and {Du{\c{t}}an}, Ioana and {Dvorkin}, Irina and {Fabj}, Gaia and {Foglizzo}, Thierry and {Ford}, Saavik and {Fouvry}, Jean-Baptiste and {Franchini}, Alessia and {Fragos}, Tassos and {Fryer}, Chris and {Gaspari}, Massimo and {Gerosa}, Davide and {Graziani}, Luca and {Groot}, Paul and {Habouzit}, Melanie and {Haggard}, Daryl and {Haiman}, Zoltan and {Han}, Wen-Biao and {Istrate}, Alina and {Johansson}, Peter H. and {Khan}, Fazeel Mahmood and {Kimpson}, Tomas and {Kokkotas}, Kostas and {Kong}, Albert and {Korol}, Valeriya and {Kremer}, Kyle and {Kupfer}, Thomas and {Lamberts}, Astrid and {Larson}, Shane and {Lau}, Mike and {Liu}, Dongliang and {Lloyd-Ronning}, Nicole and {Lodato}, Giuseppe and {Lupi}, Alessandro and {Ma}, Chung-Pei and {Maccarone}, Tomas and {Mandel}, Ilya and {Mangiagli}, Alberto and {Mapelli}, Michela and {Mathis}, St{\'e}phane and {Mayer}, Lucio and {McGee}, Sean and {McKernan}, Berry and {Miller}, M. Coleman and {Mota}, David F. and {Mumpower}, Matthew and {Nasim}, Syeda S. and {Nelemans}, Gijs and {Noble}, Scott and {Pacucci}, Fabio and {Panessa}, Francesca and {Paschalidis}, Vasileios and {Pfister}, Hugo and {Porquet}, Delphine and {Quenby}, John and {Ricarte}, Angelo and {R{\"o}pke}, Friedrich K. and {Regan}, John and {Rosswog}, Stephan and {Ruiter}, Ashley and {Ruiz}, Milton and {Runnoe}, Jessie and {Schneider}, Raffaella and {Schnittman}, Jeremy and {Secunda}, Amy and {Sesana}, Alberto and {Seto}, Naoki and {Shao}, Lijing and {Shapiro}, Stuart and {Sopuerta}, Carlos and {Stone}, Nicholas C. and {Suvorov}, Arthur and {Tamanini}, Nicola and {Tamfal}, Tomas and {Tauris}, Thomas and {Temmink}, Karel and {Tomsick}, John and {Toonen}, Silvia and {Torres-Orjuela}, Alejandro and {Toscani}, Martina and {Tsokaros}, Antonios and {Unal}, Caner and {V{\'a}zquez-Aceves}, Ver{\'o}nica and {Valiante}, Rosa and {van Putten}, Maurice and {van Roestel}, Jan and {Vignali}, Christian and {Volonteri}, Marta and {Wu}, Kinwah and {Younsi}, Ziri and {Yu}, Shenghua and {Zane}, Silvia and {Zwick}, Lorenz and {Antonini}, Fabio and {Baibhav}, Vishal and {Barausse}, Enrico and {Bonilla Rivera}, Alexander and {Branchesi}, Marica and {Branduardi-Raymont}, Graziella and {Burdge}, Kevin and {Chakraborty}, Srija and {Cuadra}, Jorge and {Dage}, Kristen and {Davis}, Benjamin and {de Mink}, Selma E. and {Decarli}, Roberto and {Doneva}, Daniela and {Escoffier}, Stephanie and {Gandhi}, Poshak and {Haardt}, Francesco and {Lousto}, Carlos O. and {Nissanke}, Samaya and {Nordhaus}, Jason and {O'Shaughnessy}, Richard and {Portegies Zwart}, Simon and {Pound}, Adam and {Schussler}, Fabian and {Sergijenko}, Olga and {Spallicci}, Alessandro and {Vernieri}, Daniele and {Vigna-G{\'o}mez}, Alejandro},
        title = "{Astrophysics with the Laser Interferometer Space Antenna}",
      journal = {Living Reviews in Relativity},
         year = 2023,
        month = dec,
       volume = {26},
       number = {1},
          eid = {2},
        pages = {2},
          doi = {10.1007/s41114-022-00041-y},
archivePrefix = {arXiv},
       eprint = {2203.06016},
 primaryClass = {gr-qc},
       adsurl = {https://ui.adsabs.harvard.edu/abs/2023LRR....26....2A}
}

@ARTICLE{armnat02,
   author = {{Armitage}, P.~J. and {Natarajan}, P.},
    title = "{Accretion during the Merger of Supermassive Black Holes}",
  journal = {\apjl},
   eprint = {arXiv:astro-ph/0201318},
     year = 2002,
    month = mar,
   volume = 567,
    pages = {L9-L12},
      doi = {10.1086/339770},
   adsurl = {http://adsabs.harvard.edu/abs/2002ApJ...567L...9A}
}

@ARTICLE{barausse12,
   author = {{Barausse}, E.},
    title = "{The evolution of massive black holes and their spins in their galactic hosts}",
  journal = {\mnras},
archivePrefix = "arXiv",
   eprint = {1201.5888},
     year = 2012,
    month = jul,
   volume = 423,
    pages = {2533-2557},
      doi = {10.1111/j.1365-2966.2012.21057.x},
   adsurl = {http://adsabs.harvard.edu/abs/2012MNRAS.423.2533B}
}

@ARTICLE{begelman80,
   author = {{Begelman}, M.~C. and {Blandford}, R.~D. and {Rees}, M.~J.},
    title = "{Massive black hole binaries in active galactic nuclei}",
  journal = {\nat},
     year = 1980,
    month = sep,
   volume = 287,
    pages = {307-309},
      doi = {10.1038/287307a0},
   adsurl = {http://adsabs.harvard.edu/abs/1980Natur.287..307B}
}

@ARTICLE{berczik06,
   author = {{Berczik}, P. and {Merritt}, D. and {Spurzem}, R. and {Bischof}, H.-P.
	},
    title = "{Efficient Merger of Binary Supermassive Black Holes in Nonaxisymmetric Galaxies}",
  journal = {\apjl},
   eprint = {arXiv:astro-ph/0601698},
     year = 2006,
    month = may,
   volume = 642,
    pages = {L21-L24},
      doi = {10.1086/504426},
   adsurl = {http://adsabs.harvard.edu/abs/2006ApJ...642L..21B}
}

@ARTICLE{blecha11,
   author = {{Blecha}, L. and {Cox}, T.~J. and {Loeb}, A. and {Hernquist}, L.
	},
    title = "{Recoiling black holes in merging galaxies: relationship to active galactic nucleus lifetimes, starbursts and the M$_{BH}$-{$\sigma$}$_{*}$ relation}",
  journal = {\mnras},
archivePrefix = "arXiv",
   eprint = {1009.4940},
 primaryClass = "astro-ph.CO",
     year = 2011,
    month = apr,
   volume = 412,
    pages = {2154-2182},
      doi = {10.1111/j.1365-2966.2010.18042.x},
   adsurl = {http://adsabs.harvard.edu/abs/2011MNRAS.412.2154B}
}

@ARTICLE{bode10,
       author = {{Bode}, Tanja and {Haas}, Roland and {Bogdanovi{\'c}}, Tamara and {Laguna}, Pablo and {Shoemaker}, Deirdre},
        title = "{Relativistic Mergers of Supermassive Black Holes and Their Electromagnetic Signatures}",
      journal = {\apj},
         year = 2010,
        month = jun,
       volume = {715},
       number = {2},
        pages = {1117-1131},
          doi = {10.1088/0004-637X/715/2/1117},
archivePrefix = {arXiv},
       eprint = {0912.0087},
 primaryClass = {gr-qc},
       adsurl = {https://ui.adsabs.harvard.edu/abs/2010ApJ...715.1117B}
}

@ARTICLE{bogdanovic22,
       author = {{Bogdanovi{\'c}}, Tamara and {Miller}, M. Coleman and {Blecha}, Laura},
        title = "{Electromagnetic counterparts to massive black-hole mergers}",
      journal = {Living Reviews in Relativity},
         year = 2022,
        month = dec,
       volume = {25},
       number = {1},
          eid = {3},
        pages = {3},
          doi = {10.1007/s41114-022-00037-8},
archivePrefix = {arXiv},
       eprint = {2109.03262},
 primaryClass = {astro-ph.HE},
       adsurl = {https://ui.adsabs.harvard.edu/abs/2022LRR....25....3B}
}

@ARTICLE{bonetti16,
   author = {{Bonetti}, M. and {Haardt}, F. and {Sesana}, A. and {Barausse}, E.
	},
    title = "{Post-Newtonian evolution of massive black hole triplets in galactic nuclei - I. Numerical implementation and tests}",
  journal = {\mnras},
archivePrefix = "arXiv",
   eprint = {1604.08770},
     year = 2016,
    month = oct,
   volume = 461,
    pages = {4419-4434},
      doi = {10.1093/mnras/stw1590},
   adsurl = {http://adsabs.harvard.edu/abs/2016MNRAS.461.4419B}
}

@ARTICLE{bonetti18a,
   author = {{Bonetti}, M. and {Haardt}, F. and {Sesana}, A. and {Barausse}, E.
	},
    title = "{Post-Newtonian evolution of massive black hole triplets in galactic nuclei - II. Survey of the parameter space}",
  journal = {\mnras},
archivePrefix = "arXiv",
   eprint = {1709.06088},
     year = 2018,
    month = jul,
   volume = 477,
    pages = {3910-3926},
      doi = {10.1093/mnras/sty896},
   adsurl = {http://adsabs.harvard.edu/abs/2018MNRAS.477.3910B}
}

@ARTICLE{bonetti18b,
       author = {{Bonetti}, Matteo and {Sesana}, Alberto and {Barausse}, Enrico and
         {Haardt}, Francesco},
        title = "{Post-Newtonian evolution of massive black hole triplets in galactic nuclei - III. A robust lower limit to the nHz stochastic background of gravitational waves}",
      journal = {\mnras},
         year = 2018,
        month = jun,
       volume = {477},
       number = {2},
        pages = {2599-2612},
          doi = {10.1093/mnras/sty874},
archivePrefix = {arXiv},
       eprint = {1709.06095},
 primaryClass = {astro-ph.GA},
       adsurl = {https://ui.adsabs.harvard.edu/abs/2018MNRAS.477.2599B}
}

@ARTICLE{bonetti19,
       author = {{Bonetti}, Matteo and {Sesana}, Alberto and {Haardt}, Francesco and
         {Barausse}, Enrico and {Colpi}, Monica},
        title = "{Post-Newtonian evolution of massive black hole triplets in galactic nuclei - IV. Implications for LISA}",
      journal = {\mnras},
         year = 2019,
        month = jul,
       volume = {486},
       number = {3},
        pages = {4044-4060},
          doi = {10.1093/mnras/stz903},
archivePrefix = {arXiv},
       eprint = {1812.01011},
 primaryClass = {astro-ph.GA},
       adsurl = {https://ui.adsabs.harvard.edu/abs/2019MNRAS.486.4044B}
}

@ARTICLE{bonoli25,
       author = {{Bonoli}, Silvia and {Izquierdo-Villalba}, David and {Spinoso}, Daniele and {Colpi}, Monica and {Sesana}, Alberto and {Polkas}, Markos and {Springel}, Volker},
        title = "{Constraints on the early growth of massive black holes from PTA and JWST with L-GalaxiesBH}",
      journal = {arXiv e-prints},
         year = 2025,
        month = sep,
          eid = {arXiv:2509.12325},
        pages = {arXiv:2509.12325},
          doi = {10.48550/arXiv.2509.12325},
archivePrefix = {arXiv},
       eprint = {2509.12325},
 primaryClass = {astro-ph.GA},
       adsurl = {https://ui.adsabs.harvard.edu/abs/2025arXiv250912325B}
}

@ARTICLE{bowen17,
       author = {{Bowen}, Dennis B. and {Campanelli}, Manuela and {Krolik}, Julian H. and {Mewes}, Vassilios and {Noble}, Scott C.},
        title = "{Relativistic Dynamics and Mass Exchange in Binary Black Hole Mini-disks}",
      journal = {\apj},
         year = 2017,
        month = mar,
       volume = {838},
       number = {1},
          eid = {42},
        pages = {42},
          doi = {10.3847/1538-4357/aa63f3},
archivePrefix = {arXiv},
       eprint = {1612.02373},
 primaryClass = {astro-ph.HE},
       adsurl = {https://ui.adsabs.harvard.edu/abs/2017ApJ...838...42B}
}

@ARTICLE{burke11,
   author = {{Burke-Spolaor}, S.},
    title = "{A radio Census of binary supermassive black holes}",
  journal = {\mnras},
archivePrefix = "arXiv",
   eprint = {1008.4382},
 primaryClass = "astro-ph.CO",
     year = 2011,
    month = feb,
   volume = 410,
    pages = {2113-2122},
      doi = {10.1111/j.1365-2966.2010.17586.x},
   adsurl = {http://adsabs.harvard.edu/abs/2011MNRAS.410.2113B}
}

@ARTICLE{buttigieg26,
       author = {{Buttigieg}, Stephanie and {Sijacki}, Debora and {Moore}, Christopher J. and {Bourne}, Martin A. and {Sesana}, Alberto},
        title = "{Comparing gravitational wave background predictions from cosmological simulations to pulsar timing observations}",
      journal = {\mnras},
         year = 2026,
        month = sep,
          doi = {10.1093/mnras/stag1749},
archivePrefix = {arXiv},
       eprint = {2607.05208},
 primaryClass = {astro-ph.GA},
       adsurl = {https://ui.adsabs.harvard.edu/abs/2026MNRAS.tmp.1627B}
}

@ARTICLE{charisi16,
       author = {{Charisi}, M. and {Bartos}, I. and {Haiman}, Z. and
         {Price-Whelan}, A.~M. and {Graham}, M.~J. and {Bellm}, E.~C. and
         {Laher}, R.~R. and {M{\'a}rka}, S.},
        title = "{A population of short-period variable quasars from PTF as supermassive black hole binary candidates}",
      journal = {\mnras},
         year = 2016,
        month = dec,
       volume = {463},
       number = {2},
        pages = {2145-2171},
          doi = {10.1093/mnras/stw1838},
archivePrefix = {arXiv},
       eprint = {1604.01020},
 primaryClass = {astro-ph.GA},
       adsurl = {https://ui.adsabs.harvard.edu/abs/2016MNRAS.463.2145C}
}

@ARTICLE{colpi14,
       author = {{Colpi}, Monica},
        title = "{Massive Binary Black Holes in Galactic Nuclei and Their Path to Coalescence}",
      journal = {\ssr},
         year = 2014,
        month = sep,
       volume = {183},
       number = {1-4},
        pages = {189-221},
          doi = {10.1007/s11214-014-0067-1},
archivePrefix = {arXiv},
       eprint = {1407.3102},
 primaryClass = {astro-ph.GA},
       adsurl = {https://ui.adsabs.harvard.edu/abs/2014SSRv..183..189C}
}

@ARTICLE{comerford25,
       author = {{Comerford}, Julia M. and {Simon}, Joseph},
        title = "{Preferential Accretion onto the Secondary Black Hole Strengthens Gravitational-wave Signals}",
      journal = {\apj},
         year = 2025,
        month = dec,
       volume = {994},
       number = {2},
          eid = {168},
        pages = {168},
          doi = {10.3847/1538-4357/ae1133},
archivePrefix = {arXiv},
       eprint = {2510.06325},
 primaryClass = {astro-ph.GA},
       adsurl = {https://ui.adsabs.harvard.edu/abs/2025ApJ...994..168C}
}

@ARTICLE{dorazio21,
       author = {{D'Orazio}, Daniel J. and {Duffell}, Paul C.},
        title = "{Orbital Evolution of Equal-mass Eccentric Binaries due to a Gas Disk: Eccentric Inspirals and Circular Outspirals}",
      journal = {\apjl},
         year = 2021,
        month = jun,
       volume = {914},
       number = {1},
          eid = {L21},
        pages = {L21},
          doi = {10.3847/2041-8213/ac0621},
archivePrefix = {arXiv},
       eprint = {2103.09251},
 primaryClass = {astro-ph.HE},
       adsurl = {https://ui.adsabs.harvard.edu/abs/2021ApJ...914L..21D}
}

@ARTICLE{dorazio15,
   author = {{D'Orazio}, D.~J. and {Haiman}, Z. and {Schiminovich}, D.},
    title = "{Relativistic boost as the cause of periodicity in a massive black-hole binary candidate}",
  journal = {\nat},
archivePrefix = "arXiv",
   eprint = {1509.04301},
 primaryClass = "astro-ph.HE",
     year = 2015,
    month = sep,
   volume = 525,
    pages = {351-353},
      doi = {10.1038/nature15262},
   adsurl = {http://adsabs.harvard.edu/abs/2015Natur.525..351D}
}

@ARTICLE{duffell20,
       author = {{Duffell}, Paul C. and {D'Orazio}, Daniel and {Derdzinski}, Andrea and {Haiman}, Zoltan and {MacFadyen}, Andrew and {Rosen}, Anna L. and {Zrake}, Jonathan},
        title = "{Circumbinary Disks: Accretion and Torque as a Function of Mass Ratio and Disk Viscosity}",
      journal = {\apj},
         year = 2020,
        month = sep,
       volume = {901},
       number = {1},
          eid = {25},
        pages = {25},
          doi = {10.3847/1538-4357/abab95},
archivePrefix = {arXiv},
       eprint = {1911.05506},
 primaryClass = {astro-ph.SR},
       adsurl = {https://ui.adsabs.harvard.edu/abs/2020ApJ...901...25D}
}

@ARTICLE{epta23,
       author = {{EPTA Collaboration} and {InPTA Collaboration} and {Antoniadis}, J. and {Arumugam}, P. and {Arumugam}, S. and {Babak}, S. and {Bagchi}, M. and {Bak Nielsen}, A. -S. and {Bassa}, C.~G. and {Bathula}, A. and {Berthereau}, A. and {Bonetti}, M. and {Bortolas}, E. and {Brook}, P.~R. and {Burgay}, M. and {Caballero}, R.~N. and {Chalumeau}, A. and {Champion}, D.~J. and {Chanlaridis}, S. and {Chen}, S. and {Cognard}, I. and {Dandapat}, S. and {Deb}, D. and {Desai}, S. and {Desvignes}, G. and {Dhanda-Batra}, N. and {Dwivedi}, C. and {Falxa}, M. and {Ferdman}, R.~D. and {Franchini}, A. and {Gair}, J.~R. and {Goncharov}, B. and {Gopakumar}, A. and {Graikou}, E. and {Grie{\ss}meier}, J. -M. and {Guillemot}, L. and {Guo}, Y.~J. and {Gupta}, Y. and {Hisano}, S. and {Hu}, H. and {Iraci}, F. and {Izquierdo-Villalba}, D. and {Jang}, J. and {Jawor}, J. and {Janssen}, G.~H. and {Jessner}, A. and {Joshi}, B.~C. and {Kareem}, F. and {Karuppusamy}, R. and {Keane}, E.~F. and {Keith}, M.~J. and {Kharbanda}, D. and {Kikunaga}, T. and {Kolhe}, N. and {Kramer}, M. and {Krishnakumar}, M.~A. and {Lackeos}, K. and {Lee}, K.~J. and {Liu}, K. and {Liu}, Y. and {Lyne}, A.~G. and {McKee}, J.~W. and {Maan}, Y. and {Main}, R.~A. and {Mickaliger}, M.~B. and {Ni{\c{t}}u}, I.~C. and {Nobleson}, K. and {Paladi}, A.~K. and {Parthasarathy}, A. and {Perera}, B.~B.~P. and {Perrodin}, D. and {Petiteau}, A. and {Porayko}, N.~K. and {Possenti}, A. and {Prabu}, T. and {Quelquejay Leclere}, H. and {Rana}, P. and {Samajdar}, A. and {Sanidas}, S.~A. and {Sesana}, A. and {Shaifullah}, G. and {Singha}, J. and {Speri}, L. and {Spiewak}, R. and {Srivastava}, A. and {Stappers}, B.~W. and {Surnis}, M. and {Susarla}, S.~C. and {Susobhanan}, A. and {Takahashi}, K. and {Tarafdar}, P. and {Theureau}, G. and {Tiburzi}, C. and {van der Wateren}, E. and {Vecchio}, A. and {Venkatraman Krishnan}, V. and {Verbiest}, J.~P.~W. and {Wang}, J. and {Wang}, L. and {Wu}, Z.},
        title = "{The second data release from the European Pulsar Timing Array. III. Search for gravitational wave signals}",
      journal = {\aap},
         year = 2023,
        month = oct,
       volume = {678},
          eid = {A50},
        pages = {A50},
          doi = {10.1051/0004-6361/202346844},
archivePrefix = {arXiv},
       eprint = {2306.16214},
 primaryClass = {astro-ph.HE},
       adsurl = {https://ui.adsabs.harvard.edu/abs/2023A&A...678A..50E}
}

@ARTICLE{epta24astro,
       author = {{EPTA Collaboration} and {InPTA Collaboration} and {Antoniadis}, J. and {Arumugam}, P. and {Arumugam}, S. and {Babak}, S. and {Bagchi}, M. and {Bak Nielsen}, A.-S. and {Bassa}, C.~G. and {Bathula}, A. and {Berthereau}, A. and {Bonetti}, M. and {Bortolas}, E. and {Brook}, P.~R. and {Burgay}, M. and {Caballero}, R.~N. and {Chalumeau}, A. and {Champion}, D.~J. and {Chanlaridis}, S. and {Chen}, S. and {Cognard}, I. and {Dandapat}, S. and {Deb}, D. and {Desai}, S. and {Desvignes}, G. and {Dhanda-Batra}, N. and {Dwivedi}, C. and {Falxa}, M. and {Ferdman}, R.~D. and {Franchini}, A. and {Gair}, J.~R. and {Goncharov}, B. and {Gopakumar}, A. and {Graikou}, E. and {Grie{\ss}meier}, J.-M. and {Gualandris}, A. and {Guillemot}, L. and {Guo}, Y.~J. and {Gupta}, Y. and {Hisano}, S. and {Hu}, H. and {Iraci}, F. and {Izquierdo-Villalba}, D. and {Jang}, J. and {Jawor}, J. and {Janssen}, G.~H. and {Jessner}, A. and {Joshi}, B.~C. and {Kareem}, F. and {Karuppusamy}, R. and {Keane}, E.~F. and {Keith}, M.~J. and {Kharbanda}, D. and {Kikunaga}, T. and {Kolhe}, N. and {Kramer}, M. and {Krishnakumar}, M.~A. and {Lackeos}, K. and {Lee}, K.~J. and {Liu}, K. and {Liu}, Y. and {Lyne}, A.~G. and {McKee}, J.~W. and {Maan}, Y. and {Main}, R.~A. and {Mickaliger}, M.~B. and {Ni{\c{t}}u}, I.~C. and {Nobleson}, K. and {Paladi}, A.~K. and {Parthasarathy}, A. and {Perera}, B.~B.~P. and {Perrodin}, D. and {Petiteau}, A. and {Porayko}, N.~K. and {Possenti}, A. and {Prabu}, T. and {Quelquejay Leclere}, H. and {Rana}, P. and {Samajdar}, A. and {Sanidas}, S.~A. and {Sesana}, A. and {Shaifullah}, G. and {Singha}, J. and {Speri}, L. and {Spiewak}, R. and {Srivastava}, A. and {Stappers}, B.~W. and {Surnis}, M. and {Susarla}, S.~C. and {Susobhanan}, A. and {Takahashi}, K. and {Tarafdar}, P. and {Theureau}, G. and {Tiburzi}, C. and {van der Wateren}, E. and {Vecchio}, A. and {Venkatraman Krishnan}, V. and {Verbiest}, J.~P.~W. and {Wang}, J. and {Wang}, L. and {Wu}, Z. and {Auclair}, P. and {Barausse}, E. and {Caprini}, C. and {Crisostomi}, M. and {Fastidio}, F. and {Khizriev}, T. and {Middleton}, H. and {Neronov}, A. and {Postnov}, K. and {Roper Pol}, A. and {Semikoz}, D. and {Smarra}, C. and {Steer}, D.~A. and {Truant}, R.~J. and {Valtolina}, S.},
        title = "{The second data release from the European Pulsar Timing Array. IV. Implications for massive black holes, dark matter, and the early Universe}",
      journal = {\aap},
         year = 2024,
        month = may,
       volume = {685},
          eid = {A94},
        pages = {A94},
          doi = {10.1051/0004-6361/202347433},
archivePrefix = {arXiv},
       eprint = {2306.16227},
 primaryClass = {astro-ph.CO},
       adsurl = {https://ui.adsabs.harvard.edu/abs/2024A&A...685A..94E}
}

@ARTICLE{eracleous12,
   author = {{Eracleous}, M. and {Boroson}, T.~A. and {Halpern}, J.~P. and 
	{Liu}, J.},
    title = "{A Large Systematic Search for Close Supermassive Binary and Rapidly Recoiling Black Holes}",
  journal = {\apjs},
     year = 2012,
    month = aug,
   volume = 201,
      eid = {23},
    pages = {23},
      doi = {10.1088/0067-0049/201/2/23},
   adsurl = {http://adsabs.harvard.edu/abs/2012ApJS..201...23E}
}

@ARTICLE{escala05,
   author = {{Escala}, A. and {Larson}, R.~B. and {Coppi}, P.~S. and {Mardones}, D.
	},
    title = "{The Role of Gas in the Merging of Massive Black Holes in Galactic Nuclei. II. Black Hole Merging in a Nuclear Gas Disk}",
  journal = {\apj},
   eprint = {arXiv:astro-ph/0406304},
     year = 2005,
    month = sep,
   volume = 630,
    pages = {152-166},
      doi = {10.1086/431747},
   adsurl = {http://adsabs.harvard.edu/abs/2005ApJ...630..152E}
}

@ARTICLE{farris14,
   author = {{Farris}, B.~D. and {Duffell}, P. and {MacFadyen}, A.~I. and 
	{Haiman}, Z.},
    title = "{Binary Black Hole Accretion from a Circumbinary Disk: Gas Dynamics inside the Central Cavity}",
  journal = {\apj},
archivePrefix = "arXiv",
   eprint = {1310.0492},
 primaryClass = "astro-ph.HE",
     year = 2014,
    month = mar,
   volume = 783,
      eid = {134},
    pages = {134},
      doi = {10.1088/0004-637X/783/2/134},
   adsurl = {http://adsabs.harvard.edu/abs/2014ApJ...783..134F}
}

@ARTICLE{farris15,
       author = {{Farris}, B.~D. and {Duffell}, P. and {MacFadyen}, A.~I. and {Haiman}, Z.},
        title = "{Binary black hole accretion during inspiral and merger.}",
      journal = {\mnras},
         year = 2015,
        month = feb,
       volume = {447},
        pages = {L80-L84},
          doi = {10.1093/mnrasl/slu184},
archivePrefix = {arXiv},
       eprint = {1409.5124},
 primaryClass = {astro-ph.HE},
       adsurl = {https://ui.adsabs.harvard.edu/abs/2015MNRAS.447L..80F}
}

@ARTICLE{farris10,
       author = {{Farris}, Brian D. and {Liu}, Yuk Tung and {Shapiro}, Stuart L.},
        title = "{Binary black hole mergers in gaseous environments: ``Binary Bondi`` and ``binary Bondi-Hoyle-Lyttleton'' accretion}",
      journal = {\prd},
         year = 2010,
        month = apr,
       volume = {81},
       number = {8},
          eid = {084008},
        pages = {084008},
          doi = {10.1103/PhysRevD.81.084008},
archivePrefix = {arXiv},
       eprint = {0912.2096},
 primaryClass = {astro-ph.HE},
       adsurl = {https://ui.adsabs.harvard.edu/abs/2010PhRvD..81h4008F}
}

@ARTICLE{fastidio24,
       author = {{Fastidio}, F. and {Gualandris}, A. and {Sesana}, A. and {Bortolas}, E. and {Dehnen}, W.},
        title = "{Eccentricity evolution of PTA sources from cosmological initial conditions}",
      journal = {\mnras},
         year = 2024,
        month = jul,
       volume = {532},
       number = {1},
        pages = {295-304},
          doi = {10.1093/mnras/stae1411},
archivePrefix = {arXiv},
       eprint = {2406.02710},
 primaryClass = {astro-ph.GA},
       adsurl = {https://ui.adsabs.harvard.edu/abs/2024MNRAS.532..295F}
}

@ARTICLE{fastidio25,
       author = {{Fastidio}, F. and {Bortolas}, E. and {Gualandris}, A. and {Sesana}, A. and {Read}, J.~I. and {Dehnen}, W.},
        title = "{Realistic consecutive galaxy mergers form eccentric pulsar timing array sources}",
      journal = {\aap},
         year = 2025,
        month = nov,
       volume = {703},
          eid = {A86},
        pages = {A86},
          doi = {10.1051/0004-6361/202556035},
archivePrefix = {arXiv},
       eprint = {2506.16539},
 primaryClass = {astro-ph.GA},
       adsurl = {https://ui.adsabs.harvard.edu/abs/2025A&A...703A..86F}
}

@ARTICLE{francois26,
       author = {{Fran{\c{c}}ois}, Thibaut L. and {Gualandris}, Alessia and {Dehnen}, Walter},
        title = "{Scaling relations for binary black hole merger times from cosmological initial conditions}",
      journal = {\mnras},
         year = 2026,
        month = sep,
       volume = {551},
       number = {3},
          eid = {stag1298},
        pages = {stag1298},
          doi = {10.1093/mnras/stag1298},
archivePrefix = {arXiv},
       eprint = {2607.07813},
 primaryClass = {astro-ph.GA},
       adsurl = {https://ui.adsabs.harvard.edu/abs/2026MNRAS.551g1298F}
}

@ARTICLE{gardiner24,
       author = {{Gardiner}, Emiko C. and {Kelley}, Luke Zoltan and {Lemke}, Anna-Malin and {Mitridate}, Andrea},
        title = "{Beyond the Background: Gravitational-wave Anisotropy and Continuous Waves from Supermassive Black Hole Binaries}",
      journal = {\apj},
         year = 2024,
        month = apr,
       volume = {965},
       number = {2},
          eid = {164},
        pages = {164},
          doi = {10.3847/1538-4357/ad2be8},
archivePrefix = {arXiv},
       eprint = {2309.07227},
 primaryClass = {astro-ph.HE},
       adsurl = {https://ui.adsabs.harvard.edu/abs/2024ApJ...965..164G}
}

@ARTICLE{gardiner25,
       author = {{Gardiner}, Emiko C. and {B{\'e}csy}, Bence and {Kelley}, Luke Zoltan and {Cornish}, Neil J.},
        title = "{Characterizing Continuous Gravitational Waves from Supermassive Black Hole Binaries in Realistic Pulsar Timing Array Data}",
      journal = {\apj},
         year = 2025,
        month = aug,
       volume = {988},
       number = {2},
          eid = {222},
        pages = {222},
          doi = {10.3847/1538-4357/ade4c2},
archivePrefix = {arXiv},
       eprint = {2502.16016},
 primaryClass = {astro-ph.CO},
       adsurl = {https://ui.adsabs.harvard.edu/abs/2025ApJ...988..222G}
}

@ARTICLE{gardiner26,
       author = {{Gardiner}, Emiko C.},
       title = "{Constraints on Supermassive Black Hole Binaries from the Lack of Resolvable Sources in the NANOGrav 15-yr Dataset}",
      journal = {in preparation},
         year = {2026}
}

@ARTICLE{gold14,
       author = {{Gold}, Roman and {Paschalidis}, Vasileios and {Etienne}, Zachariah B. and {Shapiro}, Stuart L. and {Pfeiffer}, Harald P.},
        title = "{Accretion disks around binary black holes of unequal mass: General relativistic magnetohydrodynamic simulations near decoupling}",
      journal = {\prd},
         year = 2014,
        month = mar,
       volume = {89},
       number = {6},
          eid = {064060},
        pages = {064060},
          doi = {10.1103/PhysRevD.89.064060},
archivePrefix = {arXiv},
       eprint = {1312.0600},
 primaryClass = {astro-ph.HE},
       adsurl = {https://ui.adsabs.harvard.edu/abs/2014PhRvD..89f4060G}
}

@article{gualandris16,
    author = {Gualandris, Alessia and Read, Justin I. and Dehnen, Walter and Bortolas, Elisa},
    title = {Collisionless loss-cone refilling: there is no final parsec problem},
    journal = {Monthly Notices of the Royal Astronomical Society},
    volume = {464},
    number = {2},
    pages = {2301-2310},
    year = {2017},
    month = {01},
    issn = {0035-8711},
    doi = {10.1093/mnras/stw2528},
    url = {https://doi.org/10.1093/mnras/stw2528},
    eprint = {https://academic.oup.com/mnras/article-pdf/464/2/2301/18518741/stw2528.pdf},
}

@ARTICLE{gualandris22,
       author = {{Gualandris}, Alessia and {Khan}, Fazeel Mahmood and {Bortolas}, Elisa and {Bonetti}, Matteo and {Sesana}, Alberto and {Berczik}, Peter and {Holley-Bockelmann}, Kelly},
        title = "{Eccentricity evolution of massive black hole binaries from formation to coalescence}",
      journal = {\mnras},
         year = 2022,
        month = apr,
       volume = {511},
       number = {4},
        pages = {4753-4765},
          doi = {10.1093/mnras/stac241},
archivePrefix = {arXiv},
       eprint = {2201.08646},
 primaryClass = {astro-ph.GA},
       adsurl = {https://ui.adsabs.harvard.edu/abs/2022MNRAS.511.4753G}
}

@ARTICLE{haiman09,
       author = {{Haiman}, Zolt{\'a}n and {Kocsis}, Bence and {Menou}, Kristen},
        title = "{The Population of Viscosity- and Gravitational Wave-driven Supermassive Black Hole Binaries Among Luminous Active Galactic Nuclei}",
      journal = {\apj},
         year = 2009,
        month = aug,
       volume = {700},
       number = {2},
        pages = {1952-1969},
          doi = {10.1088/0004-637X/700/2/1952},
archivePrefix = {arXiv},
       eprint = {0904.1383},
 primaryClass = {astro-ph.CO},
       adsurl = {https://ui.adsabs.harvard.edu/abs/2009ApJ...700.1952H}
}

@ARTICLE{harris26,
       author = {{Harris}, C.~J. and {G{\"u}ltekin}, Kayhan and {Blecha}, Laura},
        title = "{Core Scouring Dynamics and Gravitational Wave Consequences: Constraints on Supermassive Black Hole Binary Hardening}",
      journal = {arXiv e-prints},
         year = 2026,
        month = jan,
          eid = {arXiv:2601.07762},
        pages = {arXiv:2601.07762},
          doi = {10.48550/arXiv.2601.07762},
archivePrefix = {arXiv},
       eprint = {2601.07762},
 primaryClass = {astro-ph.GA},
       adsurl = {https://ui.adsabs.harvard.edu/abs/2026arXiv260107762H}
}

@ARTICLE{hofloe07,
   author = {{Hoffman}, L. and {Loeb}, A.},
    title = "{Dynamics of triple black hole systems in hierarchically merging massive galaxies}",
  journal = {\mnras},
   eprint = {arXiv:astro-ph/0612517},
     year = 2007,
    month = may,
   volume = 377,
    pages = {957-976},
      doi = {10.1111/j.1365-2966.2007.11694.x},
   adsurl = {http://adsabs.harvard.edu/abs/2007MNRAS.377..957H}
}

@ARTICLE{holley25,
       author = {{Holley-Bockelmann}, Kelly and {Khan}, Fazeel Mahmood and {Williams}, Isaiah and {Roth}, Jaelyn and {Rizzo Smith}, Michael and {Porter}, Kaitlin and {Bellovary}, Jillian and {Derdzinski}, Andrea and {Macci{\`o}}, Andrea V.},
        title = "{Handy Relation between Binary Black Hole Merger Times and Host Galaxy Properties}",
      journal = {\apjl},
         year = 2025,
        month = dec,
       volume = {995},
       number = {1},
          eid = {L32},
        pages = {L32},
          doi = {10.3847/2041-8213/ae1ccd},
archivePrefix = {arXiv},
       eprint = {2508.14253},
 primaryClass = {astro-ph.GA},
       adsurl = {https://ui.adsabs.harvard.edu/abs/2025ApJ...995L..32H}
}

@ARTICLE{holley06,
       author = {{Holley-Bockelmann}, Kelly and {Sigurdsson}, Steinn},
        title = "{A Full Loss Cone For Triaxial Galaxies}",
      journal = {arXiv e-prints},
         year = 2006,
        month = jan,
          eid = {astro-ph/0601520},
        pages = {astro-ph/0601520},
archivePrefix = {arXiv},
       eprint = {astro-ph/0601520},
 primaryClass = {astro-ph},
       adsurl = {https://ui.adsabs.harvard.edu/abs/2006astro.ph..1520H}
}

@ARTICLE{holley08,
       author = {{Holley-Bockelmann}, Kelly and {G{\"u}ltekin}, Kayhan and
         {Shoemaker}, Deirdre and {Yunes}, Nicolas},
        title = "{Gravitational Wave Recoil and the Retention of Intermediate-Mass Black Holes}",
      journal = {\apj},
         year = 2008,
        month = oct,
       volume = {686},
       number = {2},
        pages = {829-837},
          doi = {10.1086/591218},
archivePrefix = {arXiv},
       eprint = {0707.1334},
 primaryClass = {astro-ph},
       adsurl = {https://ui.adsabs.harvard.edu/abs/2008ApJ...686..829H}
}

@ARTICLE{holley15,
       author = {{Holley-Bockelmann}, Kelly and {Khan}, Fazeel Mahmood},
        title = "{Galaxy Rotation and Rapid Supermassive Binary Coalescence}",
      journal = {\apj},
         year = 2015,
        month = sep,
       volume = {810},
       number = {2},
          eid = {139},
        pages = {139},
          doi = {10.1088/0004-637X/810/2/139},
archivePrefix = {arXiv},
       eprint = {1505.06203},
 primaryClass = {astro-ph.GA},
       adsurl = {https://ui.adsabs.harvard.edu/abs/2015ApJ...810..139H}
}

@ARTICLE{izquierdo22,
       author = {{Izquierdo-Villalba}, David and {Sesana}, Alberto and {Bonoli}, Silvia and {Colpi}, Monica},
        title = "{Massive black hole evolution models confronting the n-Hz amplitude of the stochastic gravitational wave background}",
      journal = {\mnras},
         year = 2022,
        month = jan,
       volume = {509},
       number = {3},
        pages = {3488-3503},
          doi = {10.1093/mnras/stab3239},
archivePrefix = {arXiv},
       eprint = {2108.11671},
 primaryClass = {astro-ph.GA},
       adsurl = {https://ui.adsabs.harvard.edu/abs/2022MNRAS.509.3488I}
}

@ARTICLE{katz20,
       author = {{Katz}, Michael L. and {Kelley}, Luke Zoltan and {Dosopoulou}, Fani and
         {Berry}, Samantha and {Blecha}, Laura and {Larson}, Shane L.},
        title = "{Probing massive black hole binary populations with LISA}",
      journal = {\mnras},
         year = 2020,
        month = jan,
       volume = {491},
       number = {2},
        pages = {2301-2317},
          doi = {10.1093/mnras/stz3102},
archivePrefix = {arXiv},
       eprint = {1908.05779},
 primaryClass = {astro-ph.HE},
       adsurl = {https://ui.adsabs.harvard.edu/abs/2020MNRAS.491.2301K}
}

@ARTICLE{kelley17a,
   author = {{Kelley}, L.~Z. and {Blecha}, L. and {Hernquist}, L.},
    title = "{Massive black hole binary mergers in dynamical galactic environments}",
  journal = {\mnras},
archivePrefix = "arXiv",
   eprint = {1606.01900},
 primaryClass = "astro-ph.HE",
     year = 2017,
    month = jan,
   volume = 464,
    pages = {3131-3157},
      doi = {10.1093/mnras/stw2452},
   adsurl = {http://adsabs.harvard.edu/abs/2017MNRAS.464.3131K}
}

@ARTICLE{kelley17b,
   author = {{Kelley}, L.~Z. and {Blecha}, L. and {Hernquist}, L. and {Sesana}, A. and 
	{Taylor}, S.~R.},
    title = "{The gravitational wave background from massive black hole binaries in Illustris: spectral features and time to detection with pulsar timing arrays}",
  journal = {\mnras},
archivePrefix = "arXiv",
   eprint = {1702.02180},
 primaryClass = "astro-ph.HE",
     year = 2017,
    month = nov,
   volume = 471,
    pages = {4508-4526},
      doi = {10.1093/mnras/stx1638},
   adsurl = {http://adsabs.harvard.edu/abs/2017MNRAS.471.4508K}
}

@ARTICLE{kelley18,
   author = {{Kelley}, L.~Z. and {Blecha}, L. and {Hernquist}, L. and {Sesana}, A. and 
	{Taylor}, S.~R.},
    title = "{Single sources in the low-frequency gravitational wave sky: properties and time to detection by pulsar timing arrays}",
  journal = {\mnras},
archivePrefix = "arXiv",
   eprint = {1711.00075},
 primaryClass = "astro-ph.HE",
     year = 2018,
    month = jun,
   volume = 477,
    pages = {964-976},
      doi = {10.1093/mnras/sty689},
   adsurl = {http://adsabs.harvard.edu/abs/2018MNRAS.477..964K}
}

@misc{kelley24,
  author       = {Luke Zoltan Kelley and
                  Emiko C Gardiner and
                  David Wright and
                  Magdalena Siwek and
                  Cayenne Matt and
                  Jeff Hazboun and
                  A. Cingoranelli and
                  J. Andrew Casey-Clyde and
                  tingtingliu-astro and
                  Kayhan Gültekin and
                  Siddharth Mohite and
                  siyuan-chen and
                  Jeremy M. Wachter and
                  Stephen Taylor},
  title        = {nanograv/holodeck: v1.5.2},
  month        = apr,
  year         = 2024,
  publisher    = {Zenodo},
  version      = {v1.5.2},
  doi          = {10.5281/zenodo.10966412},
  url          = {https://doi.org/10.5281/zenodo.10966412},
}

@ARTICLE{khan11,
       author = {{Khan}, Fazeel Mahmood and {Just}, Andreas and {Merritt}, David},
        title = "{Efficient Merger of Binary Supermassive Black Holes in Merging Galaxies}",
      journal = {\apj},
         year = 2011,
        month = may,
       volume = {732},
       number = {2},
          eid = {89},
        pages = {89},
          doi = {10.1088/0004-637X/732/2/89},
archivePrefix = {arXiv},
       eprint = {1103.0272},
 primaryClass = {astro-ph.CO},
       adsurl = {https://ui.adsabs.harvard.edu/abs/2011ApJ...732...89K}
}

@ARTICLE{khan12b,
       author = {{Khan}, Fazeel Mahmood and {Preto}, Miguel and {Berczik}, Peter and {Berentzen}, Ingo and {Just}, Andreas and {Spurzem}, Rainer},
        title = "{Mergers of Unequal-mass Galaxies: Supermassive Black Hole Binary Evolution and Structure of Merger Remnants}",
      journal = {\apj},
         year = 2012,
        month = apr,
       volume = {749},
       number = {2},
          eid = {147},
        pages = {147},
          doi = {10.1088/0004-637X/749/2/147},
archivePrefix = {arXiv},
       eprint = {1202.2124},
 primaryClass = {astro-ph.CO},
       adsurl = {https://ui.adsabs.harvard.edu/abs/2012ApJ...749..147K}
}

@ARTICLE{khan13,
       author = {{Khan}, Fazeel Mahmood and {Holley-Bockelmann}, Kelly and
         {Berczik}, Peter and {Just}, Andreas},
        title = "{Supermassive Black Hole Binary Evolution in Axisymmetric Galaxies: The Final Parsec Problem is Not a Problem}",
      journal = {\apj},
         year = 2013,
        month = aug,
       volume = {773},
       number = {2},
          eid = {100},
        pages = {100},
          doi = {10.1088/0004-637X/773/2/100},
archivePrefix = {arXiv},
       eprint = {1302.1871},
 primaryClass = {astro-ph.GA},
       adsurl = {https://ui.adsabs.harvard.edu/abs/2013ApJ...773..100K}
}

@ARTICLE{khan18a,
       author = {{Khan}, Fazeel Mahmood and {Berczik}, Peter and {Just}, Andreas},
        title = "{Gravitational wave driven mergers and coalescence time of supermassive black holes}",
      journal = {\aap},
         year = 2018,
        month = jul,
       volume = {615},
          eid = {A71},
        pages = {A71},
          doi = {10.1051/0004-6361/201730489},
archivePrefix = {arXiv},
       eprint = {1803.11394},
 primaryClass = {astro-ph.GA},
       adsurl = {https://ui.adsabs.harvard.edu/abs/2018A&A...615A..71K}
}

@ARTICLE{kocsis11,
       author = {{Kocsis}, B. and {Sesana}, A.},
        title = "{Gas-driven massive black hole binaries: signatures in the nHz gravitational wave background}",
      journal = {\mnras},
         year = 2011,
        month = mar,
       volume = {411},
       number = {3},
        pages = {1467-1479},
          doi = {10.1111/j.1365-2966.2010.17782.x},
archivePrefix = {arXiv},
       eprint = {1002.0584},
 primaryClass = {astro-ph.CO},
       adsurl = {https://ui.adsabs.harvard.edu/abs/2011MNRAS.411.1467K}
}

@ARTICLE{kormendy13,
       author = {{Kormendy}, John and {Ho}, Luis C.},
        title = "{Coevolution (Or Not) of Supermassive Black Holes and Host Galaxies}",
      journal = {\araa},
         year = 2013,
        month = aug,
       volume = {51},
       number = {1},
        pages = {511-653},
          doi = {10.1146/annurev-astro-082708-101811},
archivePrefix = {arXiv},
       eprint = {1304.7762},
 primaryClass = {astro-ph.CO},
       adsurl = {https://ui.adsabs.harvard.edu/abs/2013ARA&A..51..511K}
}

@article{laal2026,
   title={Multimessenger Probes of the Supermassive Black Hole Binary Population: The Role of Pulsar Timing Arrays},
   volume={1004},
   ISSN={1538-4357},
   url={http://dx.doi.org/10.3847/1538-4357/ae6cea},
   DOI={10.3847/1538-4357/ae6cea},
   number={1},
   journal={The Astrophysical Journal},
   publisher={American Astronomical Society},
   author={Laal, Nima and Taylor, Stephen R. and Matt, Cayenne and Gültekin, Kayhan},
   year={2026},
   month=jun, pages={90} }

@ARTICLE{laal25,
       author = {{Laal}, Nima and {Taylor}, Stephen R. and {Kelley}, Luke Zoltan and {Simon}, Joseph and {G{\"u}ltekin}, Kayhan and {Wright}, David and {B{\'e}csy}, Bence and {Casey-Clyde}, J. Andrew and {Chen}, Siyuan and {Cingoranelli}, Alexander and {D'Orazio}, Daniel J. and {Gardiner}, Emiko C. and {Lamb}, William G. and {Matt}, Cayenne and {Siwek}, Magdalena S. and {Wachter}, Jeremy M.},
        title = "{Deep Neural Emulation of the Supermassive Black Hole Binary Population}",
      journal = {\apj},
         year = 2025,
        month = mar,
       volume = {982},
       number = {1},
          eid = {55},
        pages = {55},
          doi = {10.3847/1538-4357/adb4ef},
archivePrefix = {arXiv},
       eprint = {2411.10519},
 primaryClass = {astro-ph.IM},
       adsurl = {https://ui.adsabs.harvard.edu/abs/2025ApJ...982...55L}
}

@ARTICLE{lai23,
       author = {{Lai}, Dong and {Mu{\~n}oz}, Diego J.},
        title = "{Circumbinary Accretion: From Binary Stars to Massive Binary Black Holes}",
      journal = {\araa},
         year = 2023,
        month = aug,
       volume = {61},
        pages = {517-560},
          doi = {10.1146/annurev-astro-052622-022933},
archivePrefix = {arXiv},
       eprint = {2211.00028},
 primaryClass = {astro-ph.HE},
       adsurl = {https://ui.adsabs.harvard.edu/abs/2023ARA&A..61..517L}
}

@ARTICLE{leja20,
       author = {{Leja}, Joel and {Speagle}, Joshua S. and {Johnson}, Benjamin D. and {Conroy}, Charlie and {van Dokkum}, Pieter and {Franx}, Marijn},
        title = "{A New Census of the 0.2 < z < 3.0 Universe. I. The Stellar Mass Function}",
      journal = {\apj},
         year = 2020,
        month = apr,
       volume = {893},
       number = {2},
          eid = {111},
        pages = {111},
          doi = {10.3847/1538-4357/ab7e27},
archivePrefix = {arXiv},
       eprint = {1910.04168},
 primaryClass = {astro-ph.GA},
       adsurl = {https://ui.adsabs.harvard.edu/abs/2020ApJ...893..111L}
}

@ARTICLE{li23,
       author = {{Li}, Kunyang and {Bogdanovi{\'c}}, Tamara and {Ballantyne}, David R. and {Bonetti}, Matteo},
        title = "{Massive Black Hole Binaries from the TNG50-3 Simulation. II. Using Dual AGNs to Predict the Rate of Black Hole Mergers}",
      journal = {\apj},
         year = 2023,
        month = dec,
       volume = {959},
       number = {1},
          eid = {3},
        pages = {3},
          doi = {10.3847/1538-4357/ad04d2},
archivePrefix = {arXiv},
       eprint = {2207.14231},
 primaryClass = {astro-ph.GA},
       adsurl = {https://ui.adsabs.harvard.edu/abs/2023ApJ...959....3L}
}

@ARTICLE{li2026_ramcoal,
       author = {{Li}, Kunyang and {Beckmann}, Ricarda S. and {Dubois}, Yohan and {Volonteri}, Marta},
        title = "{Set them free: extending RAMCOAL to model massive black hole triplets in hydrodynamical simulations of galaxies}",
      journal = {arXiv e-prints},
         year = 2026,
        month = jul,
          eid = {arXiv:2607.04121},
        pages = {arXiv:2607.04121},
          doi = {10.48550/arXiv.2607.04121},
archivePrefix = {arXiv},
       eprint = {2607.04121},
 primaryClass = {astro-ph.GA},
       adsurl = {https://ui.adsabs.harvard.edu/abs/2026arXiv260704121L}
}

@ARTICLE{li2025_ramcoal,
       author = {{Li}, Kunyang and {Volonteri}, Marta and {Dubois}, Yohan and {Beckmann}, Ricarda and {Trebitsch}, Maxime},
        title = "{RAMCOAL: Tracking on-the-fly massive black hole binary evolution and coalescence in galaxy simulations}",
      journal = {\aap},
         year = 2025,
        month = sep,
       volume = {701},
          eid = {A232},
        pages = {A232},
          doi = {10.1051/0004-6361/202452562},
archivePrefix = {arXiv},
       eprint = {2410.07856},
 primaryClass = {astro-ph.GA},
       adsurl = {https://ui.adsabs.harvard.edu/abs/2025A&A...701A.232L}
}

@ARTICLE{liepold24,
       author = {{Liepold}, Emily R. and {Ma}, Chung-Pei},
        title = "{Big Galaxies and Big Black Holes: The Massive Ends of the Local Stellar and Black Hole Mass Functions and the Implications for Nanohertz Gravitational Waves}",
      journal = {\apjl},
         year = 2024,
        month = aug,
       volume = {971},
       number = {2},
          eid = {L29},
        pages = {L29},
          doi = {10.3847/2041-8213/ad66b8},
archivePrefix = {arXiv},
       eprint = {2407.14595},
 primaryClass = {astro-ph.GA},
       adsurl = {https://ui.adsabs.harvard.edu/abs/2024ApJ...971L..29L}
}

@ARTICLE{macfadyen08,
       author = {{MacFadyen}, Andrew I. and {Milosavljevi{\'c}}, Milo{\v{s}}},
        title = "{An Eccentric Circumbinary Accretion Disk and the Detection of Binary Massive Black Holes}",
      journal = {\apj},
         year = 2008,
        month = jan,
       volume = {672},
       number = {1},
        pages = {83-93},
          doi = {10.1086/523869},
archivePrefix = {arXiv},
       eprint = {astro-ph/0607467},
 primaryClass = {astro-ph},
       adsurl = {https://ui.adsabs.harvard.edu/abs/2008ApJ...672...83M}
}

@ARTICLE{mannerkoski23,
       author = {{Mannerkoski}, Matias and {Rawlings}, Alexander and {Johansson}, Peter H. and {Naab}, Thorsten and {Rantala}, Antti and {Springel}, Volker and {Irodotou}, Dimitrios and {Liao}, Shihong},
        title = "{KETJU - resolving small-scale supermassive black hole dynamics in GADGET-4}",
      journal = {\mnras},
         year = 2023,
        month = sep,
       volume = {524},
       number = {3},
        pages = {4062-4082},
          doi = {10.1093/mnras/stad2139},
archivePrefix = {arXiv},
       eprint = {2306.04963},
 primaryClass = {astro-ph.IM},
       adsurl = {https://ui.adsabs.harvard.edu/abs/2023MNRAS.524.4062M}
}

@ARTICLE{matt26a,
       author = {{Matt}, Cayenne and {G{\"u}ltekin}, Kayhan and {Kelley}, Luke Zoltan and {Blecha}, Laura and {Simon}, Joseph and {Agazie}, Gabriella and {Anumarlapudi}, Akash and {Archibald}, Anne M. and {Arzoumanian}, Zaven and {Baier}, Jeremy G. and {Baker}, Paul T. and {B{\'e}csy}, Bence and {Brazier}, Adam and {Brook}, Paul R. and {Burke-Spolaor}, Sarah and {Burnette}, Rand and {Case}, Robin and {Casey-Clyde}, J. Andrew and {Charisi}, Maria and {Chatterjee}, Shami and {Cohen}, Tyler and {Cordes}, James M. and {Cornish}, Neil J. and {Crawford}, Fronefield and {Cromartie}, H. Thankful and {Crowter}, Kathryn and {DeCesar}, Megan E. and {Demorest}, Paul B. and {Deng}, Heling and {Dey}, Lankeswar and {Dolch}, Timothy and {Ferrara}, Elizabeth C. and {Fiore}, William and {Fonseca}, Emmanuel and {Freedman}, Gabriel E. and {Gardiner}, Emiko C. and {Garver-Daniels}, Nate and {Gentile}, Peter A. and {Gersbach}, Kyle A. and {Glaser}, Joseph and {Good}, Deborah C. and {Harris}, C.~J. and {Hazboun}, Jeffrey S. and {Jennings}, Ross J. and {Johnson}, Aaron D. and {Jones}, Megan L. and {Kaplan}, David L. and {Kerr}, Matthew and {Key}, Joey S. and {Laal}, Nima and {Lam}, Michael T. and {Lamb}, William G. and {Larsen}, Bjorn and {Lazio}, T. Joseph W. and {Lewandowska}, Natalia and {Liu}, Tingting and {Lorimer}, Duncan R. and {Luo}, Jing and {Lynch}, Ryan S. and {Ma}, Chung-Pei and {Madison}, Dustin R. and {McEwen}, Alexander and {McKee}, James W. and {McLaughlin}, Maura A. and {McMann}, Natasha and {Meyers}, Bradley W. and {Meyers}, Patrick M. and {Mingarelli}, Chiara M.~F. and {Mitridate}, Andrea and {Ng}, Cherry and {Nice}, David J. and {Ocker}, Stella Koch and {Olum}, Ken D. and {Pennucci}, Timothy T. and {Perera}, Benetge B.~P. and {Petrov}, Polina and {Pol}, Nihan S. and {Radovan}, Henri A. and {Ransom}, Scott M. and {Ray}, Paul S. and {Romano}, Joseph D. and {Runnoe}, Jessie C. and {Saffer}, Alexander and {Sardesai}, Shashwat C. and {Schmiedekamp}, Ann and {Schmiedekamp}, Carl and {Schmitz}, Kai and {Shapiro-Albert}, Brent J. and {Siemens}, Xavier and {Fiscella}, Sophia V. Sosa and {Stairs}, Ingrid H. and {Stinebring}, Daniel R. and {Stovall}, Kevin and {Susobhanan}, Abhimanyu and {Swiggum}, Joseph K. and {Taylor}, Jacob and {Taylor}, Stephen R. and {Thompson}, Mercedes S. and {Turner}, Jacob E. and {Vallisneri}, Michele and {van Haasteren}, Rutger and {Vigeland}, Sarah J. and {Wahl}, Haley M. and {Wilson}, Kevin P. and {Witt}, Caitlin A. and {Wright}, David and {Young}, Olivia},
        title = "{Inferring M$_{BH}$─M$_{bulge}$ Evolution from the Gravitational-wave Background}",
      journal = {\apj},
         year = 2026,
        month = feb,
       volume = {997},
       number = {2},
          eid = {188},
        pages = {188},
          doi = {10.3847/1538-4357/ae2480},
archivePrefix = {arXiv},
       eprint = {2508.18126},
 primaryClass = {astro-ph.HE},
       adsurl = {https://ui.adsabs.harvard.edu/abs/2026ApJ...997..188M}
}

@ARTICLE{matt26b,
       author = {{Matt}, Cayenne and {G{\"u}ltekin}, Kayhan and {Agazie}, Gabriella and {Agarwal}, Nikita and {Anumarlapudi}, Akash and {Archibald}, Anne M. and {Arzoumanian}, Zaven and {Baier}, Jeremy G. and {Baker}, Paul T. and {B{\'e}csy}, Bence and {Blecha}, Laura and {Brazier}, Adam and {Brook}, Paul R. and {Burke-Spolaor}, Sarah and {Burnette}, Rand and {Case}, Robin and {Casey-Clyde}, J. Andrew and {Charisi}, Maria and {Chatterjee}, Shami and {Cohen}, Tyler and {Cordes}, James M. and {Cornish}, Neil J. and {Crawford}, Fronefield and {Cromartie}, H. Thankful and {Crowter}, Kathryn and {DeCesar}, Megan E. and {Demorest}, Paul B. and {Deng}, Heling and {Dey}, Lankeswar and {Dolch}, Timothy and {Doskoch}, Graham M. and {Ferrara}, Elizabeth C. and {Fiore}, William and {Fonseca}, Emmanuel and {Freedman}, Gabriel E. and {Gardiner}, Emiko C. and {Garver-Daniels}, Nate and {Gentile}, Peter A. and {Gersbach}, Kyle A. and {Glaser}, Joseph and {Good}, Deborah C. and {Harris}, C.~J. and {Hazboun}, Jeffrey S. and {Jennings}, Ross J. and {Johnson}, Aaron D. and {Jones}, Megan L. and {Kaplan}, David L. and {Kavumkandathil Sreekumar}, Anala and {Kelley}, Luke Zoltan and {Kerr}, Matthew and {Key}, Joey S. and {Laal}, Nima and {Lam}, Michael T. and {Lamb}, William G. and {Larsen}, Bjorn and {Lazio}, T. Joseph W. and {Lewandowska}, Natalia and {Liu}, Tingting and {Lorimer}, Duncan R. and {Luo}, Jing and {Lynch}, Ryan S. and {Ma}, Chung-Pei and {Madison}, Dustin R. and {Martsen}, Ashley and {McEwen}, Alexander and {McKee}, James W. and {McLaughlin}, Maura A. and {McMann}, Natasha and {Meyers}, Bradley W. and {Meyers}, Patrick M. and {Mingarelli}, Chiara M.~F. and {Mitridate}, Andrea and {Ng}, Cherry and {Nice}, David J. and {Nichols}, Shania and {Ocker}, Stella Koch and {Olum}, Ken D. and {Pennucci}, Timothy T. and {Perera}, Benetge B.~P. and {Petrov}, Polina and {Pol}, Nihan S. and {Radovan}, Henri A. and {Ransom}, Scott M. and {Ray}, Paul S. and {Romano}, Joseph D. and {Runnoe}, Jessie C. and {Saffer}, Alexander and {Sardesai}, Shashwat C. and {Schmiedekamp}, Ann and {Schmiedekamp}, Carl and {Schmitz}, Kai and {Shapiro-Albert}, Brent J. and {Siemens}, Xavier and {Simon}, Joseph and {Sosa Fiscella}, Sophia V. and {Stairs}, Ingrid H. and {Stinebring}, Daniel R. and {Stovall}, Kevin and {Susobhanan}, Abhimanyu and {Swiggum}, Joseph K. and {Taylor}, Jacob and {Taylor}, Stephen R. and {Thompson}, Mercedes S. and {Turner}, Jacob E. and {Vallisneri}, Michele and {van Haasteren}, Rutger and {Vigeland}, Sarah J. and {Wahl}, Haley M. and {Wilson}, Kevin P. and {Witt}, Caitlin A. and {Wright}, David and {Young}, Olivia},
        title = "{Gravitational Wave Measurement of the $M_\mathrm{BH}$-$M_\mathrm{bulge}$ Intrinsic Scatter at High Redshift}",
      journal = {arXiv e-prints},
         year = 2026,
        month = mar,
          eid = {arXiv:2603.11167},
        pages = {arXiv:2603.11167},
          doi = {10.48550/arXiv.2603.11167},
archivePrefix = {arXiv},
       eprint = {2603.11167},
 primaryClass = {astro-ph.HE},
       adsurl = {https://ui.adsabs.harvard.edu/abs/2026arXiv260311167M}
}

@ARTICLE{mayer07,
       author = {{Mayer}, L. and {Kazantzidis}, S. and {Madau}, P. and {Colpi}, M. and
         {Quinn}, T. and {Wadsley}, J.},
        title = "{Rapid Formation of Supermassive Black Hole Binaries in Galaxy Mergers with Gas}",
      journal = {Science},
         year = 2007,
        month = jun,
       volume = {316},
       number = {5833},
        pages = {1874},
          doi = {10.1126/science.1141858},
archivePrefix = {arXiv},
       eprint = {0706.1562},
 primaryClass = {astro-ph},
       adsurl = {https://ui.adsabs.harvard.edu/abs/2007Sci...316.1874M}
}

@ARTICLE{merritt05,
       author = {{Merritt}, David and {Milosavljevi{\'c}}, Milos},
        title = "{Massive Black Hole Binary Evolution}",
      journal = {Living Reviews in Relativity},
         year = 2005,
        month = nov,
       volume = {8},
        pages = {8},
          doi = {10.12942/lrr-2005-8},
archivePrefix = {arXiv},
       eprint = {astro-ph/0410364},
 primaryClass = {astro-ph},
       adsurl = {https://ui.adsabs.harvard.edu/abs/2005LRR.....8....8M}
}

@ARTICLE{milphi05,
   author = {{Milosavljevi{\'c}}, M. and {Phinney}, E.~S.},
    title = "{The Afterglow of Massive Black Hole Coalescence}",
  journal = {\apjl},
   eprint = {arXiv:astro-ph/0410343},
     year = 2005,
    month = apr,
   volume = 622,
    pages = {L93-L96},
      doi = {10.1086/429618},
   adsurl = {http://adsabs.harvard.edu/abs/2005ApJ...622L..93M}
}

@ARTICLE{moody19,
       author = {{Moody}, Mackenzie S.~L. and {Shi}, Ji-Ming and {Stone}, James M.},
        title = "{Hydrodynamic Torques in Circumbinary Accretion Disks}",
      journal = {\apj},
         year = 2019,
        month = apr,
       volume = {875},
       number = {1},
          eid = {66},
        pages = {66},
          doi = {10.3847/1538-4357/ab09ee},
archivePrefix = {arXiv},
       eprint = {1903.00008},
 primaryClass = {astro-ph.HE},
       adsurl = {https://ui.adsabs.harvard.edu/abs/2019ApJ...875...66M}
}

@ARTICLE{munoz19,
       author = {{Mu{\~n}oz}, Diego J. and {Miranda}, Ryan and {Lai}, Dong},
        title = "{Hydrodynamics of Circumbinary Accretion: Angular Momentum Transfer and Binary Orbital Evolution}",
      journal = {\apj},
         year = 2019,
        month = jan,
       volume = {871},
       number = {1},
          eid = {84},
        pages = {84},
          doi = {10.3847/1538-4357/aaf867},
archivePrefix = {arXiv},
       eprint = {1810.04676},
 primaryClass = {astro-ph.HE},
       adsurl = {https://ui.adsabs.harvard.edu/abs/2019ApJ...871...84M}
}

@ARTICLE{munoz20,
       author = {{Mu{\~n}oz}, Diego J. and {Lai}, Dong and {Kratter}, Kaitlin and {Mirand
        a}, Ryan},
        title = "{Circumbinary Accretion from Finite and Infinite Disks}",
      journal = {\apj},
         year = 2020,
        month = feb,
       volume = {889},
       number = {2},
          eid = {114},
        pages = {114},
          doi = {10.3847/1538-4357/ab5d33},
archivePrefix = {arXiv},
       eprint = {1910.04763},
 primaryClass = {astro-ph.HE},
       adsurl = {https://ui.adsabs.harvard.edu/abs/2020ApJ...889..114M}
}

@ARTICLE{nelson15b,
   author = {{Nelson}, D. and {Pillepich}, A. and {Genel}, S. and {Vogelsberger}, M. and 
	{Springel}, V. and {Torrey}, P. and {Rodriguez-Gomez}, V. and 
	{Sijacki}, D. and {Snyder}, G.~F. and {Griffen}, B. and {Marinacci}, F. and 
	{Blecha}, L. and {Sales}, L. and {Xu}, D. and {Hernquist}, L.
	},
    title = "{The illustris simulation: Public data release}",
  journal = {Astronomy and Computing},
archivePrefix = "arXiv",
   eprint = {1504.00362},
     year = 2015,
    month = nov,
   volume = 13,
    pages = {12-37},
      doi = {10.1016/j.ascom.2015.09.003},
   adsurl = {http://adsabs.harvard.edu/abs/2015A%26C....13...12N}
}

@ARTICLE{noble12,
       author = {{Noble}, Scott C. and {Mundim}, Bruno C. and {Nakano}, Hiroyuki and {Krolik}, Julian H. and {Campanelli}, Manuela and {Zlochower}, Yosef and {Yunes}, Nicol{\'a}s},
        title = "{Circumbinary Magnetohydrodynamic Accretion into Inspiraling Binary Black Holes}",
      journal = {\apj},
         year = 2012,
        month = aug,
       volume = {755},
       number = {1},
          eid = {51},
        pages = {51},
          doi = {10.1088/0004-637X/755/1/51},
archivePrefix = {arXiv},
       eprint = {1204.1073},
 primaryClass = {astro-ph.HE},
       adsurl = {https://ui.adsabs.harvard.edu/abs/2012ApJ...755...51N}
}

@ARTICLE{paschalidis21,
       author = {{Paschalidis}, Vasileios and {Bright}, Jane and {Ruiz}, Milton and {Gold}, Roman},
        title = "{Minidisk Dynamics in Accreting, Spinning Black Hole Binaries: Simulations in Full General Relativity}",
      journal = {\apjl},
         year = 2021,
        month = apr,
       volume = {910},
       number = {2},
          eid = {L26},
        pages = {L26},
          doi = {10.3847/2041-8213/abee21},
archivePrefix = {arXiv},
       eprint = {2102.06712},
 primaryClass = {astro-ph.HE},
       adsurl = {https://ui.adsabs.harvard.edu/abs/2021ApJ...910L..26P}
}

@ARTICLE{peters64,
       author = {{Peters}, P.~C.},
        title = "{Gravitational Radiation and the Motion of Two Point Masses}",
      journal = {Physical Review},
         year = 1964,
        month = nov,
       volume = {136},
       number = {4B},
        pages = {1224-1232},
          doi = {10.1103/PhysRev.136.B1224},
       adsurl = {https://ui.adsabs.harvard.edu/abs/1964PhRv..136.1224P}
}

@ARTICLE{petrov24,
       author = {{Petrov}, Polina and {Taylor}, Stephen R. and {Charisi}, Maria and {Ma}, Chung-Pei},
        title = "{Identifying Host Galaxies of Supermassive Black Hole Binaries Found by PTAs}",
      journal = {arXiv e-prints},
         year = 2024,
        month = jun,
          eid = {arXiv:2406.04409},
        pages = {arXiv:2406.04409},
          doi = {10.48550/arXiv.2406.04409},
archivePrefix = {arXiv},
       eprint = {2406.04409},
 primaryClass = {astro-ph.GA},
       adsurl = {https://ui.adsabs.harvard.edu/abs/2024arXiv240604409P}
}

@ARTICLE{phinney01,
       author = {{Phinney}, E.~S.},
        title = "{A Practical Theorem on Gravitational Wave Backgrounds}",
      journal = {arXiv e-prints},
         year = 2001,
        month = aug,
          eid = {astro-ph/0108028},
        pages = {astro-ph/0108028},
          doi = {10.48550/arXiv.astro-ph/0108028},
archivePrefix = {arXiv},
       eprint = {astro-ph/0108028},
 primaryClass = {astro-ph},
       adsurl = {https://ui.adsabs.harvard.edu/abs/2001astro.ph..8028P}
}

@ARTICLE{quinlan96,
       author = {{Quinlan}, Gerald D.},
        title = "{The dynamical evolution of massive black hole binaries I. Hardening in a fixed stellar background}",
      journal = {\na},
         year = 1996,
        month = jul,
       volume = {1},
       number = {1},
        pages = {35-56},
          doi = {10.1016/S1384-1076(96)00003-6},
archivePrefix = {arXiv},
       eprint = {astro-ph/9601092},
 primaryClass = {astro-ph},
       adsurl = {https://ui.adsabs.harvard.edu/abs/1996NewA....1...35Q}
}

@ARTICLE{quinlan97,
       author = {{Quinlan}, Gerald D. and {Hernquist}, Lars},
        title = "{The dynamical evolution of massive black hole binaries {\textemdash} II. Self-consistent N-body integrations}",
      journal = {\na},
         year = 1997,
        month = dec,
       volume = {2},
       number = {6},
        pages = {533-554},
          doi = {10.1016/S1384-1076(97)00039-0},
archivePrefix = {arXiv},
       eprint = {astro-ph/9706298},
 primaryClass = {astro-ph},
       adsurl = {https://ui.adsabs.harvard.edu/abs/1997NewA....2..533Q}
}

@ARTICLE{rantala17,
       author = {{Rantala}, Antti and {Pihajoki}, Pauli and {Johansson}, Peter H. and {Naab}, Thorsten and {Lah{\'e}n}, Natalia and {Sawala}, Till},
        title = "{Post-Newtonian Dynamical Modeling of Supermassive Black Holes in Galactic-scale Simulations}",
      journal = {\apj},
         year = 2017,
        month = may,
       volume = {840},
       number = {1},
          eid = {53},
        pages = {53},
          doi = {10.3847/1538-4357/aa6d65},
archivePrefix = {arXiv},
       eprint = {1611.07028},
 primaryClass = {astro-ph.GA},
       adsurl = {https://ui.adsabs.harvard.edu/abs/2017ApJ...840...53R}
}

@ARTICLE{ravi12,
       author = {{Ravi}, V. and {Wyithe}, J.~S.~B. and {Hobbs}, G. and {Shannon}, R.~M. and {Manchester}, R.~N. and {Yardley}, D.~R.~B. and {Keith}, M.~J.},
        title = "{Does a ``Stochastic'' Background of Gravitational Waves Exist in the Pulsar Timing Band?}",
      journal = {\apj},
         year = 2012,
        month = dec,
       volume = {761},
       number = {2},
          eid = {84},
        pages = {84},
          doi = {10.1088/0004-637X/761/2/84},
archivePrefix = {arXiv},
       eprint = {1210.3854},
 primaryClass = {astro-ph.CO},
       adsurl = {https://ui.adsabs.harvard.edu/abs/2012ApJ...761...84R}
}

@ARTICLE{reardon23,
       author = {{Reardon}, Daniel J. and {Zic}, Andrew and {Shannon}, Ryan M. and {Hobbs}, George B. and {Bailes}, Matthew and {Di Marco}, Valentina and {Kapur}, Agastya and {Rogers}, Axl F. and {Thrane}, Eric and {Askew}, Jacob and {Bhat}, N.~D. Ramesh and {Cameron}, Andrew and {Cury{\l}o}, Ma{\l}gorzata and {Coles}, William A. and {Dai}, Shi and {Goncharov}, Boris and {Kerr}, Matthew and {Kulkarni}, Atharva and {Levin}, Yuri and {Lower}, Marcus E. and {Manchester}, Richard N. and {Mandow}, Rami and {Miles}, Matthew T. and {Nathan}, Rowina S. and {Os{\l}owski}, Stefan and {Russell}, Christopher J. and {Spiewak}, Ren{\'e}e and {Zhang}, Songbo and {Zhu}, Xing-Jiang},
        title = "{Search for an Isotropic Gravitational-wave Background with the Parkes Pulsar Timing Array}",
      journal = {\apjl},
         year = 2023,
        month = jul,
       volume = {951},
       number = {1},
          eid = {L6},
        pages = {L6},
          doi = {10.3847/2041-8213/acdd02},
archivePrefix = {arXiv},
       eprint = {2306.16215},
 primaryClass = {astro-ph.HE},
       adsurl = {https://ui.adsabs.harvard.edu/abs/2023ApJ...951L...6R}
}

@ARTICLE{ricarte18,
       author = {{Ricarte}, Angelo and {Natarajan}, Priyamvada},
        title = "{The observational signatures of supermassive black hole seeds}",
      journal = {\mnras},
         year = 2018,
        month = dec,
       volume = {481},
       number = {3},
        pages = {3278-3292},
          doi = {10.1093/mnras/sty2448},
archivePrefix = {arXiv},
       eprint = {1809.01177},
 primaryClass = {astro-ph.GA},
       adsurl = {https://ui.adsabs.harvard.edu/abs/2018MNRAS.481.3278R}
}

@ARTICLE{rodriguez15,
   author = {{Rodriguez-Gomez}, V. and {Genel}, S. and {Vogelsberger}, M. and 
	{Sijacki}, D. and {Pillepich}, A. and {Sales}, L.~V. and {Torrey}, P. and 
	{Snyder}, G. and {Nelson}, D. and {Springel}, V. and {Ma}, C.-P. and 
	{Hernquist}, L.},
    title = "{The merger rate of galaxies in the Illustris simulation: a comparison with observations and semi-empirical models}",
  journal = {\mnras},
archivePrefix = "arXiv",
   eprint = {1502.01339},
     year = 2015,
    month = may,
   volume = 449,
    pages = {49-64},
      doi = {10.1093/mnras/stv264},
   adsurl = {http://adsabs.harvard.edu/abs/2015MNRAS.449...49R}
}

@ARTICLE{rodriguez06,
   author = {{Rodriguez}, C. and {Taylor}, G.~B. and {Zavala}, R.~T. and 
	{Peck}, A.~B. and {Pollack}, L.~K. and {Romani}, R.~W.},
    title = "{A Compact Supermassive Binary Black Hole System}",
  journal = {\apj},
   eprint = {arXiv:astro-ph/0604042},
     year = 2006,
    month = jul,
   volume = 646,
    pages = {49-60},
      doi = {10.1086/504825},
   adsurl = {http://adsabs.harvard.edu/abs/2006ApJ...646...49R}
}

@ARTICLE{roedig11,
       author = {{Roedig}, C. and {Dotti}, M. and {Sesana}, A. and {Cuadra}, J. and {Colpi}, M.},
        title = "{Limiting eccentricity of subparsec massive black hole binaries surrounded by self-gravitating gas discs}",
      journal = {\mnras},
         year = 2011,
        month = aug,
       volume = {415},
       number = {4},
        pages = {3033-3041},
          doi = {10.1111/j.1365-2966.2011.18927.x},
archivePrefix = {arXiv},
       eprint = {1104.3868},
 primaryClass = {astro-ph.CO},
       adsurl = {https://ui.adsabs.harvard.edu/abs/2011MNRAS.415.3033R}
}

@ARTICLE{roebber16,
       author = {{Roebber}, Elinore and {Holder}, Gilbert and {Holz}, Daniel E. and {Warren}, Michael},
        title = "{Cosmic Variance in the Nanohertz Gravitational Wave Background}",
      journal = {\apj},
         year = 2016,
        month = mar,
       volume = {819},
       number = {2},
          eid = {163},
        pages = {163},
          doi = {10.3847/0004-637X/819/2/163},
archivePrefix = {arXiv},
       eprint = {1508.07336},
 primaryClass = {astro-ph.CO},
       adsurl = {https://ui.adsabs.harvard.edu/abs/2016ApJ...819..163R}
}

@ARTICLE{sah24,
       author = {{Sah}, Mohit Raj and {Mukherjee}, Suvodip and {Saeedzadeh}, Vida and {Babul}, Arif and {Tremmel}, Michael and {Quinn}, Thomas R.},
        title = "{Imprints of supermassive black hole evolution on the spectral and spatial anisotropy of nano-hertz stochastic gravitational-wave background}",
      journal = {\mnras},
         year = 2024,
        month = sep,
       volume = {533},
       number = {2},
        pages = {1568-1582},
          doi = {10.1093/mnras/stae1930},
archivePrefix = {arXiv},
       eprint = {2404.14508},
 primaryClass = {astro-ph.CO},
       adsurl = {https://ui.adsabs.harvard.edu/abs/2024MNRAS.533.1568S}
}

@ARTICLE{sampson15,
       author = {{Sampson}, Laura and {Cornish}, Neil J. and {McWilliams}, Sean T.},
        title = "{Constraining the solution to the last parsec problem with pulsar timing}",
      journal = {\prd},
         year = 2015,
        month = apr,
       volume = {91},
       number = {8},
          eid = {084055},
        pages = {084055},
          doi = {10.1103/PhysRevD.91.084055},
archivePrefix = {arXiv},
       eprint = {1503.02662},
 primaryClass = {gr-qc},
       adsurl = {https://ui.adsabs.harvard.edu/abs/2015PhRvD..91h4055S}
}

@ARTICLE{satheesh25,
       author = {{Satheesh}, Pranav and {Blecha}, Laura and {Kelley}, Luke Zoltan},
        title = "{Mergers and Recoil in Triple Massive Black Hole Systems from Illustris}",
      journal = {\apj},
         year = 2025,
        month = nov,
       volume = {993},
       number = {2},
          eid = {222},
        pages = {222},
          doi = {10.3847/1538-4357/ae0a48},
archivePrefix = {arXiv},
       eprint = {2506.04369},
 primaryClass = {astro-ph.GA},
       adsurl = {https://ui.adsabs.harvard.edu/abs/2025ApJ...993..222S}
}

@ARTICLE{sato25b,
       author = {{Sato-Polito}, Gabriela and {Zaldarriaga}, Matias},
        title = "{Uncertainties in the supermassive black hole abundance and implications for the GW background}",
      journal = {arXiv e-prints},
         year = 2025,
        month = sep,
          eid = {arXiv:2509.08041},
        pages = {arXiv:2509.08041},
          doi = {10.48550/arXiv.2509.08041},
archivePrefix = {arXiv},
       eprint = {2509.08041},
 primaryClass = {astro-ph.GA},
       adsurl = {https://ui.adsabs.harvard.edu/abs/2025arXiv250908041S}
}

@ARTICLE{sato25a,
       author = {{Sato-Polito}, Gabriela and {Zaldarriaga}, Matias and {Quataert}, Eliot},
        title = "{Evolution of supermassive black holes in light of PTA measurements: Implications for growth by mergers and accretion}",
      journal = {\prd},
         year = 2025,
        month = dec,
       volume = {112},
       number = {12},
          eid = {123018},
        pages = {123018},
          doi = {10.1103/1br7-s1rc},
archivePrefix = {arXiv},
       eprint = {2501.09786},
 primaryClass = {astro-ph.CO},
       adsurl = {https://ui.adsabs.harvard.edu/abs/2025PhRvD.112l3018S}
}

@ARTICLE{sato23,
       author = {{Sato-Polito}, Gabriela and {Zaldarriaga}, Matias and {Quataert}, Eliot},
        title = "{Where are NANOGrav's big black holes?}",
      journal = {arXiv e-prints},
         year = 2023,
        month = dec,
          eid = {arXiv:2312.06756},
        pages = {arXiv:2312.06756},
          doi = {10.48550/arXiv.2312.06756},
archivePrefix = {arXiv},
       eprint = {2312.06756},
 primaryClass = {astro-ph.CO},
       adsurl = {https://ui.adsabs.harvard.edu/abs/2023arXiv231206756S}
}

@ARTICLE{sesana18,
   author = {{Sesana}, A. and {Haiman}, Z. and {Kocsis}, B. and {Kelley}, L.~Z.
	},
    title = "{Testing the Binary Hypothesis: Pulsar Timing Constraints on Supermassive Black Hole Binary Candidates}",
  journal = {\apj},
archivePrefix = "arXiv",
   eprint = {1703.10611},
 primaryClass = "astro-ph.HE",
     year = 2018,
    month = mar,
   volume = 856,
      eid = {42},
    pages = {42},
      doi = {10.3847/1538-4357/aaad0f},
   adsurl = {http://adsabs.harvard.edu/abs/2018ApJ...856...42S}
}

@ARTICLE{sesana10,
       author = {{Sesana}, Alberto},
        title = "{Self Consistent Model for the Evolution of Eccentric Massive Black Hole Binaries in Stellar Environments: Implications for Gravitational Wave Observations}",
      journal = {\apj},
         year = 2010,
        month = aug,
       volume = {719},
       number = {1},
        pages = {851-864},
          doi = {10.1088/0004-637X/719/1/851},
archivePrefix = {arXiv},
       eprint = {1006.0730},
 primaryClass = {astro-ph.CO},
       adsurl = {https://ui.adsabs.harvard.edu/abs/2010ApJ...719..851S}
}

@ARTICLE{sesana09,
   author = {{Sesana}, A. and {Volonteri}, M. and {Haardt}, F.},
    title = "{LISA detection of massive black hole binaries: imprint of seed populations and extreme recoils}",
  journal = {Classical and Quantum Gravity},
archivePrefix = "arXiv",
   eprint = {0810.5554},
     year = 2009,
    month = may,
   volume = 26,
   number = 9,
      eid = {094033},
    pages = {094033},
      doi = {10.1088/0264-9381/26/9/094033},
   adsurl = {http://adsabs.harvard.edu/abs/2009CQGra..26i4033S}
}

@ARTICLE{sesana08,
   author = {{Sesana}, A. and {Vecchio}, A. and {Colacino}, C.~N.},
    title = "{The stochastic gravitational-wave background from massive black hole binary systems: implications for observations with Pulsar Timing Arrays}",
  journal = {\mnras},
archivePrefix = "arXiv",
   eprint = {0804.4476},
     year = 2008,
    month = oct,
   volume = 390,
    pages = {192-209},
      doi = {10.1111/j.1365-2966.2008.13682.x},
   adsurl = {http://adsabs.harvard.edu/abs/2008MNRAS.390..192S}
}

@ARTICLE{sesana15,
       author = {{Sesana}, Alberto and {Khan}, Fazeel Mahmood},
        title = "{Scattering experiments meet N-body - I. A practical recipe for the evolution of massive black hole binaries in stellar environments}",
      journal = {\mnras},
         year = 2015,
        month = nov,
       volume = {454},
       number = {1},
        pages = {L66-L70},
          doi = {10.1093/mnrasl/slv131},
archivePrefix = {arXiv},
       eprint = {1505.02062},
 primaryClass = {astro-ph.GA},
       adsurl = {https://ui.adsabs.harvard.edu/abs/2015MNRAS.454L..66S}
}

@ARTICLE{shi12,
       author = {{Shi}, Ji-Ming and {Krolik}, Julian H. and {Lubow}, Stephen H. and {Hawley}, John F.},
        title = "{Three-dimensional Magnetohydrodynamic Simulations of Circumbinary Accretion Disks: Disk Structures and Angular Momentum Transport}",
      journal = {\apj},
         year = 2012,
        month = apr,
       volume = {749},
       number = {2},
          eid = {118},
        pages = {118},
          doi = {10.1088/0004-637X/749/2/118},
archivePrefix = {arXiv},
       eprint = {1110.4866},
 primaryClass = {astro-ph.HE},
       adsurl = {https://ui.adsabs.harvard.edu/abs/2012ApJ...749..118S}
}

@ARTICLE{siwek24,
       author = {{Siwek}, Magdalena and {Kelley}, Luke Zoltan and {Hernquist}, Lars},
        title = "{Signatures of circumbinary disc dynamics in multimessenger population studies of massive black hole binaries}",
      journal = {\mnras},
         year = 2024,
        month = nov,
       volume = {534},
       number = {3},
        pages = {2609-2620},
          doi = {10.1093/mnras/stae2251},
archivePrefix = {arXiv},
       eprint = {2403.08871},
 primaryClass = {astro-ph.HE},
       adsurl = {https://ui.adsabs.harvard.edu/abs/2024MNRAS.534.2609S}
}

@ARTICLE{siwek23a,
       author = {{Siwek}, Magdalena and {Weinberger}, Rainer and {Mu{\~n}oz}, Diego J. and {Hernquist}, Lars},
        title = "{Preferential accretion and circumbinary disc precession in eccentric binary systems}",
      journal = {\mnras},
         year = 2023,
        month = feb,
       volume = {518},
       number = {4},
        pages = {5059-5071},
          doi = {10.1093/mnras/stac3263},
archivePrefix = {arXiv},
       eprint = {2203.02514},
 primaryClass = {astro-ph.HE},
       adsurl = {https://ui.adsabs.harvard.edu/abs/2023MNRAS.518.5059S}
}

@ARTICLE{siwek23b,
       author = {{Siwek}, Magdalena and {Weinberger}, Rainer and {Hernquist}, Lars},
        title = "{Orbital evolution of binaries in circumbinary discs}",
      journal = {\mnras},
         year = 2023,
        month = jun,
       volume = {522},
       number = {2},
        pages = {2707-2717},
          doi = {10.1093/mnras/stad1131},
archivePrefix = {arXiv},
       eprint = {2302.01785},
 primaryClass = {astro-ph.HE},
       adsurl = {https://ui.adsabs.harvard.edu/abs/2023MNRAS.522.2707S}
}

@ARTICLE{tang18,
       author = {{Tang}, Yike and {Haiman}, Zolt{\'a}n and {MacFadyen}, Andrew},
        title = "{The late inspiral of supermassive black hole binaries with circumbinary gas discs in the LISA band}",
      journal = {\mnras},
         year = 2018,
        month = may,
       volume = {476},
       number = {2},
        pages = {2249-2257},
          doi = {10.1093/mnras/sty423},
archivePrefix = {arXiv},
       eprint = {1801.02266},
 primaryClass = {astro-ph.HE},
       adsurl = {https://ui.adsabs.harvard.edu/abs/2018MNRAS.476.2249T}
}

@ARTICLE{tiede20,
       author = {{Tiede}, Christopher and {Zrake}, Jonathan and {MacFadyen}, Andrew and {Haiman}, Zoltan},
        title = "{Gas-driven Inspiral of Binaries in Thin Accretion Disks}",
      journal = {\apj},
         year = 2020,
        month = sep,
       volume = {900},
       number = {1},
          eid = {43},
        pages = {43},
          doi = {10.3847/1538-4357/aba432},
archivePrefix = {arXiv},
       eprint = {2005.09555},
 primaryClass = {astro-ph.GA},
       adsurl = {https://ui.adsabs.harvard.edu/abs/2020ApJ...900...43T}
}

@ARTICLE{volonteri07,
   author = {{Volonteri}, M.},
    title = "{Gravitational Recoil: Signatures on the Massive Black Hole Population}",
  journal = {\apjl},
   eprint = {arXiv:astro-ph/0703180},
     year = 2007,
    month = jul,
   volume = 663,
    pages = {L5-L8},
      doi = {10.1086/519525},
   adsurl = {http://adsabs.harvard.edu/abs/2007ApJ...663L...5V}
}

@ARTICLE{volonteri03,
   author = {{Volonteri}, M. and {Haardt}, F. and {Madau}, P.},
    title = "{The Assembly and Merging History of Supermassive Black Holes in Hierarchical Models of Galaxy Formation}",
  journal = {\apj},
   eprint = {astro-ph/0207276},
     year = 2003,
    month = jan,
   volume = 582,
    pages = {559-573},
      doi = {10.1086/344675},
   adsurl = {http://adsabs.harvard.edu/abs/2003ApJ...582..559V}
}

@ARTICLE{wyithe03a,
       author = {{Wyithe}, J. Stuart B. and {Loeb}, Abraham},
        title = "{Low-Frequency Gravitational Waves from Massive Black Hole Binaries: Predictions for LISA and Pulsar Timing Arrays}",
      journal = {\apj},
         year = 2003,
        month = jun,
       volume = {590},
       number = {2},
        pages = {691-706},
          doi = {10.1086/375187},
archivePrefix = {arXiv},
       eprint = {astro-ph/0211556},
 primaryClass = {astro-ph},
       adsurl = {https://ui.adsabs.harvard.edu/abs/2003ApJ...590..691W}
}

@ARTICLE{xu23,
       author = {{Xu}, Heng and {Chen}, Siyuan and {Guo}, Yanjun and {Jiang}, Jinchen and {Wang}, Bojun and {Xu}, Jiangwei and {Xue}, Zihan and {Nicolas Caballero}, R. and {Yuan}, Jianping and {Xu}, Yonghua and {Wang}, Jingbo and {Hao}, Longfei and {Luo}, Jingtao and {Lee}, Kejia and {Han}, Jinlin and {Jiang}, Peng and {Shen}, Zhiqiang and {Wang}, Min and {Wang}, Na and {Xu}, Renxin and {Wu}, Xiangping and {Manchester}, Richard and {Qian}, Lei and {Guan}, Xin and {Huang}, Menglin and {Sun}, Chun and {Zhu}, Yan},
        title = "{Searching for the Nano-Hertz Stochastic Gravitational Wave Background with the Chinese Pulsar Timing Array Data Release I}",
      journal = {Research in Astronomy and Astrophysics},
         year = 2023,
        month = jul,
       volume = {23},
       number = {7},
          eid = {075024},
        pages = {075024},
          doi = {10.1088/1674-4527/acdfa5},
archivePrefix = {arXiv},
       eprint = {2306.16216},
 primaryClass = {astro-ph.HE},
       adsurl = {https://ui.adsabs.harvard.edu/abs/2023RAA....23g5024X}
}

@ARTICLE{yu02,
   author = {{Yu}, Q.},
    title = "{Evolution of massive binary black holes}",
  journal = {\mnras},
   eprint = {arXiv:astro-ph/0109530},
     year = 2002,
    month = apr,
   volume = 331,
    pages = {935-958},
      doi = {10.1046/j.1365-8711.2002.05242.x},
   adsurl = {http://adsabs.harvard.edu/abs/2002MNRAS.331..935Y}
}

@ARTICLE{zrake21,
       author = {{Zrake}, Jonathan and {Tiede}, Christopher and {MacFadyen}, Andrew and {Haiman}, Zolt{\'a}n},
        title = "{Equilibrium Eccentricity of Accreting Binaries}",
      journal = {\apjl},
         year = 2021,
        month = mar,
       volume = {909},
       number = {1},
          eid = {L13},
        pages = {L13},
          doi = {10.3847/2041-8213/abdd1c},
archivePrefix = {arXiv},
       eprint = {2010.09707},
 primaryClass = {astro-ph.HE},
       adsurl = {https://ui.adsabs.harvard.edu/abs/2021ApJ...909L..13Z}
}

@ARTICLE{chen2017,
       author = {{Chen}, Siyuan and {Sesana}, Alberto and {Del Pozzo}, Walter},
        title = "{Efficient computation of the gravitational wave spectrum emitted by eccentric massive black hole binaries in stellar environments}",
      journal = {\mnras},
         year = 2017,
        month = sep,
       volume = {470},
       number = {2},
        pages = {1738-1749},
          doi = {10.1093/mnras/stx1093},
archivePrefix = {arXiv},
       eprint = {1612.00455},
 primaryClass = {astro-ph.CO},
       adsurl = {https://ui.adsabs.harvard.edu/abs/2017MNRAS.470.1738C}
}
\bibliographystyle{aasjournalv7}

\end{document}